\documentclass[acmsmall,screen,nonacm]{acmart}
\setcopyright{none}
\acmConference{}{}{}
\acmDOI{}
\acmISBN{}
\startPage{1}
\authorsaddresses{}

\fancypagestyle{paperbody}{%
  \fancyhf{}%
  \fancyhead[C]{\footnotesize\shortauthors}%
  \fancyhead[R]{\footnotesize\thepage}%
}

\usepackage{amsmath,mathtools}
\usepackage{attachfile2}
\makeatletter
\def\cake@fileentry#1{%
  \ifcsname atfi@fsobj@#1\endcsname
    (#1) \csname atfi@fsobj@#1\endcsname\ifxetex\else\space 0 R\fi
  \fi}
\AtEndDocument{%
  \edef\cake@files{\cake@fileentry{CakeCutting-Lean-v1.zip}\cake@fileentry{envy_free_cake-v1.19-supplement.pdf}}%
  \ifx\cake@files\@empty\else
    \ifxetex
    \else\ifluatex
      \pdfextension names { /EmbeddedFiles << /Names [\cake@files] >> }%
    \else
      \pdfnames{ /EmbeddedFiles << /Names [\cake@files] >> }%
    \fi\fi
  \fi}
\newcommand{\jointproofs}{%
  \IfFileExists{envy_free_cake-v1.19-supplement.pdf}{%
    \textattachfile[mimetype={application/pdf},description={Supplement on joint complexity bounds},color={0 0.2 0.45}]{\detokenize{envy_free_cake-v1.19-supplement.pdf}}{\textit{Supplement on joint complexity bounds}}%
  }{\textit{Supplement on joint complexity bounds}}}
\makeatother
\usepackage{enumitem,array,booktabs,tabularx}
\usepackage{tikz}
\usetikzlibrary{arrows.meta,patterns,shapes.geometric,calc,decorations.pathreplacing}
\usepackage[ruled,vlined,linesnumbered]{algorithm2e}
\DontPrintSemicolon
\SetAlFnt{\small}
\SetAlCapFnt{\small}
\SetAlCapNameFnt{\small}
\SetAlCapHSkip{0pt}

\SetKwInput{KwIn}{Input}
\SetKwInput{KwOut}{Output}
\SetKw{Return}{return}
\SetKwComment{tcp}{$\triangleright$\ }{}

\SetCommentSty{algcomment}
\setcitestyle{authoryear}
\hypersetup{keeppdfinfo}
\theoremstyle{acmplain}
\newtheorem{theorem}{Theorem}
\newtheorem{lemma}{Lemma}
\newtheorem{proposition}{Proposition}
\newtheorem{corollary}{Corollary}
\theoremstyle{acmdefinition}

\newtheorem{example}{Example}
\newcommand{\Core}{\textsc{Core}}
\newcommand{\Repair}{\textsc{Repair}}
\newcommand{\Trim}{\textsc{Trim}}
\newcommand{\Divide}{\textsc{Divide}}
\newcommand{\Reassign}{\textsc{Reassign}}
\newcommand{\A}{\mathcal A}
\newcommand{\cK}{\mathcal K}
\newcommand{\rr}[2]{\lceil #1\rceil_{#2}}
\newcommand{\one}{\mathbf{1}}
\newcommand{\concat}{\mathbin{+\!+}}
\DeclareRobustCommand{\stepnum}[1]{\tikz[baseline=(c.base)]\node[circle,draw,fill=blue!6,inner sep=0.4pt,minimum size=3.3mm,font=\scriptsize\sffamily\bfseries](c){#1};}
\setlist[enumerate]{itemsep=2pt,topsep=4pt}
\setlist[itemize]{itemsep=2pt,topsep=4pt}

\begin{document}
\hypersetup{linkcolor=ACMDarkBlue,citecolor=ACMDarkBlue,urlcolor=ACMDarkBlue}
\raggedbottom

\title{A Deterministic Polynomial-Query Protocol for Envy-Free Cake Cutting}
\author{Haris Aziz}
\orcid{0000-0001-9757-4956}
\affiliation{%
  \department{School of Computer Science and Engineering}
  \institution{UNSW Sydney}
  \city{Sydney}
  \state{NSW}
  \country{Australia}}
\email{haris.aziz@unsw.edu.au}
\renewcommand{\shortauthors}{Haris Aziz}
\begin{abstract}
We give a deterministic protocol that divides a heterogeneous cake completely and without envy among $n$ agents using $O(n^6\log n)$ Robertson--Webb queries in the worst case. Valuations may be arbitrary nonnegative, nonatomic measures, and bundles may be disconnected. The protocol makes $O(n)$ calls to certified trimming, which uses $O(n^5\log n)$ queries on $n$ pieces. In a certified trim, each agent receives the end part, or suffix, of a different piece without envy, and every shortened piece has a witness tie leading back to a whole piece. At each step, the witness values the shortened suffix exactly as much as her own. For pieces of positive interval length with strictly increasing suffix values, the retained suffix lengths are unique, although their holders may differ. Candidate assignments share a table of value and cut queries. If the coordinatewise median of their value estimates fails a matching test, the test identifies an agent whose value is underestimated, ruling out at least half the remaining assignments. A weighted median of cut marks lets one repair step remove a small envy while allocating almost all of the residue. For complete envy-free chore allocation, the query bound is $O(n^7\log n)$.
\end{abstract}

\maketitle
\hypersetup{pdfauthor={Haris Aziz},pdftitle={A Deterministic Polynomial-Query Protocol for Envy-Free Cake Cutting}}
\pagestyle{paperbody}

\section{Introduction}

\begin{quote}
		``\textit{Since the 1940s, the computation of envy-free cake divisions
	has baffled many great minds across multiple disciplines}.'' ---  \citet[p.~83]{procaccia2013cake} 
\end{quote}

\medskip
\label{sec:intro}
Suppose agents must divide a heterogeneous, divisible good, such as a plot of land, a stretch of broadcast time or a cake. Agents may value the same piece differently, and each should value her own share at least as highly as anyone else's. This requirement is \emph{envy-freeness}. How many questions about their values suffice to divide the whole cake in this way?

A protocol learns about values only by asking agents questions, or \emph{queries}, in the standard model of \citet{RoWe98a}. An evaluation query asks an agent for her value of an interval. A cut query asks her to mark where, starting from a given point, the cake reaches a given value to her. For two agents a few queries suffice: one agent cuts the cake into two pieces she values equally, and the other takes the piece she prefers. \citet{GaSt58} asked whether a cake can always be divided without envy among any number of agents. For three agents, Selfridge and Conway found a protocol with a bounded number of cuts \citep{RoWe98a}.

For an arbitrary number of agents, it took over twenty years to pass from a finite protocol \citep{BrTa95a} to one whose number of queries is bounded by a function of $n$ alone \citep{AzMa16a}, and that bound was a tower of six $n$'s. At that time, Procaccia~\cite[p.~118]{procaccia2020perspective} observed the astronomical bound and wrote that  ``\textit{this begs the question of whether there exists a cake-cutting
protocol that is both envy free and computationally efficient}.'' By 2026, the deterministic bound had come down to single exponential \citep{YeBai26}.
In this paper, our goal is to obtain a  polynomial-query envy-free protocol. Following \citet[pp.~450--451]{Edmonds65}, who singled out polynomial
(as opposed to exponential) growth as the mark of a `good' algorithm, we
seek protocols whose number of queries is polynomial in the number of agents. Mackenzie and Suzuki~\cite[p.~1]{MaSu26a} write,
``\textit{Whether polynomially many queries suffice is a central open
question in fair division}.'' This paper solves the open question. 
Algorithm~\ref{alg:divide} gives the following guarantee.\footnote{In the Lean archive \url{https://github.com/harisaziz/EFcake/releases/download/v1.0/CakeCutting-Lean-v1.zip}, the theorem \texttt{CakeCutting.mainClaim\_proof} establishes \texttt{CakeCutting.MainClaim} for every $n \ge 2$, all finite atomless Borel measures on $[0,1]$, and every valid cut reply. The archive also contains the verification record and build instructions.}

\begin{theorem}\label{thm:main}
For $n\ge2$ agents with finite, nonnegative, nonatomic, additive valuations, a deterministic protocol returns a complete, exactly envy-free cake allocation in finite unions of intervals using $O(n^7\log n)$ Robertson--Webb queries. The bound holds for every choice of valid cut answers.
\end{theorem}

The protocol can be implemented within the same query bound using only queries on single intervals whose endpoints are $0$, $1$, or points returned by earlier cut queries (Appendix~\ref{app:query-endpoints}).
Appendix~\ref{app:batched-cake} gives a variant that batches calls to the same cutter. It makes only $O(n)$ calls to certified trimming and uses $O(n^6\log n)$ Robertson--Webb queries in total.

For complete envy-free chore allocation with arbitrary finite, nonnegative, nonatomic costs, the worst-case query bound is $O(n^7\log n)$. Appendix~\ref{app:chores} explains how to use the trimming and repair bounds in the chore protocol of \citet{Aziz26chores}.

\paragraph{Envy-freeness and domination.}
Envy-freeness is easy to maintain for partial allocations. Giving everyone nothing is envy-free, and adding to each agent's bundle her part of an envy-free division of some of the remaining cake keeps the allocation envy-free, because values are additive. The difficulty is the cake not yet allocated, the \emph{residue}. Careful division steps tend to leave some of it, and the last of it must also be allocated.
The classical way out is an irrevocable advantage \citep{BrTa95a,AzMa16a}. Write $v_a(B)$ for agent $a$'s value of a piece $B$. Agent $a$'s \emph{advantage} over agent $b$ is $v_a(X_a)-v_a(X_b)$, the amount by which she values her own bundle $X_a$ above $b$'s bundle $X_b$; envy-freeness says exactly that every advantage is nonnegative. We say that $a$ \emph{dominates} $b$ when this advantage is at least her value of the whole residue. Then $b$ could receive all of the residue and $a$ would still not envy her, however the rest is divided. Once everyone outside a nonempty proper group of agents dominates everyone inside it, the residue can be divided among the group's members alone, recursively. The outsiders do not mind what the members receive, and the members only gain. A protocol therefore aims to create dominations, one at a time, while keeping its allocation envy-free.

\begin{figure}[t]
\centering
\begin{tikzpicture}[font=\footnotesize,
  box/.style={draw,rounded corners=2pt,fill=blue!5,align=center,inner sep=3pt,text width=#1,minimum height=1.05cm,
    execute at begin node={\hyphenpenalty=10000\exhyphenpenalty=10000}},
  box/.default=3.1cm,
  dec/.style={draw,diamond,aspect=2.2,fill=orange!10,align=center,inner sep=0.5pt},
  arr/.style={-{Stealth[length=1.7mm]},semithick},
  stc/.style={circle,draw,fill=white,inner sep=0.4pt,minimum size=4mm,font=\scriptsize\sffamily\bfseries}]
\node[box=1.9cm,fill=gray!8] (start) at (0,0) {start: nothing allocated};
\node[box] (core) at (3.5,0) {\Core{}: the cutter cuts the residue, \Trim{} allocates suffixes; the cutter gains an advantage (Sec.~\ref{sec:trim})};
\node[dec] (test) at (7.35,0) {dominated\\ group $G$?};
\node[box=2.7cm,fill=green!8] (rec) at (11.15,0) {recurse on $G$ with its current bundles and the residue};
\node[box] (rep) at (3.5,-2.3) {\Repair{}: remove the envy, allocate nearly all of the residue (Sec.~\ref{sec:repair})};
\node[box] (re) at (7.35,-2.3) {find an advantage over an undominated agent, reassigning new suffixes if needed, with a little envy (Sec.~\ref{sec:advantage})};
\node[stc] at (core.north west) {1};
\node[stc] at ($(test.north west)!0.5!(test.north)$) {2};
\node[stc] at (re.north east) {3};
\node[stc] at (rep.north east) {4};
\node[stc] at (rec.north west) {5};
\draw[arr] (start)--(core);
\draw[arr] (core)--(test);
\draw[arr] (test)--node[above]{yes}(rec);
\draw[arr] (test)--node[right]{no}(re);
\draw[arr] (re)--(rep);
\draw[arr] (rep)--node[left,align=right]{advantage becomes a\\ new domination;\\ at most $n^2$ repairs}(core);
\end{tikzpicture}
\caption{The main loop of \Divide{} (Algorithm~\ref{alg:divide}), with the step numbers used throughout. The allocation stays envy-free after every step except the reassignment \stepnum{3}, and \Repair{} \stepnum{4} restores it. Table~\ref{tab:round} states what each step guarantees.}
\Description{Flowchart of one round: Core, then a test for a dominated group; if yes, the group takes its current bundles into the recursive call; if no, find an advantage (reassigning if needed), then Repair, and loop back to Core.}
\label{fig:flow}
\end{figure}
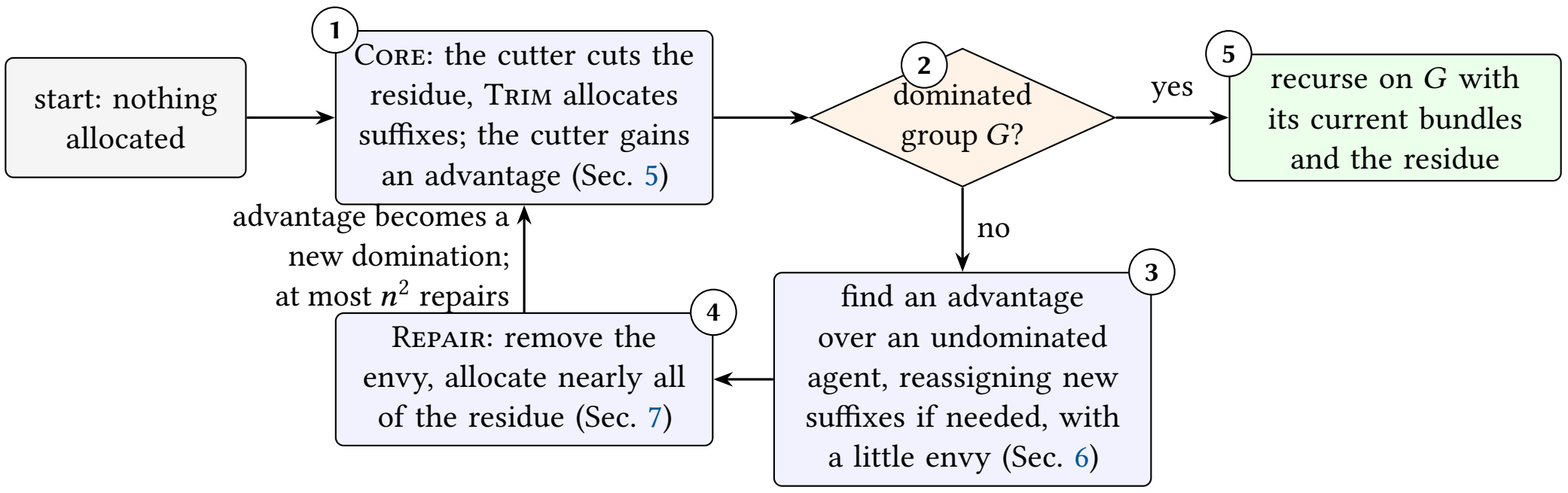

\paragraph{The protocol in brief.}
Our protocol keeps one envy-free partial allocation and repeats a round of four steps (Figure~\ref{fig:flow}) until a dominated group appears.
\begin{itemize}
\item[\stepnum{1}] In \Core{}, a fixed agent, the cutter, divides the residue into $n$ pieces she values equally, and a trimming routine gives each agent the end part of a different piece without envy. One piece stays whole, and the cutter's share is worth exactly one piece to her. If she values the new residue positively, she has a strict advantage over the holder of some trimmed piece.
\item[\stepnum{2}] If everyone outside some group now dominates everyone inside it, the group's members carry their current bundles into the recursive call and divide the residue (step~\stepnum{5}).
\item[\stepnum{3}] Otherwise the cutter may already dominate that holder, and her advantage is then wasted. We look for an advantage of some agent over a bundle she does not yet dominate. If necessary, we pass the suffixes that \Core{} has just allocated around a cycle of agents: every agent keeps her earlier bundle and receives one of the new suffixes on top of it. This may create a little envy.
\item[\stepnum{4}] \Repair{} puts $n-1$ small \emph{buffers} in front of each bundle and a large share of the residue behind it. The target for each addition is a fraction of each agent's own value of the residue. \Trim{} then gives each agent a suffix of a different one of these pieces without envy. One piece stays whole. If a cut entered a bundle, there would be too few other pieces to realize every intermediate buffer level. Each original owner above the missing level would strictly prefer the suffix on her label to every suffix below it, leaving too few agents to witness the trims below the gap (Section~\ref{sec:protect}). Thus the cuts consume only buffers. The advantage found in step~\stepnum{3} survives with only a small loss, while the residue shrinks below it. The agent now dominates whoever holds that bundle.
\end{itemize}

Old dominations follow their bundles even when the holders change, and each round that proceeds past step~\stepnum{2} increases the number of dominated pairs. Keeping the current bundles when the group shrinks preserves these dominations. There are at most $n^2$ ordered pairs in the original population, so at most $n^2$ rounds require repair over the entire execution.

\paragraph{Trimming is the bottleneck.}
Every round calls the trimming routine in \Core{}. A round that reaches \Repair{} calls it a second time. The other steps use polynomially many queries. The cutter makes $n-1$ cuts, the reassignment only reads values already recorded, and repair, apart from its call to \Trim{}, needs a number of queries logarithmic in its precision. Trimming is therefore the step that decides whether the whole protocol is polynomial. A call to \Trim{} receives $n$ ordered pieces, cuts some of them from the front, and gives each agent the remaining \emph{suffix} of a different piece, so that nobody envies anybody. Giving everyone an empty suffix would be envy-free but useless, so every trim must be justified: in a chain starting at a whole piece, each trimmed suffix is valued by an earlier holder exactly as much as her own (Figure~\ref{fig:example}). Earlier protocols find such a \emph{certified} allocation by searching over orders of agents and pieces, using exponentially many queries in the worst case \citep{AzMa16a,AzMa20a,Psom26a}.

\paragraph{Finding the trim with medians.}
Suppose the input pieces consist of intervals of positive length and every agent assigns positive value to every interval of positive length. Then all certified allocations retain the same suffixes. On every piece these are the longest suffixes possible in any envy-free suffix allocation (Proposition~\ref{prop:greatest}). Knowing one \emph{supporting assignment}, which gives the holders in some certified allocation, would suffice to recover these suffixes. Start with an infinite upper bound on each agent's value. Repeatedly cut each piece to the longest suffix, up to the whole piece, that no agent values above her current bound, and lower the bound of the piece's assigned holder to her value of what is left. Much as in the Bellman--Ford shortest-path algorithm, the bounds settle at the true values within $n$ rounds.

We do not know a supporting assignment, so we run this relaxation for all $n!$ candidate assignments simultaneously, and then discard candidates in bulk. Three observations make this possible with polynomially many queries (Figure~\ref{fig:median}).
\begin{itemize}
\item \emph{One table serves all candidates.} Different candidates ask for values and cuts at different points. We round every request \emph{up} to one of a few shared grid points, chosen as quantiles of the candidates' current bounds, so that one table of $O(n^3)$ queries answers the whole pool in each round, however many candidates remain. Rounding up only weakens an upper bound, so supporting candidates never fall below the truth.
\item \emph{The median is a lower bound.} The ties that justify the true allocation pull down the bounds of \emph{every} candidate, supporting or not: in each round, until all of them get there, at least one more of a candidate's bounds reaches the true value or falls below it. Rounding can stall this progress, when an agent's value estimate rounds above her true value or a piece's length estimate rounds above its true length. At each threshold only estimates in the grid gap containing it can cross. After $4n$ rounds a majority of the candidates are at or below the truth on every coordinate, so on each coordinate the lower median of the candidates' bounds is at most the true value.
\item \emph{A matching test names a coordinate to filter.} At the median values, a bipartite matching computation either certifies that they are the true values, or names an agent whose median value is strictly too low. Supporting candidates bound this agent's value from above, so their bounds lie strictly above the median, whereas at least half of the candidates have bounds at or below it. Those are discarded.
\end{itemize}
Each phase uses $O(n^4)$ queries, however close the candidates' values are. Each failed test eliminates at least half the pool, so $1+\log_2(n!)=O(n\log n)$ phases suffice. This gives $O(n^5\log n)$ queries per trim. Repair has the same query bound per round, and Algorithm~\ref{alg:divide} uses $O(n^2)$ rounds in total (Section~\ref{sec:loop}).

\paragraph{Repairing envy cheaply.}
The other new ingredient concerns \Repair{}, which gives each bundle a large share of the residue that every agent values nearly the same, with small \emph{buffers} placed in front. A second call to \Trim{} can then remove a small envy by consuming buffers while leaving every bundle intact. Only buffers and a small rounding error remain unallocated, so a single repair shrinks the residue enough for an advantage to become a domination. 
To build the shares, we adapt the offline $\varepsilon$-consensus splitting algorithm of \citet{AlonGraur21}. It divides an interval into $k$ parts, each worth nearly $1/k$ of each of $t$ measures, using $O(tk\log(k/\varepsilon))$ cuts. Their algorithm can query the sum of the measures. Here we have access only to individual agents' Robertson--Webb queries.
We refine only the cells of the residue that are still split between several shares, and we cut each such cell at a weighted median of the agents' half-value marks. One cut then reduces the cell's total normalized value by a constant factor on both sides, so the cost is logarithmic in the required precision.

\paragraph{Organization.}
Section~\ref{sec:related} discusses related work, and Section~\ref{sec:preliminaries} sets up the model and domination. Section~\ref{sec:glance} presents the whole protocol and proves Theorem~\ref{thm:main}, taking as given one lemma for each of its three subroutines. Sections~\ref{sec:trim}--\ref{sec:repair} prove these lemmas in the order in which the steps run. Section~\ref{sec:trim} gives the median algorithm for trimming and the cutter's guarantee. Sections~\ref{sec:advantage} and~\ref{sec:repair} cover reassignment and repair. A three-agent example runs throughout. Appendix~\ref{app:matching} proves the matching test, and Appendix~\ref{sec:ties} handles zero values and empty pieces. Appendix~\ref{app:square} states the median algorithm for general curves. Appendix~\ref{sec:trim-reductions} relates certified cake trimming to chore trimming and minimum-price equilibrium. Appendix~\ref{app:market} applies the median algorithm to assignment markets with nonlinear utilities. Appendix~\ref{app:chores} carries the trimming and repair bounds over to chores. Appendix~\ref{app:query-endpoints} gives the interval implementation, and Appendix~\ref{app:batched-cake} proves the sharper query bound for cake.

\section{Related Work}\label{sec:related}
\paragraph{Worst-case bounds for cake.}
The three-agent procedure of Selfridge and Conway uses only a bounded number of cuts \citep{RoWe98a}. For an arbitrary number of agents, the 1995 protocol of \citet{BrTa95a} terminates on every input, with no uniform bound for a fixed number of agents. In 2009, \citet{Proc09} proved a general $\Omega(n^2)$ query lower bound. In 2016, \citet{AzMa16a} gave the first bounded discrete protocol for all $n$, with an upper bound that is a tower of six $n$'s. In 2023, \citet{Sokolov23} modified and reanalyzed that protocol to obtain $n^{8n^2(1+o(1))}$ queries. In September 2026, \citet{YeBai26} reduced the bound to $n^{O(1)}2^n$ by pruning partial allocations and making polynomially many calls to their variant of \textsc{SubCore}, the trimming subroutine of \citet{AzMa16a}. The elementary protocol of \citet{Psom26a}, posted in October, uses $2^{O(n\log n)}$ queries and organizes the construction around trimming, repair and domination. Theorem~\ref{thm:main} gives a deterministic bound on every transcript. Bundles can be disconnected throughout. If each agent must receive a single interval, then for three or more agents no finite protocol guarantees a connected envy-free cake allocation \citep{Stromquist08}.

\paragraph{The parallel history for chores.}
Oskui's three-person procedures are described by \citet{RoWe98a}, and \citet{PeSu02} gave a four-person moving-knife procedure. \citet{PeSu09a} used irrevocable advantages and reserves to obtain an unbounded discrete procedure for any $n$, and \citet{DFHY18a} gave the first bounded one. \citet{FH18lower} proved an $\Omega(n\log n)$ lower bound already for proportional chore allocation. Concurrently, the November 2026 revision of \citet{FaHa26} adds a deterministic $2^{n\log_2 n+O(n)}$ chore protocol based on deferred compensation, and \citet{Aziz26chores} proposes a single-exponential one. Appendix~\ref{app:chores} shows that our trimming and repair bounds give an $O(n^7\log n)$ query bound in the chore protocol of \citet{Aziz26chores}.

\paragraph{Probabilistic bounds and the lower bound.}
\citet{Cheze26} proved a polynomial bound with high probability over a distribution of preference profiles. In October 2026, \citet{MaSu26a} gave a randomized exact protocol for every profile with $\widetilde O(n^{30})$ expected queries. Their \href{https://github.com/SimonWMackenzie/envy-free-query-exponent-five}{research note of 12 November} reports an improved bound of $\widetilde O(n^5)$ expected queries. In November, \citet{FaHa26} obtained $O(n^{12}\log n)$ expected queries for both cake and chores. They also proved a quadratic expected-query lower bound for protocols with fixed positive success probability, even when output endpoints can be chosen freely.  

\paragraph{Origins of the protocol.}
Table~\ref{tab:ingredients} summarizes which ingredients come from earlier work. The framework is that of \citet{AzMa16a,AzMa20a}: a cutter divides the residue into equal pieces, trimming allocates part of it and leaves some piece whole, domination measures progress, and a dominated group takes the residue recursively. We follow \citet{Psom26a} in organizing a round into trimming, reassignment and repair. Our repair differs in two ways: a single call protects the existing bundles and shrinks the residue enough to preserve the advantage just created, and its cost is logarithmic rather than linear in the inverse tolerance. For this we adapt the fractional rounding method of \citet{AlonGraur21} to prescribed shares and cut shared cells at weighted medians. The trimming routine is new. Its matching formulation is closely related to the extremal competitive equilibria of \citet{AJM16} (Appendix~\ref{sec:cake-minimum-equilibrium}). We compute the extremal solution with polynomially many queries by sharing grid queries among candidate assignments and filtering them with medians.

\begin{table}[t]
\caption{Ingredients and their use in this paper.}
\label{tab:ingredients}
\centering\footnotesize
\setlength{\tabcolsep}{4pt}
\begin{tabular}{@{}>{\raggedright\arraybackslash}p{0.19\linewidth}>{\raggedright\arraybackslash}p{\dimexpr0.39\linewidth-2\tabcolsep}>{\raggedright\arraybackslash}p{\dimexpr0.42\linewidth-2\tabcolsep}@{}}
\toprule
Ingredient & Earlier work & This paper\\
\midrule
Core, domination, recursion & cutter, trimming, domination as progress \citep{AzMa16a,AzMa20a} & current bundles pass into recursive calls (Sec.~\ref{sec:glance})\\
Certified trimming & \textsc{SubCore} and certified orders \citep{AzMa16a,Psom26a} & unique suffixes on positive inputs. Median filtering with shared grid queries takes $O(n^5\log n)$ queries (Sec.~\ref{sec:trim})\\
Reassignment & exchanges create an advantage \citep{AzMa16a,AzMa20a}; cake lemma \citep[Lemma~5]{Psom26a} & two-case form (Lemma~\ref{lem:reassign})\\
Repair & buffered; cost linear in the inverse tolerance \citep{Psom26a} & nearly common shares by fractional rounding \citep{AlonGraur21} with weighted-median cuts; cost logarithmic (Sec.~\ref{sec:repair})\\
\bottomrule
\end{tabular}
\end{table}

\paragraph{Medians in voting, facility location and division.}
For single-peaked preferences on a line and an odd electorate, the median peak is a majority winner \citep{Black48}. Choosing a median reported location for one facility on a line is strategyproof and minimizes total 
distance \citep{Moulin80,PT09}. 
In proportional cake cutting, \citet{EvenPaz84} split the agents at a median cut mark, and \citet{CsehFleiner20} cut at a weighted median of near-half marks, weighted by the agents' entitlements.
Trimming ranks value estimates from candidate assignments. Repair weights cut marks by each agent's normalized value of the cell.

\section{Preliminaries}
\label{sec:preliminaries}
An instance is a tuple $(N,C,(v_i)_{i\in N})$, where $N=[n]$ is a finite set of agents, $C=[0,1]$ is the cake, and each $v_i$ is a finite, nonnegative, nonatomic, additive value measure. A nonnegative integrable value density is a special case. A \textbf{piece} is a finite union of intervals. Values need not be normalized. Multiplying an agent's values by a positive constant preserves all her comparisons. Within a recursive call, $n$ denotes the number of participating agents, relabeled as $[n]$. Its input cake may be disconnected. We state asymptotic query bounds for $n\ge2$. With one agent, division and trimming both give her the whole piece.

An \textbf{allocation} $X=(X_i)_{i\in N}$ consists of pairwise disjoint pieces, one per agent. Its \textbf{residue} is $R=C\setminus\bigcup_iX_i$. The allocation is \textbf{complete} if the residue is empty. The empty piece is permitted in a partial allocation. We consistently assign endpoints using half-open intervals, except for the right endpoint of the cake. Nonatomicity makes these endpoint choices irrelevant to values.
An allocation $X$ is \textbf{envy-free} if $v_i(X_i)\geq v_i(X_j)$ for all agents $i,j\in N$.
Thus each agent receives a bundle she values most. Envy-freeness compares allocated bundles and places no restriction on how agents value the residue. For partial allocations $X$ and $Y$ on disjoint parts of the cake, write $X\oplus Y$ for the allocation giving $X_i\cup Y_i$ to each agent. Additivity implies that the union of two envy-free allocations for the same agents is envy-free.

\paragraph{Domination.} We measure progress by domination \citep{AzMa16a,AzMa20a}, a form of irrevocable advantage \citep{BrTa95a}.
In an envy-free partial allocation $X$ with residue $R$, agent $i$ \textbf{dominates} agent $j$ if $v_i(X_i)-v_i(X_j)\geq v_i(R)$. A \textbf{dominated group} is a set $\varnothing\ne G\subsetneq N$ such that every agent outside $G$ dominates every member of $G$.

If $i$ dominates $j$, giving the whole residue to $j$ still creates no envy from $i$. Domination is also permanent under the operations we use. Adding an envy-free allocation $Y$ of part of the residue preserves every existing domination: $i$'s advantage over $j$ changes by $v_i(Y_i)-v_i(Y_j)\ge0$, while her value of the residue can only fall. Section~\ref{sec:glance} shows that domination also survives when bundles change holders. Once a dominated group is found, the residue can be divided recursively among its members.

\paragraph{Queries and ordered pieces.}
An evaluation query returns $v_i([x,y])$. A cut query specifies an agent $i$, a starting point $x$, and a feasible value $w$, and returns any $y\ge x$ with $v_i([x,y])=w$ \citep{RoWe98a}. All computation uses exact arithmetic on query answers. Ties in choices made by the protocol follow fixed orders.

A piece is an ordered list of disjoint component intervals. A prefix follows this order and a suffix is what remains after removing a prefix. The order may differ from the physical order on $[0,1]$. We write $P\concat Q$ for concatenation. This lets \Repair{} place buffers before a bundle, wherever those pieces lie physically. To cut a prefix with specified value, locate its component from the values already recorded. If the target is at an endpoint, we may select that endpoint. In particular, if the cutter assigns zero mass to her whole residue, she can take the first $n-1$ pieces empty and the last piece equal to the residue. Otherwise one ordinary cut within the component suffices. A suffix mark can be found in the same way by subtracting the target from the total value.

The protocol records every active agent's value of each component interval, starting with at most $n$ evaluations for the initial totals. At each new cut to be retained, one evaluation per agent gives the value of one part, and subtraction gives the other. Including the cut itself costs at most $n+1$ queries. During \Trim{}, the input records stay fixed. Each trial suffix needs at most one query on its boundary component, and only the returned suffix boundaries enter the records for later calls.\looseness=-1

\section{The Protocol at a Glance}
\label{sec:glance}
This section presents the whole protocol and proves Theorem~\ref{thm:main}. It treats the three subroutines \Core{}, \Reassign{} and \Repair{} as black boxes. Each is described by a lemma stating exactly what the next step needs, and Sections~\ref{sec:trim}--\ref{sec:repair} prove these lemmas. Table~\ref{tab:round} summarizes one round, and Table~\ref{tab:notation} collects the main notation.

\subsection{The main loop}
\label{sec:mainloop}
\Divide{} (Algorithm~\ref{alg:divide}) starts with empty bundles and residue $C$. When a dominated group takes over the residue, its members carry their current bundles and component records into the recursive call. Everyone outside the group keeps her bundle. With one or two agents, the call divides the residue directly and adds the shares to their bundles. A larger group fixes a cutter for the call and repeats the steps of Figure~\ref{fig:flow}.

\paragraph{Labels and holders.}
Inside \Core{}, a \emph{piece label} $j$ names the cutter's piece $P_j$. Across a round of \Divide{}, a \emph{bundle label} $j$ follows the old bundle $X_j$ of agent $j$, even if it changes hands. Write $\sigma(j)$ for its final holder. The two additions to that bundle are
\begin{equation}\label{eq:life}
 X_j\ \xrightarrow{\ \stepnum{3}\ \text{add }Y_{\pi(j)}\ }\ D_j=X_j\cup Y_{\pi(j)}\ \xrightarrow{\ \stepnum{4}\ \text{add }Z_j\ }\ W_{\sigma(j)}=D_j\cup Z_j .
\end{equation}
Here $Y_i$ is the suffix that \Core{} gives \emph{agent} $i$. Reassignment sends $Y_{\pi(j)}$ to bundle label $j$, and \Repair{} adds $Z_j$ to it. At the next round the bundles are relabeled by their current holders.

\begin{algorithm}[ht]
\caption{$\Divide(N,R,X)$}\label{alg:divide}
\KwIn{A nonempty agent set $N$, an ordered residue $R$ and current envy-free bundles $X$. Initially $R=C$ and all bundles are empty.}
\KwOut{A complete envy-free allocation of $R\cup\bigcup_{i\in N}X_i$ among $N$.}
\lIf{$|N|=1$}{\Return the sole bundle with $R$ added}
\If{$|N|=2$}{Divide $R$ by cut-and-choose. \Return $X$ with each share added to its recipient's bundle\;}
Choose a cutter $r\in N$ using the fixed agent order\;
\While{true}{
  \stepnum{1}\ $(Y,U)\gets\Core(r,N,R)$\tcp*[r]{Lemma~\ref{lem:core}}
  \stepnum{2}\ \If(\tcp*[f]{graph search, Section~\ref{sec:guarantees}}){$(X\oplus Y,U)$ has a dominated group $G$}{
    \stepnum{5}\ Set $V\gets X\oplus Y$ and $W\gets\Divide(G,U,V|_G)$\;
    \Return $W$ on $G$ and $V$ on $N\setminus G$\;
  }
  \stepnum{3}\ $(a,b,\pi,\alpha)\gets\Reassign(r,X,Y,U)$, and set $D_i\gets X_i\cup Y_{\pi(i)}$ for every $i$\tcp*[r]{Lemma~\ref{lem:reassign}}
  \stepnum{4}\ $(W,\sigma,R')\gets\Repair(N,D,U,\alpha)$, and set $(X,R)\gets(W,R')$\tcp*[r]{Lemmas~\ref{lem:repair} and~\ref{lem:progress}}
}
\end{algorithm}

\begin{table}[t]
\caption{One round of \Divide. The column ``afterwards'' lists everything that the next step relies on.}
\label{tab:round}
\centering\small
\renewcommand{\arraystretch}{1.15}
\begin{tabularx}{\linewidth}{@{}>{\raggedright\arraybackslash}p{1.65cm}>{\raggedright\arraybackslash}X>{\raggedright\arraybackslash}X@{\hspace{4pt}}l@{}}
\toprule
Step & What it does & What holds afterwards & By\\
\midrule
start & --- & $X$ envy-free, with residue $R$ & ---\\[2pt]
\stepnum{1}~\Core & $\Core(r,N,R)$: the cutter $r$ cuts $R$ into $n$ pieces of equal value to her; \Trim{} gives each agent a suffix $Y_i$ of a different piece & $X\oplus Y$ envy-free, with residue $U\subseteq R$; some $h\ne r$ has $v_r(Y_r)-v_r(Y_h)\ge \frac{v_r(U)}{n-1}$ & L\ref{lem:core}\\[2pt]
\stepnum{2}~test & build the non-domination graph $\Gamma$ of $(X\oplus Y,U)$ & a dominated group $G$ exists (go to \stepnum{5}), \emph{or} $\Gamma$ is strongly connected and every $v_i(U)>0$ & \S\ref{sec:guarantees}\\[2pt]
\stepnum{3}~\Reassign & pick an arc $a\to b$ of $\Gamma$; if needed, rotate only the \emph{new} suffixes: $D_i=X_i\cup Y_{\pi(i)}$ & a little envy: $v_i(D_i)\ge v_i(D_j)-\alpha\, v_i(U)$; \newline an advantage: $v_a(D_a)-v_a(D_b)\ge (K-1)\alpha\, v_a(U)$ & L\ref{lem:reassign}\\[2pt]
\stepnum{4}~\Repair & $\Repair(N,D,U,\alpha)$: place $n-1$ buffers before each $D_j$ and a large share $T_j$ after it; \Trim{} again & $W$ envy-free; each $D_j$ intact, now held by $\sigma(j)$; additions differ by $\le 2n\alpha\, v_i(U)$; residue $v_i(R')\le 2n^2\alpha\, v_i(U)$ & L\ref{lem:repair}\\[2pt]
end & $(X,R)\gets(W,R')$ & $a$ dominates $\sigma(b)$, though she did not dominate $b$; old dominations follow their labels; so the number of dominated pairs rises & L\ref{lem:follow}, L\ref{lem:progress}\\[2pt]
\stepnum{5}~recurse & recurse on $G$ with its bundles in $X\oplus Y$ and residue $U$ & complete and envy-free & Thm~\ref{thm:main}\\
\bottomrule
\end{tabularx}
\end{table}

\subsection{What each step guarantees}
\label{sec:guarantees}

\paragraph{Step \stepnum{1}: the cutter.}
\Core{} is the Core protocol \citep{AzMa16a,AzMa20a}, with our trimming routine in place of \textsc{SubCore}. In $\Core(r,N,C)$, agent $r$ cuts $C$ into $n$ pieces of equal value to her, and \Trim{} gives each agent a suffix of a different piece without envy (Section~\ref{sec:trim}). \Trim{} uses $O(n^5\log n)$ queries (Lemma~\ref{lem:trim}). A certified trim keeps some piece whole, though not necessarily the cutter's. That piece is worth $v_r(C)/n$ to her and no suffix is worth more, so by envy-freeness her own suffix is worth exactly $v_r(C)/n$. Her advantage over the holder of any other piece is therefore her value of what was trimmed off that piece. These trimmed-off parts make up the residue $U$, so if $v_r(U)>0$ she has an advantage of at least $v_r(U)/(n-1)$ over some holder of a trimmed piece.

\begin{lemma}[Cutter's advantage; proved in Section~\ref{sec:core}]\label{lem:core}
For every agent $r$ and piece $C$, $\Core(r,N,C)$ returns an envy-free allocation $Y$ of parts of $C$, with residue $U\subseteq C$, using one call to \Trim{} and $O(n^2)$ other queries, so $O(n^5\log n)$ queries in all. The cutter's suffix satisfies $v_r(Y_r)=v_r(C)/n$. If $v_r(U)>0$, then $v_r(Y_r)-v_r(Y_h)\ge v_r(U)/(n-1)$ for some agent $h\ne r$.
\end{lemma}

\paragraph{Step \stepnum{2}: testing for a dominated group.}
The \textbf{non-domination graph} $\Gamma$ of an envy-free allocation has an arc $i\to j$ whenever $i\ne j$ and $i$ does not dominate $j$ (it is the complement of the domination graph of \citealp[Fig.~4]{AzMa20a}). A proper source strongly connected component of $\Gamma$ is a dominated group, because no arc enters it. Conversely, a dominated group has no incoming arc, so $\Gamma$ cannot be strongly connected. We can therefore find such a group, or establish that none exists, using a standard graph search. If none exists, every agent values the residue positively. Indeed, an agent with zero value for it dominates everyone else, and the other agents then form a dominated group.

\paragraph{Step \stepnum{3}: a new advantage.}
The cutter's advantage may be useless for progress, because she may already dominate the agent $h$ of Lemma~\ref{lem:core}. Step~\stepnum{3} finds an advantage of some agent $a$ over the bundle on a label $b$ that she does not yet dominate. It may pass the new suffixes around a cycle to do so, at the price of a little envy, measured by a tolerance $\alpha$. With $K=4n^2$, the advantage is at least $K-1$ times this envy, which is what step~\stepnum{4} needs.

\begin{lemma}[A useful reassignment; proved in Section~\ref{sec:advantage}]\label{lem:reassign}
Let $(X,R)$ be envy-free and let $(Y,U)=\Core(r,N,R)$. If $(X\oplus Y,U)$ has no dominated group, then $v_i(U)>0$ for every $i$, and $\Reassign$, using only recorded values, finds agents $a,b$ with $a$ not dominating $b$ in $(X,R)$, a permutation $\pi$, and $K^{-(n+1)}\le\alpha\le K^{-2}$ such that the bundles $D_i=X_i\cup Y_{\pi(i)}$ satisfy
\begin{equation}\label{eq:reassign-output}
 \begin{aligned}
 v_i(D_i)&\ge v_i(D_j)-\alpha v_i(U) &&(i,j\in N),\\
 v_a(D_a)-v_a(D_b)&\ge(K-1)\alpha v_a(U).
 \end{aligned}
\end{equation}
\end{lemma}

\paragraph{Step \stepnum{4}: repair.}
\Repair{} receives the slightly envious bundles $D_j$ and the residue $U$. It removes the envy without cutting into any $D_j$, and allocates nearly all of $U$, adding nearly the same value to every bundle in every agent's eyes. Its output may move a bundle to a new holder.

\begin{lemma}[Repair; proved in Section~\ref{sec:repair}]
Let $D$ be a partial allocation with residue $U$, and let $0<\alpha\le1/(8n^2)$.\label{lem:repair} Suppose $v_i(U)>0$ and $v_i(D_i)\ge v_i(D_j)-\alpha v_i(U)$ for all $i,j$. Then $\Repair(N,D,U,\alpha)$ returns an envy-free allocation $W$, a bijection $\sigma$ from labels to final holders, and disjoint additions $Z_j\subseteq U$ such that $W_{\sigma(j)}=D_j\cup Z_j$. Its residue $R'$ satisfies, for every agent $i$ and labels $j,l$,
\[
 v_i(R')\le 2n^2\alpha\, v_i(U),\qquad
 \bigl|v_i(Z_j)-v_i(Z_l)\bigr|\le 2n\alpha\, v_i(U).
\]
The procedure uses one call to \Trim{} and $O(n^4\log(n^4/\alpha))$ other queries.
\end{lemma}

\subsection{Domination follows its label}
\label{sec:test}
The same agent need not keep a bundle throughout the protocol. We therefore follow the \emph{label} of each old bundle when counting progress. The next lemma shows that a domination of a bundle passes to whoever holds it next, as long as the bundle only grows by cake taken from the residue.

\begin{lemma}[Domination follows its label]\label{lem:follow}
Let $(X,R)$ be envy-free, and let $W$ be an envy-free allocation with residue $R'\subseteq R$. Suppose that, for a bijection $\sigma$, each bundle $W_{\sigma(j)}$ consists of $X_j$ and pieces of $R\setminus R'$. If agent $i$ dominates $j$ in $(X,R)$, then $i$ dominates $\sigma(j)$ in $(W,R')$.
\end{lemma}
\begin{proof}
Envy-freeness and the inclusion $X_i\subseteq W_{\sigma(i)}$ give $v_i(W_i)\ge v_i(W_{\sigma(i)})\ge v_i(X_i)$. Since $W_{\sigma(j)}$ consists of $X_j$ and pieces of $R\setminus R'$, we obtain
\[
 v_i(W_i)-v_i(W_{\sigma(j)})
 \geq v_i(X_i)-v_i(W_{\sigma(j)})
 \geq v_i(X_i)-v_i(X_j)-v_i(R\setminus R').
\]
If $i$ dominates $j$ in $(X,R)$, the right-hand side is at least $v_i(R)-v_i(R\setminus R')=v_i(R')$.
\end{proof}
Distinct labels have distinct holders, so no agent's number of dominated agents decreases when bundles change hands.

\subsection{Progress, and the proof of Theorem~\ref{thm:main}}
\label{sec:loop}
Each round turns the advantage from step~\stepnum{3} into a domination. The choice $K=4n^2$ leaves enough of the advantage to cover the final residue. From agent $a$'s perspective, in units of $\alpha v_a(U)$, the bounds of Lemmas~\ref{lem:reassign} and~\ref{lem:repair} are
\begin{center}\small
\begin{tabular}{@{}ll@{}}
\toprule
Quantity & Bound\\
\midrule
Advantage before repair & at least $K-1$\\
Possible loss from unequal additions & at most $2n$\\
Remaining residue & at most $2n^2$\\
\bottomrule
\end{tabular}
\end{center}
Since $K-1-2n\ge2n^2$ for $n\ge2$, the advantage becomes a domination. We count it alongside the dominations inherited by the final holders of the old bundles.

\begin{lemma}[Progress in one round]\label{lem:progress}
Every iteration of \Divide{} that proceeds past step~\stepnum{2} preserves envy-freeness and strictly increases the number of dominated pairs
\[
 \Phi(X,R)=\bigl|\{(i,j)\in N^2:
 v_i(X_i)-v_i(X_j)\ge v_i(R)\}\bigr|.
\]
Consequently, before recursing, \Divide{} makes $O(n^2)$ calls each to \Core{} and \Repair{}.
\end{lemma}
\begin{proof}
\emph{Repair applies and restores envy-freeness.} Lemma~\ref{lem:reassign} gives exactly the input envy bound of Lemma~\ref{lem:repair}, with the same $\alpha$. It also gives $v_i(U)>0$ and $\alpha\le K^{-2}\le1/(8n^2)$. Thus repair returns an envy-free allocation $W$ with residue $R'$ and keeps each old bundle on its label, as in \eqref{eq:life}.

\emph{The surviving advantage creates a new domination.} Let $\sigma(j)$ be the final holder of label $j$. Agent $a$ may change labels, but envy-freeness gives $v_a(W_a)\ge v_a(W_{\sigma(a)})$. Hence Lemmas~\ref{lem:reassign} and~\ref{lem:repair} imply
\[
 \begin{aligned}
 v_a(W_a)-v_a(W_{\sigma(b)})
 &\ge v_a(D_a)-v_a(D_b)+v_a(Z_a)-v_a(Z_b)\\
 &\ge(K-1-2n)\alpha v_a(U)\\
 &\ge2n^2\alpha v_a(U)\ge v_a(R').
 \end{aligned}
\]
Agent $a$ therefore dominates the final holder of label $b$, although she did not dominate $b$ in $(X,R)$.

\emph{The domination count increases.} Each final bundle $W_{\sigma(j)}=X_j\cup Y_{\pi(j)}\cup Z_j$ consists of $X_j$ and pieces of $R\setminus R'$. By Lemma~\ref{lem:follow}, the map $(i,j)\mapsto(i,\sigma(j))$ sends old dominated pairs to final dominated pairs. It is injective because $\sigma$ is a bijection. The final pair $(a,\sigma(b))$ lies outside its image, since its only possible preimage is $(a,b)$ and $a$ did not dominate $b$ in $(X,R)$. Hence $\Phi(W,R')\ge\Phi(X,R)+1$. The count includes self-pairs and is at most $n^2$. Each iteration that proceeds past step~\stepnum{2} calls each subroutine once, and the returning iteration makes one final \Core{} call.
\end{proof}

\begin{proof}[Proof of Theorem~\ref{thm:main}]
\phantomsection\label{par:retained-bundles}
Call an agent \emph{frozen} once a recursive call proceeds without her. We show that the full allocation is envy-free and every frozen agent dominates every active agent after each completed update. Initially all bundles are empty and all agents are active. Write $X,R$ and $W,R'$ for the full allocations and residues before and after an update. By~\eqref{eq:life}, each active label $j$ passes to a holder $\sigma(j)$ with $W_{\sigma(j)}=X_j\cup B_j$ for some $B_j\subseteq R\setminus R'$. For a \Core{} step alone or the final division, take $\sigma$ to be the identity. Among the active agents, adding the envy-free shares from \Core{} or the final division preserves envy-freeness by additivity, and Lemma~\ref{lem:repair} gives it after a repaired round. Hence $v_i(W_i)\ge v_i(W_{\sigma(i)})\ge v_i(X_i)$, so an active agent $i$ still prefers her bundle weakly to every frozen bundle. For a frozen agent $f$ and an active label $j$, prior domination gives
\[
 v_f(W_f)-v_f(W_{\sigma(j)})=v_f(X_f)-v_f(X_j)-v_f(B_j)
 \ge v_f(R)-v_f(B_j)\ge v_f(R').
\]
When the active set shrinks, the group test gives the same domination for each newly frozen agent. Comparisons between frozen agents stay fixed. The final one- or two-agent call allocates the entire residue.

Let $\Phi$ count dominated pairs among the original $n$ agents throughout, as in Lemma~\ref{lem:progress}. Extend $\sigma$ by the identity on frozen agents. Lemma~\ref{lem:follow} maps old dominated pairs injectively to final ones by $(i,j)\mapsto(i,\sigma(j))$. Each repaired round adds the pair $(a,\sigma(b))$ outside this image, by the proof of Lemma~\ref{lem:progress} on the active set. A round that shrinks the active set cannot decrease $\Phi$. Thus there are at most $n^2$ repaired rounds in total and at most $n-1$ rounds that shrink the active set.

\emph{Counting queries.} In a round with $m\ge3$ active agents, Lemma~\ref{lem:reassign} gives $\log(1/\alpha)\le(m+1)\log(4m^2)$. The cutter uses one certified trim and $O(m^2)$ queries for its cuts and records. Repair uses one certified trim and $O(m^4\log(m^4/\alpha))$ other queries by Lemma~\ref{lem:repair}. A round thus costs $O(m^5\log m)$ queries, including evaluations of its retained boundaries. Reassignment, the domination test and cycle searches use the recorded table. The base cases use the component records and at most a constant number of additional queries. Including the $O(n)$ initial evaluations, the overall bound is
\[
 O(n)+(n^2+n-1)\,O(n^5\log n)=O(n^7\log n).
\]
Only retained boundaries enter later calls (Section~\ref{sec:preliminaries}), so there are polynomially many physical interval records. The direct trimming implementation contributes at most $n!\,n^{O(1)}$ exact operations over the full protocol, and all remaining arithmetic is polynomial in the records and $n$.
\end{proof}

\subsection{The running example, one round}
\label{sec:example-round}

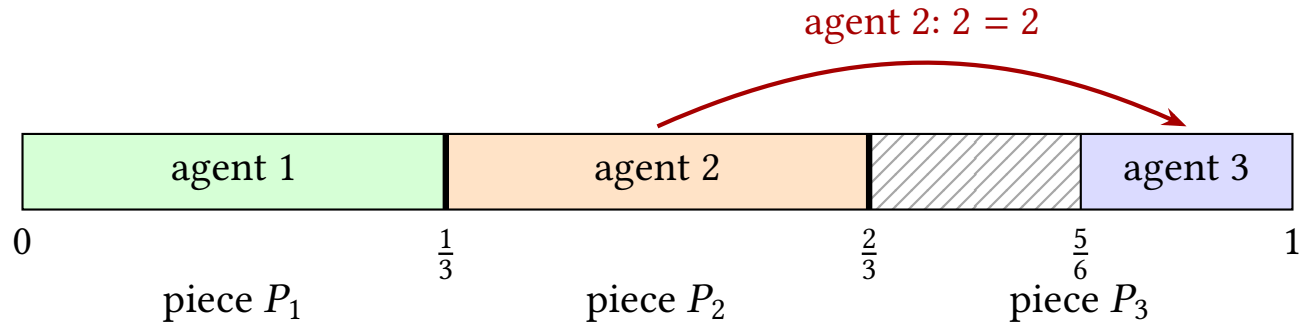
\begin{figure}[t]
\centering
\begin{minipage}[c]{0.27\textwidth}
\centering\footnotesize
\begin{tabular}{@{}lccc@{}}
\toprule
 & $P_1$ & $P_2$ & $P_3$\\
\midrule
agent 1 & 1 & 1 & 1\\
agent 2 & 1 & 2 & 4\\
agent 3 & 1 & 2 & 5\\
\bottomrule
\end{tabular}
\end{minipage}\hfill
\begin{minipage}[c]{0.7\textwidth}
\centering
\begin{tikzpicture}[x=8.4cm,y=1cm,font=\footnotesize]
\fill[green!16] (0,0) rectangle (1/3,0.5);
\fill[orange!22] (1/3,0) rectangle (2/3,0.5);
\fill[pattern=north east lines,pattern color=gray!70] (2/3,0) rectangle (5/6,0.5);
\fill[blue!14] (5/6,0) rectangle (1,0.5);
\draw (0,0) rectangle (1,0.5);
\foreach \x in {1/3,2/3,5/6} \draw (\x,0)--(\x,0.5);
\foreach \x in {1/3,2/3} \draw[very thick] (\x,0)--(\x,0.5);
\node at (1/6,0.25) {agent 1};
\node at (1/2,0.25) {agent 2};
\node at (11/12,0.25) {agent 3};
\foreach \x/\l in {0/0,0.3333/\frac13,0.6667/\frac23,0.8333/\frac56,1/1} \node[below] at (\x,0) {$\l$};
\foreach \x/\l in {0.1667/1,0.5/2,0.8333/3} \node at (\x,-0.62) {piece $P_\l$};
\draw[-{Stealth[length=1.6mm]},thick,red!65!black] (1/2,0.55) to[bend left=24] node[above]{agent 2: $2=2$} (11/12,0.55);
\end{tikzpicture}
\end{minipage}
\caption{The running example. Left: each agent's value of each third, spread evenly within it. Right: the allocation $Y$ computed by \Core{} in the first round, with agent~1 as the cutter. Agents~1 and~2 hold $P_1$ and $P_2$ whole, and agent~3 holds the last half of $P_3$. The arrow is the tie that justifies the trim (Section~\ref{sec:certified}), pointing from an earlier holder to the trimmed suffix that she values exactly as much as her own piece. The hatched half forms the residue $U$.}
\Description{A table of each agent's value for three thirds of the cake, and a bar showing Core's allocation: agent 1 holds the first third whole, agent 2 the second third whole, agent 3 the last half of the final third; an arrow from agent 2's piece to the trimmed half shows the tie that justifies the trim.}
\label{fig:example}
\end{figure}

Three agents divide $C=[0,1]$, with values spread evenly within each third and totals as in Figure~\ref{fig:example}. We follow the first round, step by step.

\begin{itemize}
\item[\stepnum{1}] Agent~1 is the cutter, and her equal-value pieces are the thirds $P_1,P_2,P_3$. Agents~2 and~3 both value $P_3$ most, so no allocation of whole thirds is envy-free. \Core{} keeps $P_1$ and $P_2$ whole and the last half of $P_3$ (Figure~\ref{fig:example}; Section~\ref{sec:trim} shows how \Trim{} finds this). The residue $U$ is the first half of $P_3$, worth $1/2$, $2$ and $5/2$ to the three agents. Agent~1 values her own piece above agent~3's half by $1/2$, twice the bound $v_1(U)/2$ of Lemma~\ref{lem:core}.
\item[\stepnum{2}] Agent~1 dominates agent~3, since her advantage of $1/2$ equals $v_1(U)$. Nobody else dominates anybody. For instance, agent~3's largest advantage, $3/2$ over agent~1, is below $v_3(U)=5/2$, and agent~2's, $1$ over agent~1, is below $v_2(U)=2$. So $\Gamma$ has every arc except $1\to3$, and it is strongly connected through the cycle $1\to2\to3\to1$.
\item[\stepnum{3}] Here $K=36$. The cutter's advantage is over agent~3, whom she already dominates, so it cannot create a new domination. The arcs $2\to1$, $3\to1$ and $3\to2$, however, carry large advantages, so \Reassign{} moves nothing (case~1 in Section~\ref{sec:advantage}). Taking the first arc in a fixed order, it selects $a=2$, $b=1$, the identity permutation and $\alpha=K^{-2}=1/1296$; the other two arcs would give the same permutation and tolerance. Thus $D=Y$: label~1 carries $P_1$, label~2 carries $P_2$, and label~3 carries agent~3's half of $P_3$.
\item[\stepnum{4}] \Repair{} adds to each label a large share and two buffers, which together are worth nearly a third of each agent's value of $U$ (Figure~\ref{fig:example-repair}). By Lemma~\ref{lem:repair}, the additions differ by at most $2n\alpha=1/216$ of each agent's value of $U$, and the new residue is worth at most $2n^2\alpha=1/72$ of that value. Since $v_2(U)=2$, agent~2's advantage of $1$ over label~1 becomes a domination:
\[
 v_2(W_2)-v_2(W_{\sigma(1)})\ \ge\ 1-\tfrac1{108}\ >\ \tfrac1{36}\ \ge\ v_2(R').
\]
In units of $v_3(U)$, agent~3's smaller advantage over labels~1 and~2 is $(5/2-2)/(5/2)=1/5$. After repair, label~3 is still ahead of both by at least $1/5-1/216>1/72$. Envy-freeness therefore gives her label~3, and she dominates both other holders. The displayed inequality excludes agent~2 from label~1, so she holds label~2 and agent~1 holds label~1. Agent~1's earlier domination of agent~3 persists by Lemma~\ref{lem:follow} applied to $(Y,U)$.
\item[\stepnum{5}] Hence $\{1\}$ is a dominated group: agents~2 and~3 both dominate agent~1. Adding the envy-free allocation of the next \Core{} call preserves these dominations, so the test in the second round finds a dominated group, and \Divide{} recurses after that call.
\end{itemize}

\section{Trimming and the Cutter}
\label{sec:trim}
This section proves Lemma~\ref{lem:core} (step~\stepnum{1}). Its main part is the trimming routine \Trim{}, which \Core{} uses in step~\stepnum{1} and \Repair{} uses again in step~\stepnum{4}. Trimming a piece of cake makes it less attractive to everyone. We therefore seek an envy-free allocation in which each trim is justified by an indifference: some earlier holder values the trimmed suffix exactly as much as her own.

The section is organized as follows. Section~\ref{sec:certified} defines certified allocations and gives a counting test for them, and Section~\ref{sec:unique} shows that their suffixes are unique on positive inputs. Section~\ref{sec:matching-view} turns the counting test into a matching condition by adding one extra node, the \emph{pin}, that stands for ``whole''. Section~\ref{sec:pin} restates this condition in terms of the agents' values, and Section~\ref{sec:median} computes the certified values with the median algorithm. Section~\ref{sec:core} returns to the cutter.

\subsection{Certified suffix allocations}
\label{sec:certified}
A \emph{suffix allocation} gives each agent a suffix of a different labeled piece. A \emph{label} names the input piece, and its \emph{holder} is the agent who receives the suffix. We index suffixes by label: $S_j$ is the suffix on label $j$, and $\sigma(j)$ is its holder, so agent $\sigma(j)$'s bundle is $S_j$. A suffix is \emph{trimmed} if some of its piece was cut off, and \emph{whole} otherwise. The allocation is \emph{certified} if it is envy-free and the holders can be ordered so that each receives a whole piece or a suffix that some earlier holder values exactly as much as the suffix held by that earlier holder. Following these indifferences backwards leads to a whole piece.

\paragraph{Running example.} Inside \Core{}, label~$j$ denotes the cutter's $j$th piece. In Figure~\ref{fig:example} this is the $j$th third, $P_j$. (Inside \Repair{}, and in Section~\ref{sec:example-round}, labels are old bundles instead.) The certified allocation keeps $P_1$ and $P_2$ whole and the last half of $P_3$. Agents~1 and~2 hold labels~1 and~2 whole. Agent~3 holds label~3, and agent~2 values its half exactly as much as $P_2$. So the order $1,2,3$ certifies the allocation.

\paragraph{A counting test.} Whether an allocation is certified can be read off from the suffixes alone. For an envy-free suffix allocation $S$, let $\mathcal{M}_i(S)$ be the set of labels $j$ maximizing $v_i(S_j)$, agent $i$'s \emph{favorite} labels. The next lemma characterizes certified suffix allocations. 

\begin{lemma}[Certification test]\label{lem:test}
An envy-free suffix allocation $S$ is certified if and only if every nonempty set $H$ of trimmed labels satisfies
\begin{equation}\label{eq:escape}
 \bigl|\{i:\mathcal{M}_i(S)\cap H\ne\varnothing\}\bigr|>|H|.
\end{equation}
\end{lemma}
In words, every set of trimmed labels contains a favorite label of more agents than it has labels.
\begin{proof}
Suppose $S$ is certified, and fix $H$. By envy-freeness, the $|H|$ holders of $H$-labels are counted on the left of \eqref{eq:escape}. In a certifying order, the earliest holder of an $H$-label holds a trimmed suffix, so an earlier holder values it as much as her own suffix, and it is one of her favorites. She holds a label outside $H$ and is counted too, giving the strict inequality. Conversely, draw an arc from each holder to the holders of the labels she ties with her own. If some holders cannot be reached from a whole piece, their labels form a nonempty set $H$ of trimmed labels that no arc enters. An agent with a favorite label in $H$ must then hold a label in $H$, so only $|H|$ agents are counted, contrary to \eqref{eq:escape}. Ordering holders by distance from a whole piece now gives a certificate.
\end{proof}
If all $n$ labels were trimmed, taking $H$ to be all labels in \eqref{eq:escape} would require more than $n$ agents. Thus at least one piece stays whole. In the running example, $\mathcal{M}_1=\{1,2\}$, $\mathcal{M}_2=\{2,3\}$ and $\mathcal{M}_3=\{3\}$. The only nonempty set of trimmed labels is $H=\{3\}$, which contains a favorite label of agents~2 and~3: two agents for one label.

\begin{lemma}[Polynomial-query certified trimming]\label{lem:trim}
Given each agent's component values on $n$ disjoint ordered pieces, \Trim{} returns an envy-free suffix allocation and a certifying order in $O(n^5\log n)$ deterministic Robertson--Webb queries. The bound includes empty pieces and zero-valued intervals.
\end{lemma}

Call an input \emph{positive} if each piece consists of one or more intervals of positive length and every agent assigns positive value to every interval of positive length. Each piece then has positive total interval length. We first prove the bound on these inputs. Appendix~\ref{sec:ties} handles general valuations and pieces, including isolated endpoint components, by running the algorithm on virtual positive curves and projecting its certificate back to the physical pieces.

\subsection{Uniqueness of certified suffixes}
\label{sec:unique}
On positive inputs, all certified allocations have the same suffixes, although two agents with the same values may exchange them, as in the example below. This gives the algorithm a single vector of suffixes to find. Without positivity this fails: if two identical agents value each of two pieces at one and a positive-length prefix of the second at zero, cuts at different points in that prefix give distinct suffixes, each worth exactly as much as the whole first piece to both agents.

\begin{proposition}[Greatest suffixes]\label{prop:greatest}
Assume positive inputs, and let $S$ be a certified allocation. On every label, $S$ has the longest suffix of any envy-free suffix allocation of the same pieces to the same agents, whoever holds it. Consequently, all certified allocations have the same suffixes, though possibly different holders.
\end{proposition}

\begin{proof}
Let $S'$ be an envy-free suffix allocation, and let $H$ be the set of labels whose suffixes are longer in $S'$ than in $S$. If $H$ is nonempty, all its labels are trimmed in $S$, so Lemma~\ref{lem:test} gives more than $|H|$ agents with a favorite label in $H$. For any such agent $i$, take a favorite $k\in H$. For every $l\notin H$, strict growth gives
\[
 v_i(S'_k)>v_i(S_k)\ge v_i(S_l)\ge v_i(S'_l).
\]
Envy-freeness of $S'$ therefore requires each of these agents to hold a label in $H$, but there are only $|H|$ such labels. Thus no suffix is longer in $S'$. Applying this conclusion in both directions to two certified allocations gives equality on every label.
\end{proof}

\paragraph{Two agents.} Restrict the running example to agents~2 and~3 and pieces $P_1,P_3$. Both value $P_3$ most, so one takes $P_1$ whole and the other a suffix of $P_3$ that the first values exactly as much as $P_1$: the last quarter if agent~2 holds $P_1$, or the last fifth if agent~3 does. Only the quarter works: if agent~2 held the fifth, she would envy agent~3, since the fifth is worth $4/5<1$ to her. The quarter is also the longer suffix, as Proposition~\ref{prop:greatest} predicts. On $P_1,P_2$ instead, the two agents have the same values, so the last half of $P_2$ is worth exactly as much as $P_1$ to both, and either agent can hold either piece: two certified allocations with the same suffixes.

\subsection{Certification as a matching condition}
\label{sec:matching-view}
Lemma~\ref{lem:test} is a counting condition of the kind that appears in Hall's marriage theorem, and it becomes a matching condition once whole pieces are marked explicitly. Let $S$ be an envy-free suffix allocation, and form its \emph{tie graph}. This bipartite graph has a node for each label on one side and, on the other side, a node for each agent and one extra node $o$, the \emph{pin}. Join label $j$ to agent $i$ if $j$ is one of her favorite labels, and join $j$ to the pin if $S_j$ is whole. The holders form a perfect matching between labels and agents, which leaves the pin uncovered.

Now count neighbors in this graph. By Lemma~\ref{lem:test}, $S$ is certified exactly when every nonempty set of trimmed labels has more neighbors than members. A set of labels that contains a whole label has this property anyway, since its neighbors include its holders and the pin. Thus $S$ is certified exactly when \emph{every} nonempty set of labels has more neighbors than members. By Hall's theorem, this holds exactly when, for each node $c$ among the agents and the pin, the labels can be matched to the other $n$ nodes along edges of the tie graph. Indeed, the strict neighbor inequality survives the deletion of any one node as Hall's weak inequality. Conversely, if such a matching exists after every deletion, choose one neighbor $c$ of a fixed nonempty set of labels and apply Hall's inequality after deleting $c$ to recover the strict inequality.

\begin{figure}[t]
\centering
\begin{tikzpicture}[font=\footnotesize,
  lab/.style={draw,rounded corners=2pt,fill=orange!14,minimum width=2.05cm,minimum height=5mm,inner sep=1pt},
  agt/.style={circle,draw,thick,fill=white,minimum size=6mm,inner sep=0pt},
  pinn/.style={draw,thick,fill=green!16,minimum size=5.5mm,inner sep=0pt},
  tie/.style={semithick,gray!55},
  hold/.style={very thick,blue!70!black},
  alt/.style={very thick,densely dashed,red!70!black}]
\foreach \dx/\p/\ttl in {0/a/{(a) The holders leave the pin uncovered},6.4/b/{(b) A matching that leaves agent~3 uncovered}}{
\begin{scope}[xshift=\dx cm]
  \node[anchor=west,font=\footnotesize\bfseries] at (-1.05,3.95) {\ttl};
  \node[lab] (\p-P1) at (0,2.2) {$P_1$, whole};
  \node[lab] (\p-P2) at (0,1.1) {$P_2$, whole};
  \node[lab] (\p-P3) at (0,0)   {$P_3$, last half};
  \node[pinn] (\p-o) at (2.9,3.3) {$o$};
  \node[agt] (\p-1) at (2.9,2.2) {1};
  \node[agt] (\p-2) at (2.9,1.1) {2};
  \node[agt] (\p-3) at (2.9,0)   {3};
  \draw[tie] (\p-P1.east)--(\p-1) (\p-P1.east)--(\p-o) (\p-P2.east)--(\p-1) (\p-P2.east)--(\p-2)
             (\p-P2.east)--(\p-o) (\p-P3.east)--(\p-2) (\p-P3.east)--(\p-3);
\end{scope}}
\draw[hold] (a-P1.east)--(a-1) (a-P2.east)--(a-2) (a-P3.east)--(a-3);
\node[anchor=west] at (3.3,2.2) {value $1$};
\node[anchor=west] at (3.3,1.1) {value $2$};
\node[anchor=west] at (3.3,0)   {value $\tfrac52$};
\draw[hold,blue!45] (b-P2.east)--(b-2) (b-P3.east)--(b-3);
\draw[alt] (b-P1.east)--(b-1) (b-P2.east)--(b-o) (b-P3.east)--(b-2);
\node[agt,fill=gray!25] at (b-3) {3};
\end{tikzpicture}
\caption{The tie graph of the certified allocation of the running example (Figure~\ref{fig:example}). Each piece is joined to the agents for whom it is a favorite and, if it is whole, to the pin $o$. (a) The holders (blue) form a perfect matching that leaves the pin uncovered; the values are the agents' favorite values. (b) A matching (dashed red) that leaves agent~3 uncovered. With the holders' edges it forms the alternating path $3,P_3,2,P_2,o$: agent~3's half of $P_3$ is tied by agent~2, whose piece $P_2$ is whole. Read backwards, this is the certificate of agent~3's trim. The allocation is certified exactly because such a matching exists for every agent and for the pin.}
\Description{Two copies of a bipartite graph between the pieces P1, P2, P3 and the agents 1, 2, 3 together with a pin node o. P1 is joined to agent 1 and the pin, P2 to agents 1 and 2 and the pin, P3 to agents 2 and 3. The left copy highlights the holders' matching P1-1, P2-2, P3-3. The right copy highlights a matching P1-1, P2-o, P3-2 that leaves agent 3 uncovered.}
\label{fig:tie-graph}
\end{figure}
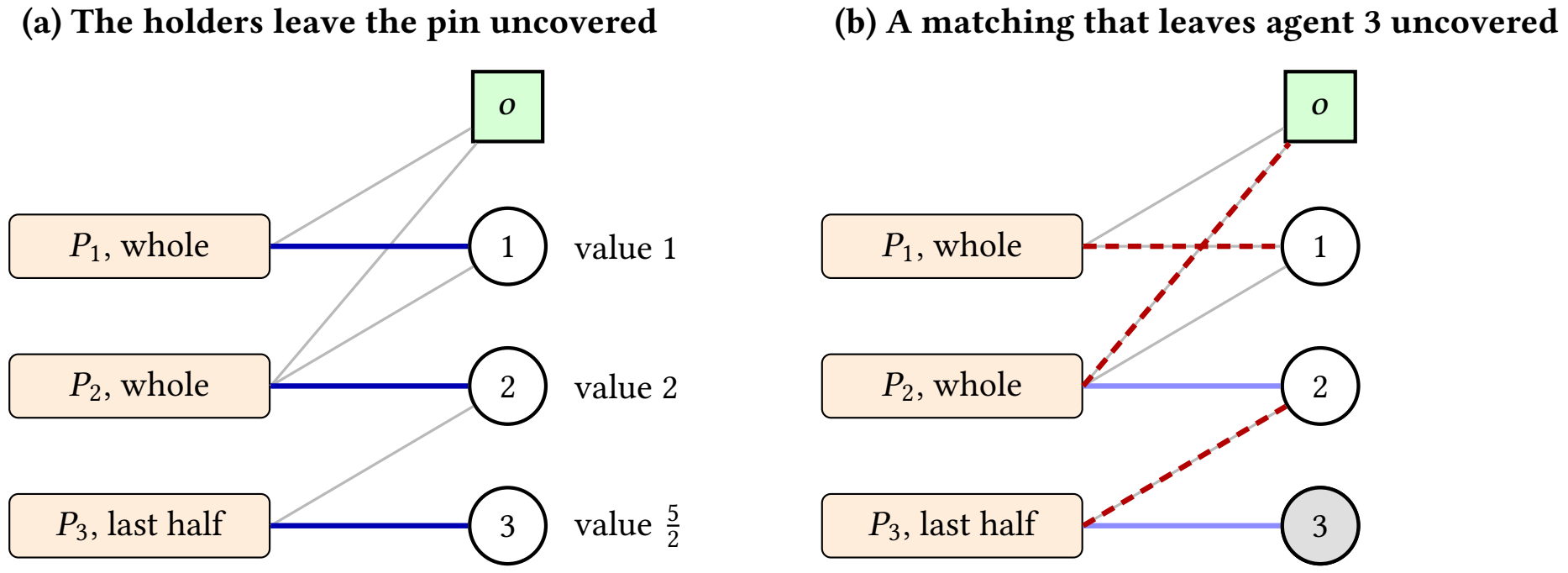

Such matchings are certificates. Superpose the holders' matching with a matching that leaves agent $i$ uncovered, and alternate between the two, starting from $i$. Agent $i$'s label is joined to a second agent, who values it exactly as much as her own suffix; that agent's label is joined to a third, and so on, until we reach a label joined to the pin, that is, a whole piece. Read backwards, this alternating path is a chain of ties from a whole piece to $i$'s suffix. Figure~\ref{fig:tie-graph} shows the tie graph of the running example. The median algorithm searches for values of the agents at which a graph of this kind admits all of these matchings.

\subsection{From suffixes to values}
\label{sec:pin}
The median algorithm searches over the agents' values rather than over suffixes, because values determine suffixes. Let $V_{ij}(t)$ be agent $i$'s value of the suffix of piece $j$ that keeps the fraction $t$ of the piece's total interval length, so that $t=0$ is the empty suffix and $t=1$ is the whole piece. On positive inputs $V_{ij}$ is continuous and strictly increasing. We extend it beyond $t=1$ with slope one, so that it becomes a continuous strictly increasing bijection of $[0,\infty)$ onto itself. Agent $i$'s \emph{mark} on piece $j$ at value $u\ge0$ is $V_{ij}^{-1}(u)$. A mark at most one is the length of a physical suffix worth $u$ to her. A mark greater than one means that even the whole piece is worth less than $u$. A \emph{forward query} returns $V_{ij}(t)$, and an \emph{inverse query} returns a mark $V_{ij}^{-1}(u)$. Each costs at most one Robertson--Webb query, an evaluation or a cut: the recorded component values locate the boundary interval of the ordered piece, and the slope-one extension needs only arithmetic.

\paragraph{Lengths from values.}
A \emph{value vector} $x\in[0,\infty)^N$ proposes a value $x_i$ for every agent $i$. If agent $i$ is to receive $x_i$, she would envy anyone holding a suffix of piece $j$ longer than her mark. So piece $j$ can keep at most the shortest mark on it, and at most the whole piece:
\begin{equation}\label{eq:recover-cake}
 t_j(x)=\min\Bigl\{1,\ \min_{i\in N}V_{ij}^{-1}(x_i)\Bigr\}.
\end{equation}
Agent $i$ is \emph{tight} on piece $j$ at $x$ if her mark attains this minimum, and piece $j$ is \emph{whole} at $x$ if $t_j(x)=1$. Since $t_j(x)$ is at most every mark,
\begin{equation}\label{eq:value-bound}
 V_{ij}\bigl(t_j(x)\bigr)\le x_i\quad\text{for every agent $i$ and piece $j$, with equality exactly when $i$ is tight on $j$.}
\end{equation}
As in Section~\ref{sec:matching-view}, with labels now called pieces, the \emph{tie graph} at $x$ joins each piece to the agents tight on it and, if it is whole, to the pin $o$. In the running example, at $x=(1,2,5/2)$ the marks on $P_3$ are $1$, $1/2$ and $1/2$. So $P_3$ keeps its last half, and agents~2 and~3 are tight on it, while $P_1$ and $P_2$ stay whole.

\paragraph{Envy-free and fully exposable value vectors.}
An \emph{assignment} is a bijection $M:N\to N$ from piece labels to agents; $M(j)$ is the agent assigned piece $j$. Call a value vector $x$ \emph{envy-free} if some assignment gives every piece to an agent tight on it. Giving each piece's suffix of length $t_j(x)$ to that agent is then an envy-free suffix allocation in which every agent $i$ receives exactly $x_i$, by~\eqref{eq:value-bound}. Conversely, let $S$ be an envy-free suffix allocation, and let $x_i$ be agent $i$'s value of her own suffix. Each holder's mark on her piece is the length of its suffix, and envy-freeness makes every other mark on that piece at least as long. So every piece $j$ has length $t_j(x)$ in $S$, and $x$ is envy-free through the holders of $S$.

Call $x$ \emph{fully exposable} if, for each node $c$ among the agents and the pin, the tie graph at $x$ matches all pieces to the other $n$ nodes. (In matching terminology, a vertex that a matching leaves uncovered is \emph{exposed}, hence the name.) Leaving the pin uncovered gives an assignment along tie-graph edges, so a fully exposable vector is envy-free. At a fully exposable $x$, any assignment that gives every piece to an agent tight on it is a \emph{supporting assignment}.

\paragraph{Fully exposable means certified.}
Let $x$ be envy-free through an assignment $M$, and let $S$ be the resulting suffix allocation. By~\eqref{eq:value-bound}, agent $i$'s favorite labels in $S$ are exactly the pieces she is tight on, and $S_j$ is whole exactly when piece $j$ is whole at $x$. So the tie graph of $S$ in Section~\ref{sec:matching-view} is the tie graph at $x$, and $S$ is certified exactly when $x$ is fully exposable. Fully exposable vectors are therefore the value vectors of certified allocations, and their supporting assignments say who holds which piece. Exposing the pin gives the holders, and exposing an agent gives the chain of ties from a whole piece to her suffix.

\paragraph{Running example.}
At the values $x^*=(1,2,5/2)$ of the certified trim of Figure~\ref{fig:example}, the marks $V_{ij}^{-1}(x^*_i)$ are as follows, using the slope-one extension beyond~$1$:
\begin{center}\small
\begin{tabular}{@{}lccc|c@{}}
\toprule
 & agent~1 ($x^*_1=1$) & agent~2 ($x^*_2=2$) & agent~3 ($x^*_3=5/2$) & length $t_j(x^*)$\\
\midrule
$P_1$ & $\mathbf{1}$ & $2$ & $5/2$ & $1$, whole\\
$P_2$ & $\mathbf{1}$ & $\mathbf{1}$ & $3/2$ & $1$, whole\\
$P_3$ & $1$ & $\mathbf{1/2}$ & $\mathbf{1/2}$ & $1/2$\\
\bottomrule
\end{tabular}
\end{center}
Bold marks attain the minimum, so they are the tight edges. Together with the edges from the whole pieces $P_1$ and $P_2$ to the pin, they form the tie graph of Figure~\ref{fig:tie-graph}. The only assignment along tight edges is $P_j\mapsto j$, so it is the unique supporting assignment. Leaving agent~3 uncovered forces $P_3$ to agent~2; with $P_2$ on the pin, this is the chain of ties in Figure~\ref{fig:tie-graph}(b).

The next proposition shows that a fully exposable vector exists, is unique, and is simultaneously the best envy-free value vector for every agent.

\begin{proposition}[Certified values]
Assume positive inputs. There is a unique fully exposable value vector $x^*$,\label{pr:pinned} and it is the coordinatewise greatest envy-free value vector. In every supporting assignment, each piece has a path of ties leading to a whole piece, in which the holder of each next piece is tight on the previous one.
\end{proposition}
\begin{proof}
\emph{Existence.} Fix $\varepsilon>0$. Consider value vectors $x$ together with assignments $M$ such that, for every piece $j$,
\[
 V^{-1}_{M(j)j}\bigl(x_{M(j)}\bigr)\ \le\ \min\Bigl\{1,\ \varepsilon+\min_{i\in N}V^{-1}_{ij}(x_i)\Bigr\}.
\]
Thus each piece's assigned mark is at most the whole piece and at most $\varepsilon$ longer than the shortest mark on it. Among these pairs, maximize the sum of the values, and among maximizers minimize the sum of the assigned marks. The zero vector is feasible, the piece assigned to agent $i$ bounds her value by $\max_jV_{ij}(1)$, and the feasible set is a finite union of compact sets, so an optimum exists.

Draw an arrow from piece $j$ to piece $k$ if the agent assigned to $k$ has a strictly shorter mark on $j$ than the agent assigned to $j$. Giving each piece on a directed cycle to the agent of its successor would preserve feasibility and lower the sum of the assigned marks. Thus the graph is acyclic and every piece has a path to a sink. A sink $j$ is already assigned an agent with a shortest mark on it, of length $d$, say. If $d<1$, continuity lets us raise that agent's value so that her mark on $j$ grows by less than $\min\{\varepsilon,1-d\}$. The shortest mark on $j$ is still at least $d$, so both the bound one and the $\varepsilon$ allowance are respected. Every other piece keeps its assigned mark, and its shortest mark can only grow. This would increase the sum of the values, contradicting maximality. Each sink therefore has $d=1$.

Let $\varepsilon$ tend to zero along a subsequence with a fixed assignment, a fixed arrow graph and convergent values. In the limit every piece's assigned mark is a shortest mark on it, of length at most one, so its agent is tight on it. Each arrow from $j$ to $k$ becomes a tie, since its strict improvement was at most $\varepsilon$: the agent assigned to $k$ is also tight on $j$. Each sink is whole. To leave a given agent $i$ uncovered, start from the piece assigned to $i$, follow its path to a sink, give each piece on the path to the agent of its successor, and join the sink to the pin. The assignment itself leaves the pin uncovered. Thus the limit is fully exposable.

\emph{Greatestness and uniqueness.} Let $x^*$ be fully exposable and put $t_j^*=t_j(x^*)$. These lengths give a certified allocation. If $x$ is any envy-free value vector and $s_j=t_j(x)$, Proposition~\ref{prop:greatest} gives $s_j\le t_j^*$ for every piece $j$. Each agent holds a favorite suffix in both allocations, so
\[
 x_i=\max_j V_{ij}(s_j)\le\max_j V_{ij}(t_j^*)=x_i^* \qquad (i\in N).
\]
Thus $x^*$ is greatest. Since any fully exposable vector is itself envy-free, applying this inequality in both directions to two such vectors proves uniqueness.

\emph{Matching properties.} We record two matching properties used in the median algorithm. Write $N_x(J)$ for the agents and the pin adjacent to at least one piece of $J$ in the tie graph at $x$. At a fully exposable $x$,
\begin{equation}\label{eq:strict-hall}
              |N_x(J)|\ge |J|+1\qquad(\varnothing\ne J).
\end{equation}
Indeed, we can leave any one neighbor of $J$ uncovered and still match all of $J$. Likewise, a matching of all pieces can cover any proper subset of the agents and the pin, by leaving a node outside that subset uncovered. Finally, superpose any supporting assignment with a matching that leaves agent $i$ uncovered. Their alternating path from $i$ to the pin gives the asserted ties for the piece assigned to $i$.
\end{proof}

To compute a certified trim, it now suffices to find $x^*$ and a supporting assignment. Equation~\eqref{eq:recover-cake} then gives the suffix lengths. Appendix~\ref{app:square} extends the argument to the general curves used for chores and assignment markets.

\subsection{Computing the certified values by medians}
\label{sec:median}
We now compute $x^*$. Neither $x^*$ nor its ties are known. What we do know is the finite set of $n!$ assignments of pieces to agents, and Proposition~\ref{pr:pinned} guarantees that one of them is supporting.

\paragraph{If a supporting assignment were known.}
Let $M$ be an assignment. Start with estimates $\hat x_i=\infty$ for every agent $i$, where an infinite estimate allows the whole piece. In each round, every piece is cut to the length allowed by the current estimates, and the estimate of its assigned agent drops to her value of what is left:
\begin{equation}\label{eq:relax}
 \hat x_{M(j)}\ \gets\ \min\Bigl\{\hat x_{M(j)},\ V_{M(j)j}\bigl(t_j(\hat x)\bigr)\Bigr\}\qquad\text{for all pieces $j$ simultaneously.}
\end{equation}
That is, each piece is cut to the longest suffix, up to the whole piece, that no agent values above her current estimate. If $M$ is supporting, each piece's assigned agent is tight on it at $x^*$, and induction shows that the estimates never fall below $x^*$. On the other hand, the ties of $x^*$ pull the estimates down. Let $L$ consist of the pin and the agents whose estimates are already at most $x^*$. Since $x^*$ is fully exposable, a matching of all pieces along tie-graph edges at $x^*$ covers $L$ whenever $L$ misses some agent, and the pieces it uses push at least one further agent to $x^*$ or below in every round (the proof of Lemma~\ref{le:median} gives the details). So after $n$ rounds the estimates equal $x^*$, much as the distance estimates of the Bellman--Ford algorithm settle after a number of rounds linear in the number of vertices.

The downward pull does not depend on $M$ being supporting: for every assignment, \eqref{eq:relax} brings all estimates to $x^*$ or below within $n$ rounds. Only supporting assignments are guaranteed to stay at or above $x^*$. If we could afford separate queries for each of the $n!$ assignments, we could therefore read off $x^*$ as the coordinatewise maximum of all their estimates after $n$ rounds. We cannot, since different assignments query different points.

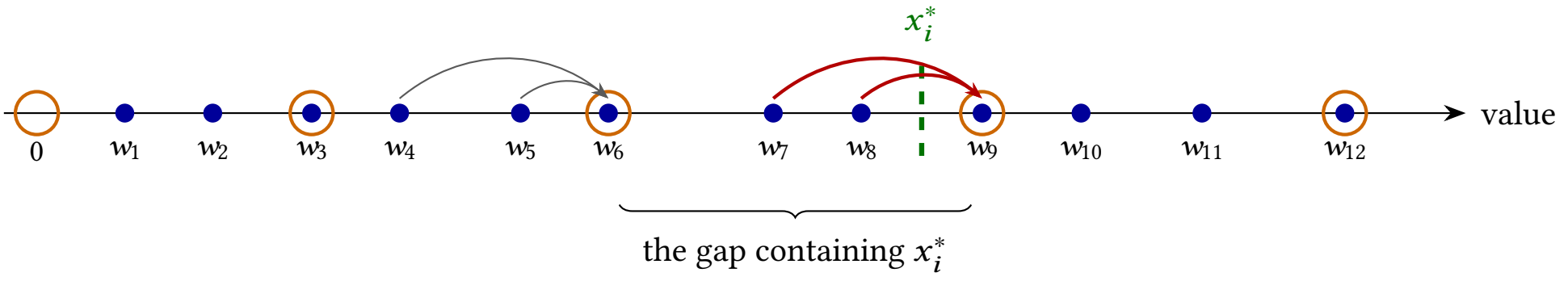
\begin{figure}[t]
\centering
\begin{tikzpicture}[x=0.9cm,y=1cm,font=\footnotesize]
\draw[-{Stealth[length=1.8mm]}] (-0.3,0)--(13.0,0) node[right]{value};
\foreach \x/\r in {0.8/1,1.6/2,2.5/3,3.3/4,4.4/5,5.2/6,6.7/7,7.5/8,8.6/9,9.5/10,10.6/11,11.9/12}{
  \fill[blue!60!black] (\x,0) circle (2.2pt);
  \node[below=3pt,font=\scriptsize] at (\x,0) {$w_{\r}$};}
\foreach \x in {0,2.5,5.2,8.6,11.9} \draw[thick,orange!80!black] (\x,0) circle (5pt);
\node[below=3pt,font=\scriptsize] at (0,0) {$0$};
\draw[very thick,dashed,green!45!black] (8.05,-0.35)--(8.05,0.45) node[above]{$x^*_i$};
\draw[-{Stealth[length=1.4mm]},gray!70!black] (3.3,0.12) to[bend left=40] (5.2,0.12);
\draw[-{Stealth[length=1.4mm]},gray!70!black] (4.4,0.12) to[bend left=40] (5.2,0.12);
\draw[-{Stealth[length=1.4mm]},red!70!black,thick] (6.7,0.12) to[bend left=40] (8.6,0.12);
\draw[-{Stealth[length=1.4mm]},red!70!black,thick] (7.5,0.12) to[bend left=40] (8.6,0.12);
\draw[decorate,decoration={brace,amplitude=3pt,mirror}] (5.3,-0.75)--(8.5,-0.75) node[midway,below=4pt]{the gap containing $x^*_i$};
\end{tikzpicture}
\caption{A grid of resolution $\gamma=4$ on a sorted list of $a=12$ estimates $w_1\le\cdots\le w_{12}$ of agent $i$'s value. The circles mark the finite grid points, namely $0$ and the entries of ranks $3,6,9,12$. The grid also contains $\infty$, omitted from the drawing. Entries between grid points round up to the next one. An entry at or below the true value $x^*_i$ is carried above it only if it lies in the open gap containing $x^*_i$ (red arrows), so at most $a/\gamma$ entries cross. Grids of lengths work in the same way.}
\Description{A number line with twelve sorted estimates of one agent's value. Every third one, and zero, is circled as a grid point. Arrows move entries up to the next grid point. The true value lies in one gap; only the two entries in that gap are carried past it.}
\label{fig:grid}
\end{figure}

\paragraph{Sharing queries by rounding.}
We keep a pool $\A$ of candidate assignments, initially all $n!$ of them, with estimates $\hat x^M$ for each $M\in\A$, and run \eqref{eq:relax} for all of them at once. To let the candidates share answers, we round every query point \emph{up} to a common grid. For a list of $a$ numbers, sorted with multiplicity, the \emph{grid of resolution $\gamma$} consists of $0$, $\infty$ and the entries of ranks $\lceil ha/\gamma\rceil$ for $h=1,\ldots,\gamma$. Round each entry to the smallest grid value weakly above it. An entry equal to a grid point stays there, even if it occurs several times (Figure~\ref{fig:grid}). Each open gap between consecutive grid points contains at most $a/\gamma$ entries of the list, because successive selected ranks skip at most $\lceil a/\gamma\rceil-1$ entries. For a grid $\mathcal G$ write $\rr{u}{\mathcal G}$ for the least grid point that is at least $u$. The values of every curve and every mark at $0$ and $\infty$ are known. The grid of agent $i$ lists the values at which she is asked to mark every piece, and the grid of piece $j$ lists the lengths at which every agent evaluates its suffix. Each candidate reads the answers at its own values and lengths, rounded up, so the number of queries depends on $n$ and $\gamma$ but not on the size of the pool.

\begin{figure}[t]
\centering
\begin{tikzpicture}[font=\footnotesize,x=1cm,y=1cm,
  cand/.style={circle,draw,thick,fill=white,minimum size=2.6mm,inner sep=0pt},
  supp/.style={rectangle,draw,thick,fill=blue!55,minimum size=2.6mm,inner sep=0pt}]
\draw[-{Stealth[length=1.8mm]}] (0,0)--(10.4,0) node[right,align=left]{estimate of\\ agent $k$'s value};
\draw[very thick,red!70!black] (4.4,-0.28)--(4.4,0.28);
\node[below,red!70!black] at (4.4,-0.28) {$y_k$};
\draw[very thick,dashed,green!45!black] (7.0,-0.28)--(7.0,0.28);
\node[below,green!35!black] at (7.0,-0.28) {$x^*_k$};
\foreach \x in {0.6,1.4,2.1,2.9,3.6,4.4,5.3,6.1,9.4} \node[cand] at (\x,0) {};
\node[supp] (s1) at (7.0,0) {};
\node[supp] (s2) at (8.3,0) {};
\draw[decorate,decoration={brace,amplitude=4pt}] (0.4,0.3)--(7.2,0.3)
  node[midway,above=5pt,align=center]{at or below $x^*_k$: at least five eighths of the pool};
\node[align=center] (lab) at (8.9,1.05) {supporting: never\\ below $x^*_k$};
\draw[-{Stealth[length=1.4mm]},thin] (lab.south)--(s2.north east);
\draw[decorate,decoration={brace,amplitude=4pt,mirror}] (0.4,-0.8)--(4.6,-0.8)
  node[midway,below=5pt,align=center]{estimate at most $y_k$: discarded,\\ at least half of the pool};
\draw[decorate,decoration={brace,amplitude=4pt,mirror}] (5.1,-0.8)--(9.6,-0.8)
  node[midway,below=5pt,align=center]{estimate above $y_k$: kept,\\ including every supporting candidate};
\end{tikzpicture}
\caption{The end of a failed phase of Algorithm~\ref{alg:median}, seen on the agent $k$ named by the matching test. Each point is one candidate assignment's estimate of her value; squares are supporting candidates. Supporting candidates never fall below the true value $x^*_k$, and at least five eighths of the pool end at or below it (Lemma~\ref{le:median}), so the lower median $y_k$ is at most $x^*_k$. When the test reports $y_k<x^*_k$, no candidate with estimate at most $y_k$ is supporting, and these candidates make up at least half of the pool.}
\Description{A number line of candidates' estimates of one agent's value. The true value and the lower median are marked, the median lying below the true value. Most candidates lie at or below the true value; the two supporting candidates lie at or above it. The candidates at or below the median, at least half of the pool, are discarded, and the rest, including the supporting ones, are kept.}
\label{fig:median}
\end{figure}
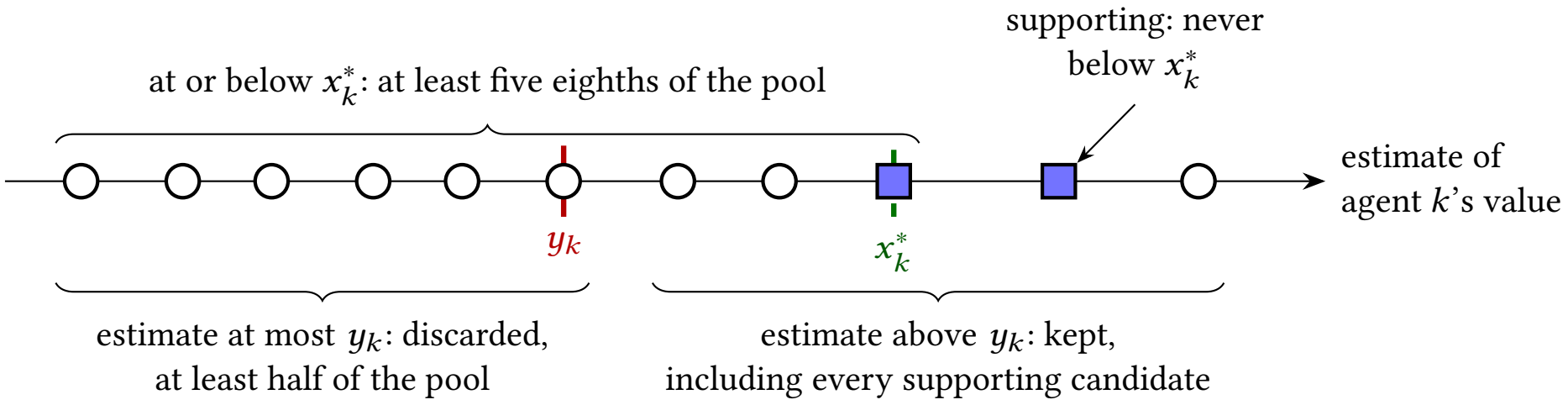

Rounding up makes an upper estimate weaker but never wrong, so supporting candidates still stay at or above $x^*$. The price is that the coordinatewise maximum is no longer reliable: a candidate whose estimate is rounded past the true value may stay above it. The \emph{median}, however, remains reliable, because, on each value or length grid, only entries in the open gap containing the true value or length can round past it (Figure~\ref{fig:grid}). We call an agent $k$ \emph{strictly short} at $y$ if $y_k<x_k^*$. Algorithm~\ref{alg:median} works in phases. Each phase runs $4n$ shared rounds on a fixed pool, takes for each agent the lower median $y_i$ of the candidates' estimates of her value, and tests $y$ by a matching computation. If the test fails, it names an agent, and comparing the candidates' estimates of her value discards at least half of the pool (Figure~\ref{fig:median}).

\begin{algorithm}[tb]
\caption{The median algorithm for certified trimming}\label{alg:median}
\KwIn{$n$ agents and $n$ pieces, with forward and inverse queries to the extended suffix curves $V_{ij}$.}
\KwOut{The fully exposable value vector $x^*$ and a supporting assignment.}
Set $\A$ to the $n!$ assignments of pieces to agents, and $\gamma\gets8(2n+1)$\;
\While(\tcp*[f]{one phase}){true}{
 For each $M\in\A$ set $\hat x_i^M\gets\infty$ for every agent $i$\;
 \For(\tcp*[f]{the pool stays fixed during a phase}){$s=1,\ldots,4n$}{
  For each agent $i$, build the grid $\mathcal G_i$ of $(\hat x_i^M)_{M\in\A}$ and query her marks $V^{-1}_{ij}$ at its points for every piece $j$\;
  For every $M\in\A$ and piece $j$, set $\hat t_j^M\gets\min\bigl\{1,\min_i V^{-1}_{ij}(\rr{\hat x_i^M}{\mathcal G_i})\bigr\}$\;
  For each piece $j$, build the grid $\mathcal G'_j$ of $(\hat t_j^M)_{M\in\A}$ and query $V_{ij}$ at its points for every agent $i$\;
  For every $M\in\A$ and piece $j$, simultaneously set $\hat x^M_{M(j)}\gets\min\{\hat x^M_{M(j)},\, V_{M(j)j}(\rr{\hat t_j^M}{\mathcal G'_j})\}$\;
 }
 For each agent $i$, let $y_i$ be the lower median, entry $\lceil|\A|/2\rceil$ of the sorted list $(\hat x^M_i)_{M\in\A}$\;
 Query the marks $V^{-1}_{ij}(y_i)$ for all $i,j$ and apply the matching test of Lemma~\ref{le:which}\;
 \lIf{the test certifies $y$}{\Return $y$ and its assignment}
 Let $k$ be the strictly short agent named by the test, and set $\A\gets\{M\in\A:\hat x_k^M>y_k\}$\;
}
\end{algorithm}

The two tables of each round are shared by the entire pool. All estimates become finite in the first round, since every length is at most one and each length grid contains its largest entry. The next lemma explains why the median can be used \emph{before} we know which candidates are correct. The constants $4n$ and $\gamma=8(2n+1)$ are chosen so that the stalled fraction of the pool, at most $1/4+1/8$, stays below one half.

\begin{lemma}[The median is a lower bound]\label{le:median}
Throughout a phase, $\hat x^M\ge x^*$ for every supporting $M$ in the pool. After the $4n$ rounds of a phase, at least five eighths of the pool satisfy $\hat x^M\le x^*$ for all agents simultaneously. In particular $y\le x^*$.
\end{lemma}
\begin{proof}
Write $t^*_j=t_j(x^*)$ for the length of piece $j$ at $x^*$.

\emph{Supporting candidates stay above.} Let $M$ be supporting. If $\hat x^M\ge x^*$, then rounding up gives $\hat t^M_j\ge t^*_j$ for every piece $j$. Since agent $M(j)$ is tight on $j$ at $x^*$, we have $V_{M(j)j}(t^*_j)=x^*_{M(j)}$, so the update and the minimum with the old estimate preserve the inequality. Induction from the initialization proves the first claim.

\emph{Most candidates fall below.} Give each candidate in the fixed pool equal weight; probabilities and expectations below are just proportions and averages over the pool, and all choices are deterministic. For a round $s=1,\ldots,4n$, let $L_{s-1}(M)$ consist of the pin and the agents $i$ with $\hat x^M_i\le x^*_i$ before its update, so that $L_0(M)=\{o\}$, and let $L_{4n}(M)$ be this set after the last round. These sets can only grow, since estimates never increase. In a given round, let $e^{\mathrm{agent}}(M)$ be the number of agents $i$ with $\hat x^M_i\le x^*_i<\rr{\hat x^M_i}{\mathcal G_i}$, whose estimate is carried above the true value by rounding, and let $e^{\mathrm{piece}}(M)$ be the number of pieces $j$ with $\hat t^M_j\le t^*_j<\rr{\hat t^M_j}{\mathcal G'_j}$. A threshold that is a grid point cannot be crossed, and otherwise only the entries in the single open grid gap containing the threshold can cross (Figure~\ref{fig:grid}). Summing their proportions over agents and over pieces gives, in every round,
\[
    \mathbb E_M e^{\mathrm{agent}}(M)\le n/\gamma,\qquad
    \mathbb E_M e^{\mathrm{piece}}(M)\le n/\gamma .
\]

Fix $M$ and a round in which $L=L_{s-1}(M)$ misses some agent, and remove from $L$ its crossing agents. A matching of all pieces along tie-graph edges at $x^*$ covers $L$, so at least $|L|-e^{\mathrm{agent}}(M)$ distinct pieces are joined at $x^*$ to a node of the remaining set. Each of them has $\hat t^M_j\le t^*_j$, and for all but at most $e^{\mathrm{piece}}(M)$ of them the rounded length also stays at most $t^*_j$. The assigned agents $M(j)$ of these pieces end the round in $L_s(M)$, since $V_{M(j)j}(t^*_j)\le x^*_{M(j)}$ by~\eqref{eq:value-bound}. They are distinct agents, and the pin is an additional member. Hence the following bound holds, trivially so when $L$ already contains every agent:
\[
 |L_s(M)|-|L_{s-1}(M)|\ge
       \one_{\{L_{s-1}(M)\ne N\cup\{o\}\}}-e^{\mathrm{agent}}(M)-e^{\mathrm{piece}}(M).
\]
Let $\eta_s$ be the proportion of the pool with $L_s(M)\ne N\cup\{o\}$; it is nonincreasing in $s$. Averaging the last inequality over the pool and summing over the $4n$ rounds gives
\[
 4n\,\eta_{4n}\le\sum_{s=0}^{4n-1}\eta_s
 \le \mathbb E_M\bigl[|L_{4n}(M)|-|L_0(M)|\bigr]+\frac{4n\cdot 2n}{\gamma}
 \le n+\frac{8n^2}{\gamma} .
\]
With $\gamma=8(2n+1)$ this gives $\eta_{4n}\le1/4+2n/\gamma<1/4+1/8=3/8$. So at least five eighths of the pool have estimates at most $x^*$ for every agent. For each agent, more than half of the entries are therefore at most her true value, so $y\le x^*$.
\end{proof}

\paragraph{The matching test.}

The median $y$ is a trial vector with $y\le x^*$. The matching test either finds a supporting assignment at $y$ or identifies an agent whose value is strictly too low. Which search the test runs depends on whether the tie graph at $y$, with the pin deleted, has a perfect matching.

\begin{lemma}[The matching test]\label{le:which}
Let $y\le x^*$. From the table of marks $V^{-1}_{ij}(y_i)$ alone, one can either certify that $y=x^*$ and obtain a supporting assignment, or identify an agent $k$ with $y_k<x^*_k$.
\end{lemma}

The test uses only the table at $y$. Form the tie graph at $y$, then delete the pin and find a maximum matching of pieces to agents. If some piece is unmatched, search from it along tie-graph edges to agents and along matching edges back to pieces. Return any reached agent. At least one is reached, and each is strictly short. If the matching is perfect, restore the pin, which is now free, and search from it along tie-graph edges to pieces and along matching edges back to agents. The agents reached are exactly those $i$ with $y_i=x^*_i$. Return the matching if every agent is reached, and otherwise any unreached agent. Fixed orders resolve every choice. Appendix~\ref{app:matching} proves these assertions using \eqref{eq:strict-hall}.

An agent whose value is already exact can even be isolated when the matching is deficient. For example, suppose $V_{ij}(t)=t$ for both agents on both pieces. Then $x^*=(1,1)$. At the trial vector $y=(1,1/2)$, both pieces are tight only for agent~2, although agent~1 already has her exact value. In either failure case the test names an agent $k$ with $y_k<x^*_k$. The filter discards every candidate $M$ with $\hat x^M_k\le y_k$, and every supporting candidate survives because its estimate is at least $x^*_k$.

\begin{theorem}[Query cost of the median algorithm]\label{th:pinned}
On positive inputs, Algorithm~\ref{alg:median} computes $x^*$ and a supporting assignment with $O(n^5\log n)$ deterministic forward and inverse queries.
\end{theorem}
\begin{proof}
If a phase fails on agent $k$, every supporting candidate survives the filter, since $\hat x^M_k\ge x^*_k>y_k$ by Lemmas~\ref{le:median} and~\ref{le:which}. By the definition of the lower median, at most $\lfloor |\A|/2\rfloor$ candidates have estimates strictly above $y_k$, so the pool shrinks by at least half. Initially the pool contains a supporting assignment by Proposition~\ref{pr:pinned}, and supporting assignments are never discarded, so within $1+\lfloor\log_2(n!)\rfloor$ phases the test succeeds. Each round queries its two grids at $O(n^2\gamma)=O(n^3)$ points, so the $4n$ rounds and the final $O(n^2)$ queries cost $O(n^4)$ queries per phase. Multiplying by the number of phases gives the bound. Sorting the lists, updating every candidate and computing the matchings costs $n!\,n^{O(1)}$ exact arithmetic operations and comparisons in total.
\end{proof}

\paragraph{A two-piece calculation.}
Take agents~2 and~3 of the running example and the pieces $P_1,P_3$, as in the two-agent example of Section~\ref{sec:unique}. Here $x^*$ gives agent~2 the value $1$ and agent~3 the value $5/4$. There are two candidates. Candidate $M_1$ assigns $P_1$ to agent~2 and $P_3$ to agent~3, and $M_2$ does the reverse. With two candidates every grid contains all entries, so nothing is rounded. The table lists each candidate's estimates for agents~2 and~3 after each round of the first phase.
\begin{center}\small
\renewcommand{\arraystretch}{1.15}
\begin{tabular}{@{}lcccccc@{}}
\toprule
round & 1 & 2 & 3 & 4 & $\cdots$ & 8\\
\midrule
$M_1$: $P_1\mapsto2$, $P_3\mapsto3$ & $(1,5)$ & $(1,\tfrac54)$ & $(1,\tfrac54)$ & $(1,\tfrac54)$ & $\cdots$ & $(1,\tfrac54)$\\
$M_2$: $P_1\mapsto3$, $P_3\mapsto2$ & $(4,1)$ & $(\tfrac45,1)$ & $(\tfrac45,\tfrac45)$ & $(\tfrac{16}{25},\tfrac45)$ & $\cdots$ & $(\tfrac{256}{625},\tfrac{64}{125})$\\
\bottomrule
\end{tabular}
\end{center}
In the first round all estimates are infinite, so both pieces stay whole, and each candidate's estimate for an agent is her value of the whole piece assigned to her. In the second round, under $M_1$, piece $P_3$ finds that agent~2 already reaches her estimate $1$ on its last quarter, so agent~3's estimate drops to her value $5/4$ of that quarter, and $M_1$ is then stable at $x^*$. Under $M_2$, piece $P_3$ finds that agent~3 reaches her estimate $1$ on its last fifth, so agent~2's estimate drops to $4/5$. In the next round piece $P_1$ pulls agent~3 down to $4/5$, and so on. Candidate $M_2$ supports no positive values at all: for agent~2 to be tight on $P_3$ she would need $x_2\le 4x_3/5$, and for agent~3 to be tight on $P_1$ she would need $x_3\le x_2$. Its estimates instead shrink by a factor $4/5$ every two rounds. After the $4n=8$ rounds both candidates are at or below $x^*$, as Lemma~\ref{le:median} predicts; already after $n=2$ rounds, the larger of the two estimates of each agent's value equals her value in $x^*$.

The lower median of two entries is the smaller one, so $y=(256/625,64/125)$. At $y$, agent~2 is the only agent tight on $P_1$, and both agents are tight on $P_3$, since they reach their values on the same suffix. The matching of pieces to agents is perfect, but no piece is whole, so the search from the free pin reaches neither agent: both values are strictly too low. The test names agent~2, the first in a fixed order. Candidate $M_1$ has estimate $1>y_2$ and survives, while $M_2$ is discarded; naming agent~3 would have discarded $M_2$ as well. In the second phase the pool is $\{M_1\}$, and $y=x^*$. Now $P_1$ is whole and agent~2 is tight on it, and both agents are tight on $P_3$, reaching their values on its last quarter. The search from the pin reaches $P_1$, then agent~2 through the matching, then $P_3$, then agent~3, so the test certifies $y$. The supporting assignment gives the whole of $P_1$ to agent~2 and the last quarter of $P_3$ to agent~3. Agent~2 values that quarter as highly as her whole piece and witnesses the trim.

\begin{proof}[Proof of Lemma~\ref{lem:trim}]
Apply Algorithm~\ref{alg:median} to the $n$ pieces. Theorem~\ref{th:pinned} counts all its forward and inverse queries, each of which costs at most one Robertson--Webb query (Section~\ref{sec:pin}). Recover the physical suffixes by \eqref{eq:recover-cake}. Evaluate each retained boundary for every agent; these $O(n^2)$ evaluations both record the boundaries for later calls and determine every agent's value of every returned suffix. A graph search then orders the holders by their distance from a whole piece, as in Lemma~\ref{lem:test}. On general measures, Appendix~\ref{sec:ties} runs these same steps on virtual positive curves, with at most one physical query per curve query. Its projection preserves the assignment, all weak comparisons, all witness ties and every whole piece.
\end{proof}

\subsection{The cutter's advantage}
\label{sec:core}
\Core{} uses \Trim{} in place of \textsc{SubCore}, the trimming routine of \citet{AzMa16a}. In $\Core(r,N,C)$, agent $r$ cuts $C$ into $n$ pieces of equal value to her. We call \Trim{} on those pieces and return its allocation $Y$, indexed by agent so that $Y_i$ is agent $i$'s suffix, together with its residue $U$.

\begin{proof}[Proof of Lemma~\ref{lem:core}]
The allocation $Y$ is envy-free by Lemma~\ref{lem:trim}. The cutter makes $n-1$ cuts, and recording a cut for every agent takes at most $n+1$ queries (Section~\ref{sec:preliminaries}), so \Core{} uses $O(n^2)$ queries besides its call to \Trim{}.

Write $\tau=v_r(C)/n$. Some piece stays whole and is worth $\tau$ to the cutter, and every suffix is worth at most $\tau$ to her. By envy-freeness, $v_r(Y_r)=\tau$, which is exactly $1/n$ of her value of $C$, also when $\tau=0$. Subtracting the values of all allocated suffixes from $n\tau$ gives
\[
 \sum_{h\ne r}\bigl(v_r(Y_r)-v_r(Y_h)\bigr)=n\tau-\sum_hv_r(Y_h)=v_r(U).
\]
These $n-1$ terms are nonnegative, so at least one of them is at least their average $v_r(U)/(n-1)$. If $v_r(U)>0$, that term is positive, and its agent $h\ne r$ holds a trimmed piece.
\end{proof}
In the running example, $\tau=1$, and the terms are $0$ for agent~2 and $1/2$ for agent~3, so $h=3$ (Section~\ref{sec:example-round}).

Calling \Core{} once with each agent as the cutter already gives a fair partial allocation \citep{AzMa16a}.

\begin{corollary}[Envy-free and proportional]\label{co:prop}
	Call \Core{} $n$ times on the successive residues, starting from the whole cake $C$, with each agent as the cutter exactly once. The union of the returned allocations is envy-free and gives every agent at least $1/n$ of her value of $C$. This takes $O(n^6\log n)$ queries.
\end{corollary}
\begin{proof}
	A union of envy-free allocations is envy-free. Fix an agent $i$, let $A_k$ be the cake allocated by the $k$th call, let $c$ be the call in which $i$ cuts, and let $R$ be the residue after the last call. In each call $k\ne c$, envy-freeness gives $i$ at least $v_i(A_k)/n$. In call $c$, Lemma~\ref{lem:core} gives her exactly $1/n$ of the residue she cuts, whose value is $\sum_{k\ge c}v_i(A_k)+v_i(R)$. Dropping the calls after $c$ from the first bound, she receives at least
	\[
	\frac1n\Bigl(\sum_{k<c}v_i(A_k)+\sum_{k\ge c}v_i(A_k)+v_i(R)\Bigr)=\frac{v_i(C)}{n}.
	\]
	Each call uses $O(n^5\log n)$ queries by Lemma~\ref{lem:core}.
\end{proof}

\section{Finding a New Advantage}
\label{sec:advantage}
This section proves Lemma~\ref{lem:reassign} (step~\stepnum{3}). Throughout, $(X,R)$ is envy-free, $(Y,U)=\Core(r,N,R)$, and $(X\oplus Y,U)$ has no dominated group, so its non-domination graph $\Gamma$ is strongly connected (Section~\ref{sec:guarantees}). We adapt the reassignment lemma of \citet[Lemma~5]{Psom26a} and the cycle exchanges of the \textsc{GoLeft} protocol \citep[Sec.~3.6]{AzMa20a}.

\subsection{What is needed, and why}
Progress requires a domination that did not exist before. \Core{} gives the cutter an advantage over the holder $h$ of a trimmed piece (Lemma~\ref{lem:core}), but she may already dominate $h$, and then nothing is gained. The arcs of $\Gamma$ show where dominations are missing. An arc $a\to b$ means that $a$ does not dominate $b$ in $(X\oplus Y,U)$, and hence not in $(X,R)$ either, since adding the envy-free allocation $Y$ preserves domination. We therefore want an arc $a\to b$ along which $a$ has an advantage.

How large must this advantage be? It need not be large in absolute terms, only large compared with the envy we create to obtain it. Step~\stepnum{4} shrinks the residue to a small multiple of that envy, and changes $a$'s advantage by at most another small multiple (Lemma~\ref{lem:repair}). An advantage $K=4n^2$ times the envy survives both (Section~\ref{sec:loop}). This relative requirement is what makes the step possible: if no arc of $\Gamma$ carries a large advantage, then moving bundles along arcs is cheap.

\paragraph{Measuring in units of the residue.}
If an agent valued $U$ at zero, she would dominate everyone else, and the others would form a dominated group. Thus $v_i(U)>0$ for every $i$, and the \emph{normalized gap}
\[
 g_{ij}=\frac{v_i(Y_i)-v_i(Y_j)}{v_i(U)}
\]
is well defined, nonnegative by envy-freeness of $Y$, and zero when $i=j$. It is agent $i$'s advantage from holding $Y_i$ instead of $Y_j$, expressed as a fraction of her value of $U$. Only the new suffixes $Y$ move; each old bundle $X_i$ stays on label $i$. We look for an arc $a\to b$, a permutation $\pi$, where agent $i$ receives $Y_{\pi(i)}$, and a tolerance $\alpha$ such that
\begin{equation}\label{eq:reassign-pieces}
 g_{i,\pi(i)}\le\alpha\quad(i\in N),
 \qquad g_{a,\pi(b)}\ge K\alpha.
\end{equation}
In words, each agent $i$ gives up at most $\alpha v_i(U)$ in moving from $Y_i$ to $Y_{\pi(i)}$. Agent $a$ values her original suffix $Y_a$ at least $K\alpha v_a(U)$ above the suffix now on label $b$.

\subsection{The algorithm}
\label{sec:reassign-alg}
\Reassign{} (Algorithm~\ref{alg:reassign}) distinguishes two cases.

\begin{algorithm}[t]
\caption{$\Reassign(r,X,Y,U)$}\label{alg:reassign}
\KwIn{The cutter $r$, an envy-free $X$ with residue $R$, and $(Y,U)=\Core(r,N,R)$, with no dominated group in $(X\oplus Y,U)$.}
\KwOut{An arc $a\to b$ of $\Gamma$, a permutation $\pi$ of the new suffixes, and a tolerance $\alpha$.}
$K\gets4n^2$, and $g_{ij}\gets\bigl(v_i(Y_i)-v_i(Y_j)\bigr)/v_i(U)$ for all $i,j$\;
$\Gamma\gets$ the non-domination graph of $(X\oplus Y,U)$\tcp*[r]{strongly connected}
\If(\tcp*[f]{case 1: a large gap; nothing moves}){some arc $a\to b$ of $\Gamma$ has $g_{ab}\ge K^{-n}$}{
  \Return $a$, $b$, $\pi=\mathrm{id}$, and $\alpha=\min\{K^{-2},\,g_{ab}/K\}$\;
}
\tcp{case 2: every arc of $\Gamma$ has gap less than $K^{-n}$}
Pick $h$ with $g_{rh}\ge 1/(n-1)$\tcp*[r]{exists by Lemma~\ref{lem:core}}
Take a simple path $r=u_0,u_1,\dots,u_\ell=h$ in $\Gamma$\;
$s\gets$ the least index with $g_{u_s h}\le K^{-(s+1)}$; set $a\gets u_{s-1}$, $b\gets u_s$, and $\alpha\gets K^{-(s+1)}$\;
Take a simple path $h=v_0,v_1,\dots,v_{\ell'}=b$ in $\Gamma$\;
Set $\pi(v_\nu)\gets v_{\nu+1}$ for $\nu<\ell'$, $\pi(b)\gets h$, and $\pi(i)\gets i$ for every other agent\tcp*[r]{rotate}
\Return $a$, $b$, $\pi$, $\alpha$\;
\end{algorithm}

\paragraph{Case 1: use an existing advantage.}
If some arc $a\to b$ already has a gap of at least $K^{-n}$, nothing needs to move. The tolerance $\alpha$ is set so that the gap is at least $K\alpha$, and capped at $K^{-2}$ so that \Repair{} applies.

\begin{figure}[t]
\centering
\begin{tikzpicture}[font=\footnotesize,
  ag/.style={circle,draw,thick,minimum size=7mm,inner sep=0pt,fill=white},
  gam/.style={-{Stealth[length=1.7mm]},semithick,gray!70!black},
  mv/.style={-{Stealth[length=1.8mm]},thick,densely dashed,orange!80!black},
  big/.style={-{Stealth[length=2mm]},very thick,red!70!black}]
\node[ag] (u0) at (0.4,0) {$r$};
\node[ag] (u1) at (1.5,0) {$u_1$};
\node (d1) at (2.7,0) {$\cdots$};
\node[ag,fill=blue!10] (a) at (3.9,0) {$a$};
\node[ag,fill=green!14] (b) at (5.6,0) {$b$};
\node (d2) at (7.0,0) {$\cdots$};
\node[ag] (h) at (8.4,0) {$h$};
\foreach \x/\y in {u0/u1,u1/d1,d1/a,a/b,b/d2,d2/h} \draw[gam] (\x)--(\y);
\draw[decorate,decoration={brace,amplitude=4pt}] (0.05,0.5)--(4.25,0.5)
  node[midway,above=5pt,align=center]{above threshold:\\ $g_{u_{s'} h}>K^{-(s'+1)}$ for $s'<s$};
\node[above=3pt,align=center] at (5.6,0.36) {first below:\\ $g_{bh}\le K^{-(s+1)}=\alpha$};
\node[above=3pt] at (8.4,0.36) {$g_{hh}=0$};
\node[below=3pt] at (0.4,-0.36) {$g_{rh}\ge\frac1{n-1}$};
\node[below=3pt] at (3.9,-0.36) {$g_{ah}>K\alpha$};
\node[ag] (v1) at (8.0,-2.4) {$v_1$};
\node[ag] (v2) at (6.0,-2.4) {$v_2$};
\draw[gam] (h)--(v1); \draw[gam] (v1)--(v2); \draw[gam] (v2)--(b);
\draw[mv] ($(v1)+(0.36,0.12)$) to[bend right=22] node[right]{$Y_{v_1}$} ($(h)+(0.36,-0.12)$);
\draw[mv] ($(v2)+(0.15,-0.4)$) to[bend right=35] node[below]{$Y_{v_2}$} ($(v1)+(-0.15,-0.4)$);
\draw[mv] ($(b)+(-0.38,-0.1)$) to[bend right=22] node[left]{$Y_b$} ($(v2)+(-0.38,0.1)$);
\draw[big] ($(h)+(-0.3,-0.3)$) to[bend left=18] node[below,pos=0.5]{$b$ takes $Y_h$} ($(b)+(0.3,-0.3)$);
\node[anchor=west,inner sep=0pt] at (-0.35,-4.0) {%
\tikz\draw[gam](0,0)--(0.5,0);\ arc of $\Gamma$ ($i$ does not dominate $j$)\qquad
\tikz\draw[mv](0,0)--(0.5,0);\ loses its taker $<K^{-n}$\qquad
\tikz\draw[big](0,0)--(0.5,0);\ loses $b$ at most $\alpha$};
\node[anchor=west,inner sep=0pt,align=left] at (-0.35,-1.6) {losses are in units\\ of the taker's $v_i(U)$};
\end{tikzpicture}
\caption{Case~2 of \Reassign. Walking from the cutter $r$ to $h$, the gap $g_{u_s h}$ starts above its threshold $K^{-(s+1)}$ and ends at $0$, so some arc $a\to b$ straddles the thresholds. Rotating the new suffixes around the cycle $h\to\cdots\to b\to h$ hands $Y_h$ to $b$. Every agent's envy is at most $\alpha\,v_i(U)$, while $a$ now values her own bundle above $b$'s by at least $(K-1)\alpha\,v_a(U)$. Old bundles $X_i$ never move.}
\Description{A path of agents from the cutter r to h with gap thresholds, the arc from a to b where the gap first falls below its threshold, and a return path from h to b along which new suffixes are passed back, with b taking h's suffix.}
\label{fig:reassign}
\end{figure}
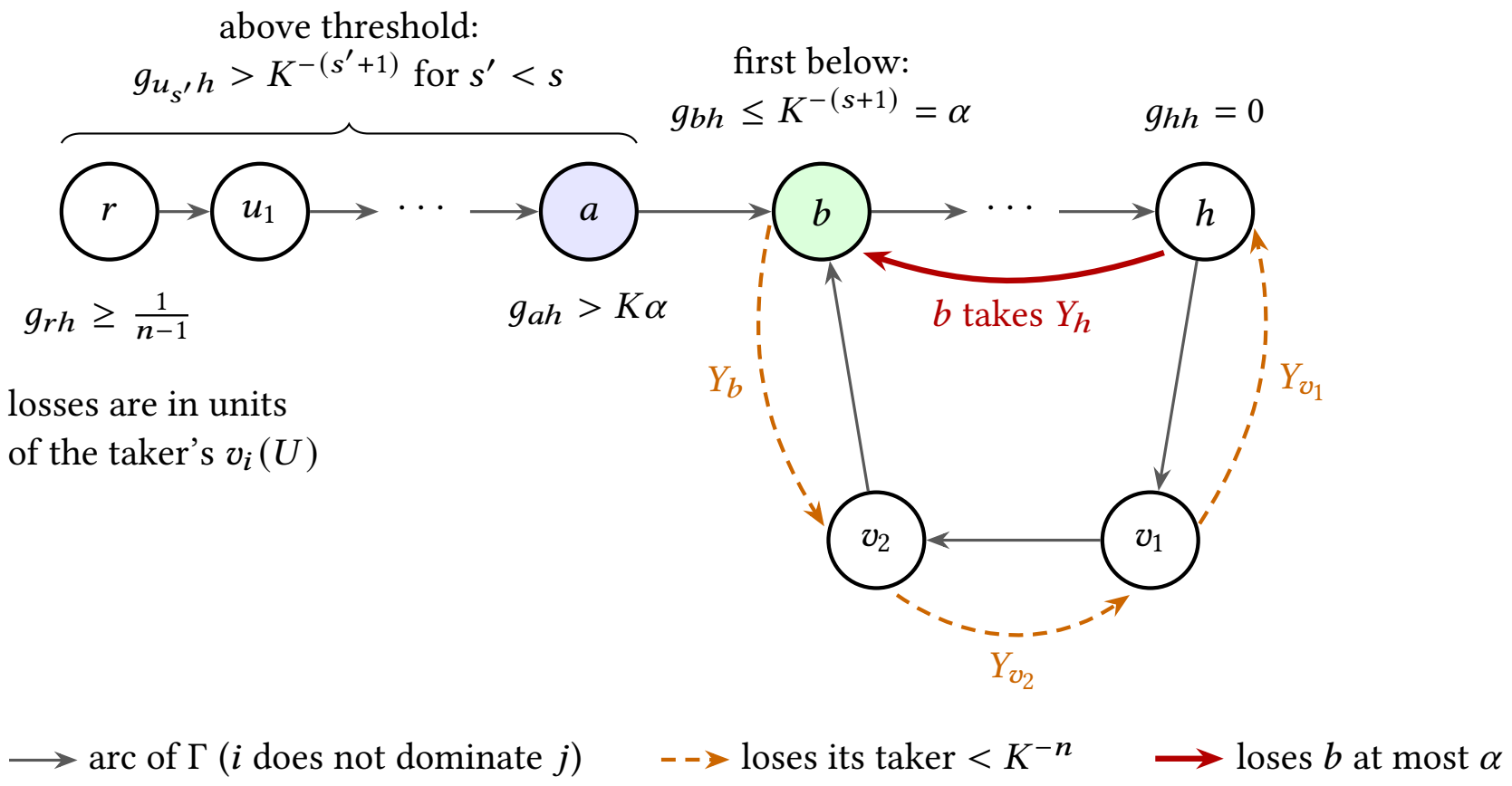

\paragraph{Case 2: rotate along cheap moves.}
Otherwise every arc of $\Gamma$ has a gap below $K^{-n}$. A small gap is a near-indifference, so passing a new suffix backwards along an arc loses its new holder very little. The cutter, however, has a large gap towards $h$. Walk along a path from $r$ to $h$ and compare each agent's gap towards $h$ with a threshold that shrinks by a factor $K$ per step (Figure~\ref{fig:reassign}). The gap starts above its threshold and ends at $g_{hh}=0$, so some arc $a\to b$ straddles the thresholds: $a$ values $Y_h$ much less than her own suffix, while $b$ would accept $Y_h$ at little loss. Now take a path from $h$ back to $b$. Each agent on it takes her successor's suffix, and $b$ takes $Y_h$. This closes a cycle. Every move loses its taker little, and $a$ now values the suffix on label $b$ much less than her own.

\paragraph{Why the thresholds shrink geometrically.}
A fixed threshold would only give an arc from an agent above it to an agent below it, and the drop across that arc could be tiny. Thresholds that shrink by a factor $K$ per step force a drop by a factor of more than $K$ across a single arc: $a$'s gap exceeds $K$ times what $b$ is allowed. The path has at most $n-1$ arcs, so the tolerance never falls below $K^{-n}$ in case~2, nor below $K^{-(n+1)}$ overall.

\paragraph{When the cutter's advantage is wasted.}
In the first round the cutter's advantage over the holder of a trimmed piece is her value of its removed prefix. She therefore dominates $h$ exactly when this prefix accounts for all her value of the residue. In the running example another arc supplies the advantage for case~1.

For an instance requiring case~2, let agent~1 cut the cake into thirds that she values at $(1,1,1)$ and that each of agents~2 and~3 values at $(1,1,4)$, with values spread evenly within each third. A certified allocation gives agents~1 and~2 the first two thirds and agent~3 the last quarter of the third piece, so the cutter's only positive gap is towards agent~3, whom she already dominates. Agents~2 and~3 are indifferent among all three suffixes, so every arc of $\Gamma$ has gap zero, and swapping their suffixes gives the cutter an advantage of $3/4$ over the previously undominated label~2.

\subsection{Proof of Lemma~\ref{lem:reassign}}
The non-domination graph $\Gamma$ of $(X\oplus Y,U)$ is strongly connected. Its arcs were also undominated in $(X,R)$, since adding an envy-free allocation preserves domination. We showed above that $v_i(U)>0$ for every $i$, so the gaps $g_{ij}$ are well defined. We first establish \eqref{eq:reassign-pieces} in each case.

\emph{Case 1.} If some arc $a\to b$ has $g_{ab}\ge K^{-n}$, take the identity permutation and $\alpha=\min\{K^{-2},g_{ab}/K\}$. Then $g_{i,\pi(i)}=0\le\alpha$ and $g_{a,\pi(b)}=g_{ab}\ge K\alpha$, so \eqref{eq:reassign-pieces} holds, and $K^{-(n+1)}\le\alpha\le K^{-2}$.

\emph{Case 2.} Every arc now has gap less than $K^{-n}$. By Lemma~\ref{lem:core}, some holder $h$ satisfies $g_{rh}\ge1/(n-1)>1/K$. Follow a simple path $r=u_0,\ldots,u_\ell=h$ in $\Gamma$. Compare $g_{u_s h}$ with $K^{-(s+1)}$: the gap starts above its threshold and ends at zero. At the first index $s$ with $g_{u_s h}\le K^{-(s+1)}$, set
\[
 a=u_{s-1},\qquad b=u_s,\qquad\beta=K^{-(s+1)}.
\]
Then $a\to b$ is an arc, $g_{ah}>K\beta$, $g_{bh}\le\beta$, and $K^{-n}\le\beta\le K^{-2}$, since $1\le s\le\ell\le n-1$.

Take a simple return path from $h$ to $b$. Give each path agent before $b$ her successor's suffix and give $b$ the suffix $Y_h$, fixing everyone else. This is a cyclic permutation. Every move along the return path loses its taker less than $K^{-n}\le\beta$ in relative units, and the closing move loses $b$ at most $\beta$. Thus \eqref{eq:reassign-pieces} holds with $\alpha=\beta$. All choices use recorded values.

\emph{Translate these bounds to the input of repair.} Envy-freeness of $X$ and $Y$ gives $v_i(D_j)\le v_i(X_i)+v_i(Y_i)$ for every $j$, whereas \eqref{eq:reassign-pieces} gives $v_i(D_i)\ge v_i(X_i)+v_i(Y_i)-\alpha v_i(U)$. This proves the first inequality in \eqref{eq:reassign-output}. Also, since $v_a(X_a)\ge v_a(X_b)$ by envy-freeness of $X$,
\[
 v_a(D_a)-v_a(D_b)
 \ge\bigl(g_{a,\pi(b)}-g_{a,\pi(a)}\bigr)v_a(U)
 \ge(K-1)\alpha v_a(U).
\]
The loss of one unit allows for $a$ herself receiving a slightly less valuable suffix during the rotation. \qed

\paragraph{How small the tolerance can be.}
The tolerance $\alpha$ may be as small as $K^{-(n+1)}=2^{-\Theta(n\log n)}$. This is why Lemma~\ref{lem:repair} depends only logarithmically on $1/\alpha$: a repair whose cost grew linearly in $1/\alpha$ would by itself need $2^{\Theta(n\log n)}$ queries per call.

\section{Repairing Envy and Shrinking the Residue}
\label{sec:repair}
This section proves Lemma~\ref{lem:repair} (step~\stepnum{4}). The input is an allocation $D$ with residue $U$ and a little envy: $v_i(D_i)\ge v_i(D_j)-\alpha v_i(U)$ for all $i,j$, where $0<\alpha\le1/(8n^2)$. Labels are named after the agents who own the input bundles, so label $j$ carries $D_j$ and agent $j$ is its \emph{owner}; after repair its \emph{holder} $\sigma(j)$ may be someone else.

\subsection{What repair must achieve}
\label{sec:repair-goal}
\Repair{} must do four things at once:
\begin{enumerate}[label=(\roman*)]
\item remove the envy, so that the output is envy-free;
\item keep every input bundle $D_j$ intact, so that old dominations survive (Lemma~\ref{lem:follow});
\item allocate almost all of $U$, so that the residue falls below agent $a$'s advantage;
\item add nearly the same value to every bundle, in every agent's eyes, so that the advantage itself survives.
\end{enumerate}
Trimming removes envy, so (i) is easy on its own. But trimming the bundles $D_j$ directly would cut into them and violate (ii). Adding an exactly equal-value share of $U$ to every bundle would serve (iii) and (iv), since it preserves every value gap, but it would leave the envy in place. \Repair{} does both. It attaches to every bundle a large share of $U$ that every agent values nearly the same, places $n-1$ small \emph{buffers} in front, and calls \Trim{} once. Trimming removes a prefix, so it must consume buffers before it can reach a bundle, and Section~\ref{sec:protect} shows that it never gets through. In the protocol of \citet{AzMa16a}, further Core calls shrink the residue until an advantage becomes a domination \citep[Sec.~3.2]{AzMa20a}. \Repair{} achieves the required shrinkage in one call.

\begin{example}[Why buffers are needed]\label{ex:toy}
Let $n=3$ and $\alpha=1/(8n^2)=1/72$, the largest tolerance allowed. The input bundles are three blocks $D_1,D_2,D_3$, and the residue $U$ is a block worth $72$ to every agent. Each agent's value is spread evenly within each block:
\begin{center}\small
\begin{tabular}{@{}l|ccc|c@{}}
\toprule
 & $D_1$ & $D_2$ & $D_3$ & $U$\\
\midrule
agent 1 & $\mathbf8$ & 5 & 4 & 72\\
agent 2 & 4 & $\mathbf7$ & 4 & 72\\
agent 3 & 4 & 5 & $\mathbf4$ & 72\\
\bottomrule
\end{tabular}
\end{center}
So $\alpha v_i(U)=1$. Agent~3 envies $D_2$ by exactly $1$, and nobody else envies anybody. Suppose we give each bundle a third of $U$, placed behind it, and call \Trim{} on the three resulting pieces. The equal thirds change no comparison, so agent~3 still envies label~2 by $1$. \Trim{} keeps labels~1 and~3 whole and trims the front of label~2 until agent~3, holding label~3, values it exactly as much as her own. The front of label~2 is $D_2$, so a fifth of an input bundle returns to the residue. We return to this example with buffers in Section~\ref{sec:construction}.
\end{example}

\subsection{Nearly common shares}
\label{sec:shares}
The construction needs large shares and small buffers whose values are close to prescribed targets for every agent. What matters is the total error over all shares, rather than the error of each one.

\begin{lemma}[Approximate common shares]\label{lem:shares}
Let $w_1,\ldots,w_m\ge0$ sum to one, let $0<\delta<1$, and normalize $v_i(U)=1$ for all $i$. We can find disjoint pieces $A_q\subseteq U$ with $v_i(A_q)\le w_q$ for every $i,q$, leaving residue $U_0$ with $v_i(U_0)\le\delta$ for every $i$. Equivalently, the total deficit is
\[
 \sum_{q=1}^m\bigl(w_q-v_i(A_q)\bigr)=v_i(U_0)\le\delta.
\]
The procedure uses $O(n^2m\log(n^2m/\delta))$ queries and polynomially many exact arithmetic operations in the size of its interval records, $n,m$, and $\log(1/\delta)$.
\end{lemma}

We prove this lemma in Section~\ref{sec:shares-proof}. A consequence used below is that, for any set $F$ of slots,
\begin{equation}\label{eq:common-union}
 \sum_{q\in F}w_q-\delta
 \le v_i\!\left(\bigcup_{q\in F}A_q\right)
 \le\sum_{q\in F}w_q.
\end{equation}
Thus a union of many shares still incurs a total error of at most $\delta$.

\subsection{The construction}
\label{sec:construction}
\begin{figure}[t]
\centering
\begin{minipage}[c]{0.7\textwidth}
\centering
\begin{tikzpicture}[x=1cm,y=1cm,font=\footnotesize]
\foreach \j/\cut/\lev in {1/0/2,2/0.22/1,3/0/2}{
  \pgfmathsetmacro{\yy}{-0.58*\j}
  \node[left] at (-0.05,\yy+0.2) {label \j};
  \fill[orange!22] (0,\yy) rectangle (1.9,\yy+0.4);
  \fill[blue!14] (1.9,\yy) rectangle (4.3,\yy+0.4);
  \fill[green!16] (4.3,\yy) rectangle (6.9,\yy+0.4);
  \ifdim\cut pt>0pt
    \fill[white] (0,\yy) rectangle (\cut,\yy+0.4);
    \fill[pattern=north east lines,pattern color=gray!70] (0,\yy) rectangle (\cut,\yy+0.4);
    \draw[red!70!black,thick] (\cut,\yy-0.05)--(\cut,\yy+0.45);
  \fi
  \foreach \x in {0,0.95,1.9,4.3,6.9} \draw (\x,\yy)--(\x,\yy+0.4);
  \draw (0,\yy)--(6.9,\yy) (0,\yy+0.4)--(6.9,\yy+0.4);
  \node[fill=orange!22,inner sep=0.5pt] at (0.58,\yy+0.2) {$Q_{\j,2}$};
  \node at (1.425,\yy+0.2) {$Q_{\j,1}$};
  \node at (3.1,\yy+0.2) {$D_\j$};
  \node at (5.6,\yy+0.2) {$T_\j$};
  \node[right] at (6.95,\yy+0.2) {level \lev};
}
\draw[decorate,decoration={brace,amplitude=3pt}] (1.9,-0.1)--(6.9,-0.1) node[midway,above=3pt]{protected part};
\draw[-{Stealth},thick] (0,-2.15)--(1.6,-2.15) node[right]{trimming starts at the front};
\end{tikzpicture}
\end{minipage}\hfill
\begin{minipage}[c]{0.24\textwidth}
\centering
\begin{tikzpicture}[font=\footnotesize,ag/.style={circle,draw,thick,minimum size=5.5mm,inner sep=0pt},arr/.style={-{Stealth[length=1.6mm]},thick}]
\draw[green!50!black,dashed,rounded corners,fill=green!6] (-0.42,0.85) rectangle (0.42,1.65);
\node[ag,fill=white] (1) at (0,1.25) {1};
\node[ag,fill=white] (2) at (-0.75,0) {2};
\node[ag,fill=white] (3) at (0.75,0) {3};
\draw[arr] (2)--(1);
\draw[arr] (3)--(1);
\end{tikzpicture}
\end{minipage}
\caption{\Repair{} in the running example (not to scale), in the first round, where $D=Y$. Label~1 carries agent~1's piece $D_1=P_1$, label~2 carries agent~2's piece $D_2=P_2$, and label~3 carries agent~3's suffix $D_3$ (the last half of $P_3$). Left: each label carries two buffers $Q_{j,2},Q_{j,1}$, its input bundle $D_j$ and a large share $T_j$. With exact common shares nothing would be trimmed. The cut shown, in an outer buffer, is one possible outcome of the rounding, and the level counts the whole buffers kept. In every outcome each protected part $D_j\cup T_j$ survives. Right: in every outcome agents~2 and~3 both dominate agent~1, so $\{1\}$ is a dominated group (Section~\ref{sec:example-round}). Here arrows show domination, unlike the arcs of the non-domination graph.}
\Description{Three labels, each laid out as two buffers, an input bundle and a large share; the outer buffer of label 2 is partly trimmed. A small graph shows agents 2 and 3 each dominating agent 1, who forms a dominated group.}
\label{fig:example-repair}
\end{figure}
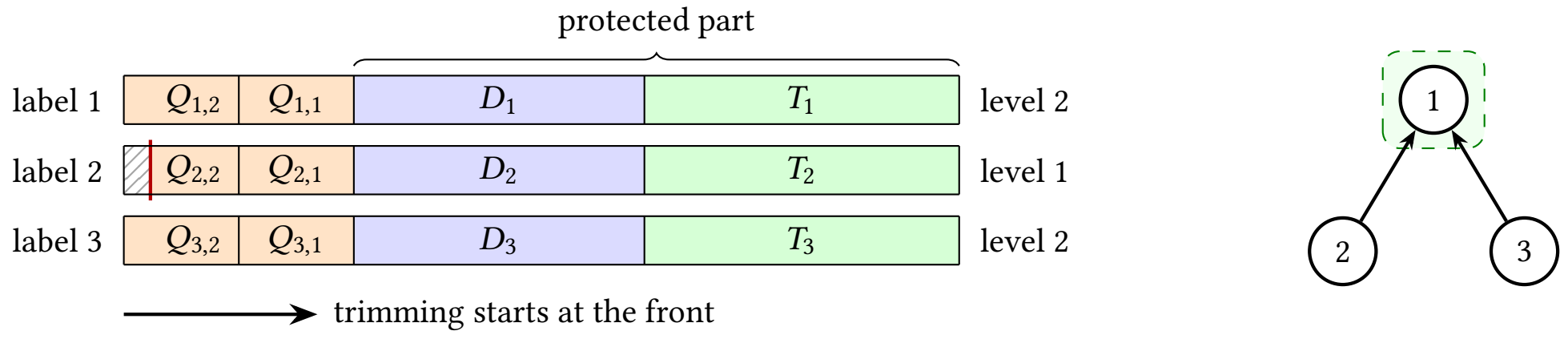

\Repair{} (Algorithm~\ref{alg:repair}) normalizes each agent's value of $U$ to $1$ and cuts $U$ into $n^2$ nearly common shares, with the targets in Table~\ref{tab:repair-params}. On label $j$ it places $n-1$ buffers, then $D_j$, then a large share $T_j$. We call $D_j\cup T_j$ the \emph{protected part} of label $j$ (Figure~\ref{fig:example-repair}). Since every agent values every buffer nearly the same, it does not matter which buffers go to which label.

\begin{table}[t]
\caption{The parameters of $\Repair(N,D,U,\alpha)$, in units where every agent's value of $U$ is $1$.}
\label{tab:repair-params}
\centering\small
\begin{tabular}{@{}lll@{}}
\toprule
Symbol & Value & Role\\
\midrule
$B$ & $n(n-1)$ & number of buffers, $n-1$ per label\\
$p$ & $2\alpha$ & target value of each buffer: more than the envy $\alpha$ plus rounding loss $\delta$\\
$t$ & $(1-Bp)/n$ & target value of each large share; the shares carry almost all of $U$\\
$\delta$ & $\alpha/2$ & total rounding loss allowed in Lemma~\ref{lem:shares}\\
\bottomrule
\end{tabular}
\end{table}

\begin{algorithm}[t]
\caption{$\Repair(N,D,U,\alpha)$}\label{alg:repair}
\KwIn{A partial allocation $D$ with residue $U$ and tolerance $\alpha$ satisfying Lemma~\ref{lem:repair}.}
\KwOut{An envy-free allocation $W$ retaining every $D_j$ intact, the holder map $\sigma$, and the residue $R'$.}
For this call, scale each value measure so that $v_i(U)=1$\;
Set $B\gets n(n-1)$, $p\gets2\alpha$, $t\gets(1-Bp)/n$, and $\delta\gets\alpha/2$\;
Use Lemma~\ref{lem:shares} to make $n$ large shares $T_j$ of target $t$ and $B$ buffers $Q_{j,k}$ ($j\in N$, $1\le k\le n-1$) of target $p$, with total deficit at most $\delta$\;
For every label $j$, form $P'_j\gets Q_{j,n-1}\concat\cdots\concat Q_{j,1}\concat(D_j\concat T_j)$\;
$W\gets\Trim(N,(P'_j)_{j\in N})$, and let $\sigma(j)$ be the holder of label $j$\;
\Return $W$, $\sigma$, and $R'\gets(U\cup\bigcup_jD_j)\setminus\bigcup_iW_i$\;
\end{algorithm}

\paragraph{How the envy is removed.}
\Repair{} does not remove the envy itself. It arranges the pieces so that one call to \Trim{}, which always returns an envy-free allocation, removes it safely. Trimming removes envy in two ways. A piece wanted by several agents can be trimmed from the front until one of them values it no more than another suffix, and because the buffers come first, this consumes buffer material. Or the envious agent can take the bundle she envied while its holder takes another, which is why the holder of a label may change.

\paragraph{Example~\ref{ex:toy}, continued.}
Here $p=1/36$ and $t=5/18$, so each buffer is worth $2$ and each large share $20$ to every agent. Every agent's value is spread evenly on $U$, so blocks of the right lengths are exact common shares. Each label then carries $24$ in addition to its input bundle, for every agent; write $P'_j$ for the piece formed on label $j$:
\begin{center}\small
\begin{tabular}{@{}lccc@{}}
\toprule
 & $P'_1$ & $P'_2$ & $P'_3$\\
\midrule
agent 1 & $\mathbf{32}$ & 29 & 28\\
agent 2 & 28 & $\mathbf{31}$ & 28\\
agent 3 & 28 & 29 & $\mathbf{28}$\\
\bottomrule
\end{tabular}
\end{center}
Agent~3 still envies label~2 by $1$. \Trim{} now removes this envy by trimming half of the outer buffer of $P'_2$, worth $1$ to everybody, so that agent~3 values label~2 at $28$, as much as her own. Every input bundle is intact. Labels~1 and~3 have level~2 and label~2 has level~1, the additions are worth $24$, $23$ and $24$, and the residue is half a buffer.

\subsection{Why every input bundle survives}
\label{sec:protect}
We now prove Lemma~\ref{lem:repair}. Use the normalized values of Algorithm~\ref{alg:repair}. Since $nt+Bp=1$ and $Bp\le1/4$, the targets are valid. The output is envy-free by Lemma~\ref{lem:trim}. It remains to show that no protected part is cut, and to bound the unequal additions and the residue.

\paragraph{Levels.}
Let $S_j$ be the returned suffix on label $j$, whoever holds it. Its \emph{level} $\ell(j)$ is $-1$ if it cuts into the protected part, and otherwise is the number of whole buffers it retains. A suffix of level $\lambda\ge0$ contains its protected part, $\lambda$ whole buffers, and at most part of one more buffer. A whole piece has level $n-1$. Repair succeeds exactly when no level is $-1$.

\begin{figure}[t]
\centering
\begin{tikzpicture}[x=1cm,y=0.8cm,font=\footnotesize]
\node[left] at (-0.2,1.3) {$S_l$, level $\ge\lambda+2$};
\node[left] at (-0.2,0) {$S_j$, level $\lambda$};
\foreach \x/\lab in {0/{Q},1.6/{\cdots},3.2/{Q},4.8/{Q}} {\fill[orange!25] (\x,1) rectangle (\x+1.5,1.6); \draw (\x,1) rectangle (\x+1.5,1.6); \node at (\x+0.75,1.3) {$\lab$};}
\fill[blue!14] (6.4,1) rectangle (9.4,1.6); \draw (6.4,1) rectangle (9.4,1.6); \node at (7.9,1.3) {$D_l\cup T_l$};
\fill[pattern=north east lines,pattern color=gray!60] (1.6,-0.3) rectangle (2.3,0.3);
\fill[orange!25] (2.3,-0.3) rectangle (3.1,0.3); \draw (1.6,-0.3) rectangle (3.1,0.3);
\draw[red!70!black,thick] (2.3,-0.38)--(2.3,0.38);
\foreach \x/\lab in {3.2/{Q},4.8/{Q}} {\fill[orange!25] (\x,-0.3) rectangle (\x+1.5,0.3); \draw (\x,-0.3) rectangle (\x+1.5,0.3); \node at (\x+0.75,0) {$\lab$};}
\fill[blue!14] (6.4,-0.3) rectangle (9.4,0.3); \draw (6.4,-0.3) rectangle (9.4,0.3); \node at (7.9,0) {$D_j\cup T_j$};
\draw[decorate,decoration={brace,amplitude=3pt}] (0,1.75)--(6.3,1.75) node[midway,above=3pt]{at least $\lambda+2$ whole buffers: together worth at least $(\lambda+2)p-\delta$};
\draw[decorate,decoration={brace,amplitude=3pt,mirror}] (1.6,-0.45)--(6.3,-0.45) node[midway,below=3pt]{$\lambda$ whole buffers and part of one more: together worth at most $(\lambda+1)p$};
\draw[red!70!black,thick,rounded corners] (-0.08,0.9) rectangle (1.58,1.7);
\node[red!70!black,font=\scriptsize] at (0.75,0.62) {extra buffer};
\end{tikzpicture}
\caption{Step~1 of the proof, through the eyes of the owner $l$ of the upper label $l$, with $v_l(U)=1$. A suffix two or more levels higher carries at least one extra whole buffer (outlined), worth about $p=2\alpha$ to her. This outweighs her input envy, $v_l(D_l)\ge v_l(D_j)-\alpha$, because by \eqref{eq:common-union} all deficits together are worth at most $\delta=\alpha/2$ to her.}
\Description{Two suffixes compared: one with at least lambda plus two whole buffers before its protected part, and one with lambda whole buffers and part of another; the extra whole buffer is outlined.}
\label{fig:extra-buffer}
\end{figure}
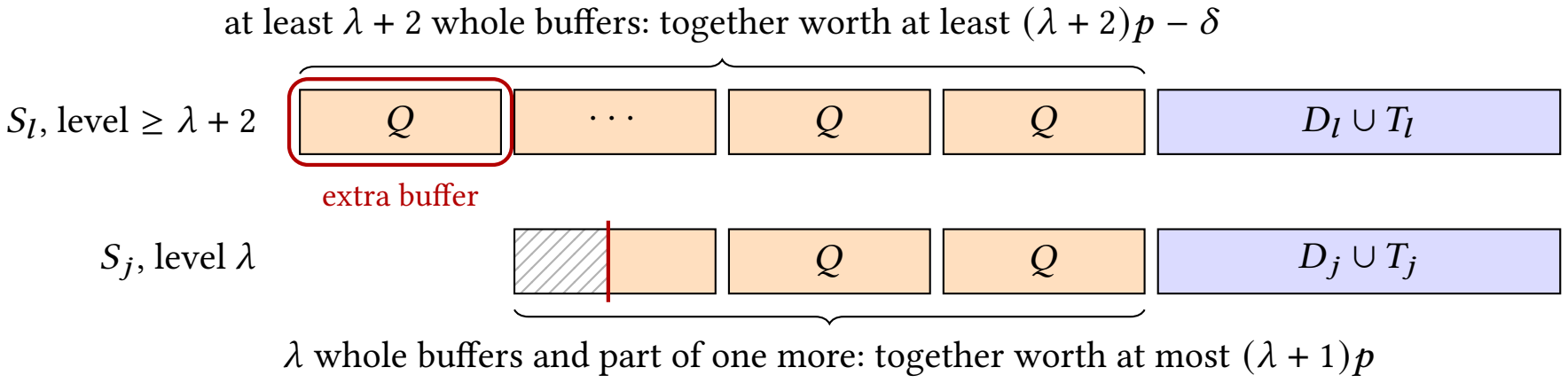

\paragraph{Step 1: two levels apart means strictly more valuable.}
Let $j$ and $l$ be labels with $\ell(j)=\lambda$ and $\ell(l)\ge\lambda+2$, and look through the eyes of the owner $l$ of the upper label $l$, who need not be its final holder. Label $l$ carries at least one more whole buffer than label $j$ could, even if $S_j$ keeps part of its next buffer (Figure~\ref{fig:extra-buffer}). By \eqref{eq:common-union},
\[
 v_l(S_l)\ge v_l(D_l)+t+(\lambda+2)p-\delta,\qquad
 v_l(S_j)\le v_l(D_j)+t+(\lambda+1)p.
\]
The second bound holds also at $\lambda=-1$, since that suffix lies inside its protected part. The input envy bound $v_l(D_l)\ge v_l(D_j)-\alpha$ therefore gives
\begin{equation}\label{eq:level-comparison}
 \ell(l)\ge\ell(j)+2
 \quad\Longrightarrow\quad
 v_l(S_l)-v_l(S_j)\ge p-\delta-\alpha=\tfrac{\alpha}{2}>0.
\end{equation}
In words: one extra buffer is worth $p=2\alpha$, which outweighs both the envy $\alpha$ and the total rounding loss $\delta=\alpha/2$. Adjacent levels may differ by only part of a buffer, which is why two levels are needed.

\paragraph{Step 2: no protected part is cut.}
Suppose some protected part were cut, so that level $-1$ occurs. Level $n-1$ also occurs, since a certified allocation keeps some piece whole. With $n$ labels and $n+1$ possible levels $-1,0,\ldots,n-1$, some intermediate level $\mu$ is missing. Let $H=\{j:\ell(j)<\mu\}$, the labels below the gap; it is nonempty. Every label outside $H$ is at least two levels above every label in $H$. By \eqref{eq:level-comparison}, the owner of each label outside $H$ values her own label's suffix strictly above every suffix in $H$, so none of her favorite labels lies in $H$. Only the $|H|$ owners of labels in $H$ can therefore have a favorite label in $H$, and every label in $H$ is trimmed, since its level is below $\mu\le n-2$. This contradicts the certification test, Lemma~\ref{lem:test}. Thus every protected part survives. The buffers are there exactly to make the owners above a gap strictly prefer their own side.

\paragraph{Step 3: only buffers and the rounding residue remain.}
Let $\sigma(j)$ be the final holder of label $j$. We now have $W_{\sigma(j)}=S_j=D_j\cup Z_j$, where $Z_j$ consists of $T_j$ and the retained buffers and buffer parts. For every $i,j$,
\[
 t-\delta\le v_i(Z_j)\le t+(n-1)p.
\]
Hence $|v_i(Z_j)-v_i(Z_l)|\le(n-1)p+\delta\le2n\alpha$. The residue contains only discarded buffer parts and $U_0$ from Lemma~\ref{lem:shares}, so $v_i(R')\le Bp+\delta\le2n^2\alpha$. Rescale to obtain the claims. There are $n+B=n^2$ slots, giving $O(n^4\log(n^4/\alpha))$ queries beyond the single call to \Trim{}. This proves Lemma~\ref{lem:repair}. \qed

\subsection{Computing the common shares}
\label{sec:shares-proof}
It remains to prove Lemma~\ref{lem:shares}. We adapt the fractional rounding method of \citet{AlonGraur21}, who split a necklace into equal shares between thieves; here the shares are prescribed and unequal, and cells are cut with Robertson--Webb queries. The idea has three steps. First, meet every target exactly with a fractional assignment, in which every agent values every slot at exactly its target. Second, make all but a few cells integral without breaking any target, using the fact that the number of split cells is bounded by the number of equations. Third, repeatedly refine only the split cells. Every child has at most three quarters of its parent's aggregate mass. After $d$ rounds, each still-shared cell therefore has aggregate mass at most $n(3/4)^d$. At most $nm$ cells are still shared, so their total mass is at most $n^2m(3/4)^d$.

How should a split cell be refined? Halving every agent's value of it separately would cut it at up to $n$ places. A single cut suffices if it is placed well, at a weighted median of the agents' half-value marks, much as \citet{EvenPaz84} split a group of agents at a median mark.

\begin{lemma}[One cut for a shared cell]\label{le:weighted}
Let the $n$ normalized measures on an ordered cell $E$ be $\mu_i$, and let $W(E)=\sum_i\mu_i(E)>0$. In $O(n)$ queries one can split $E$ and record the two children so that each has aggregate mass at most $3W(E)/4$.
\end{lemma}
\begin{proof}
For each agent with $w_i=\mu_i(E)>0$, obtain a mark for $w_i/2$ from the front. Give it weight $w_i$ and choose the first mark at which cumulative weight reaches $W(E)/2$. The marks weakly to either side have at least $W(E)/2$ total weight. Each mark weakly to the left contributes at least half its owner's mass to the left child, so that child has mass at least $W(E)/4$. The right child satisfies the same bound. Their sum is $W(E)$, giving the two upper bounds. Only the median mark is retained. Its location subdivides at most one physical component, and one evaluation per agent records the resulting masses by subtraction. The reasoning permits tied marks and zero-weight agents.
\end{proof}

\begin{proof}[Proof of Lemma~\ref{lem:shares}]
If $m=1$, assign all of $U$ to the sole slot without queries. For $m\ge2$, fractions are bookkeeping variables, not physical allocations. Think of each slot as a recipient with target $w_q$. Initially $U$ is one cell, and its fraction in slot $q$ is $w_q$.

\emph{Keep few split cells.} For a cell $E$, denote its current fraction in slot $q$ by $x_{Eq}$. Preserve the equations $\sum_q x_{Eq}=1$ for every cell $E$ and $\sum_E v_i(E)x_{Eq}=w_q$ for every $i,q$. With $\kappa$ cells there are at most $\kappa+nm$ constraints. If more than $\kappa+nm$ fractions are positive, the corresponding columns are linearly dependent. Move the fractions along a dependence until one becomes zero, keeping all fractions nonnegative. The dependence preserves each cell sum, so it has both signs. Only positive fractions are changed, and this ends with at most $\kappa+nm$ positive fractions. Cells assigned to one slot never change: their cell equation fixes their only positive fraction at one. As each cell needs at least one positive fraction, at most $nm$ cells remain split between slots.

\emph{Refine the shared cells.} Permanently assign each integral cell to its slot. A cell of zero aggregate mass can also be given outright to any slot. Split each remaining shared cell once by Lemma~\ref{le:weighted}, give its children their parent's fractions and eliminate dependent columns again. Any cell still shared after $d$ rounds has descended through $d$ splits, since integral cells never become shared again. The initial aggregate mass is $n$, so each surviving shared cell has aggregate mass at most $n(3/4)^d$. There are at most $nm$ of them. Take
\[
 d=\bigl\lceil\log_{4/3}(n^2 m/\delta)\bigr\rceil .
\]
Leaving these cells in the residue loses at most $n^2m(3/4)^d\le\delta$ aggregate mass, and hence at most $\delta$ to any agent. The integral slots remain below their targets because all discarded fractions are nonnegative. At most $nm$ cells cost $O(n)$ queries each per round, including records. Elimination is polynomial in the number of cells, targets and rounds. Fixed pivot and mark orders make all choices deterministic.
\end{proof}

%

\section{Conclusions}\label{sec:conclusion}

Envy-free cake cutting, posed by \citet{GaSt58}, has a long history: it was
solved for three agents by Selfridge and Conway, for any number of agents by
the finite but unbounded protocol of \citet{BrTa95a}, and with a bounded
number of queries by \citet{AzMa16a}.
Building on the domination framework of the latter, we have given a
deterministic protocol that divides a heterogeneous cake completely and
without envy among $n$ agents using polynomially many Robertson--Webb
queries, $O(n^6\log n)$ in our analysis. The main new ingredient is certified
trimming when the supporting assignment is unknown: a shared table of queries
serves all candidate assignments, the median of their estimates bounds each
agent's value from below, and each failed matching test rules out at least
half of the remaining candidates. A second median, of the agents' half-value
marks, lets a single repair turn an advantage into a domination.

We expect these ideas to lead to simpler and faster protocols. Certified
suffix allocations also seem to merit study in their own right: their
structure connects naturally to competitive equilibria, auctions and stable
matchings, and exploring these connections may yield new links between fair
division and market design. 

\section*{Development of This Work and Use of AI}

The author used LLMs, specifically ChatGPT and Claude, for mathematical
exploration and to help with the exposition. Like many follow-up works, the
paper builds on the framework of \citet{AzMa16a,AzMa20a}. The effort focused
on a polynomial reduction of the core protocol to computing a certified
envy-free suffix allocation and on a formalization of structural properties
of these allocations. Some of these properties were known in 2016 and went
unused because the bottleneck in \citet{AzMa16a,AzMa20a} lay elsewhere.
After identifying the new reduction and formalizing these structural
properties, the author worked extensively with LLMs, guiding them to exploit these properties algorithmically, building on the idea of query-efficient medians, used both to filter candidate assignments and to cut shared pieces at weighted medians. The results emerged from an intensive human--AI
collaboration.

\section*{Acknowledgments}
The author thanks Sarmad Abbasi (for editorial feedback), Florian Brandl (for raising the question of restricting queries to intervals with established endpoints),  and Ariel Procaccia (for a suggestion regarding a linear improvement in the bound). The argument in Appendix~\ref{app:batched-cake} comes from a note shared by Procaccia. With a small polynomial upper bound now in place, several experts are now collaborating towards a final sprint towards a final resolution of the lower and upper query bound.

\bibliographystyle{ACM-Reference-Format}
\bibliography{references}

\appendix
\begin{table}[H]
\caption{Notation for one round of \Divide{} and for certified trimming.}
\label{tab:notation}
\centering\small
\renewcommand{\arraystretch}{1.08}
\begin{tabularx}{\linewidth}{@{}l>{\raggedright\arraybackslash}X@{\hspace{5pt}}l@{}}
\toprule
Symbol & Meaning & Where\\
\midrule
$X,\,R$ & Allocation and residue at the start of a round & \S\ref{sec:mainloop}\\
$Y,\,U$ & Suffixes from \Core{}, indexed by their recipients, and its residue & \S\ref{sec:mainloop}\\
$\pi(i),\,\sigma(j)$ & Agent $i$ receives $Y_{\pi(i)}$ on reassignment. Label $j$ goes to holder $\sigma(j)$ after repair. & \S\ref{sec:mainloop}\\
$D_j,\,Z_j$ & $D_j=X_j\cup Y_{\pi(j)}$ enters \Repair{}, which adds $Z_j$ to label $j$ & \S\ref{sec:mainloop}\\
$W,\,R'$ & Repaired allocation and residue, with $W_{\sigma(j)}=D_j\cup Z_j$ and $R'\subseteq U\subseteq R$ & \S\ref{sec:guarantees}\\
$\Gamma$ & Non-domination graph & \S\ref{sec:guarantees}\\
$g_{ij}$ & Normalized gap $(v_i(Y_i)-v_i(Y_j))/v_i(U)$ & \S\ref{sec:advantage}\\
\midrule
$P_j,\,S_j$ & Input piece on label $j$ and its retained suffix & \S\ref{sec:certified}\\
$\mathcal M_i(S)$ & Agent $i$'s favorite labels in a suffix allocation $S$ & \S\ref{sec:certified}\\
$o$ & The pin, a node standing for ``whole'' in the tie graph & \S\ref{sec:matching-view}\\
$V_{ij},\,t_j(x)$ & Agent $i$'s value of a suffix of piece $j$ as a function of its length, whose inverse gives her marks; and the length of piece $j$ at values $x$ & \S\ref{sec:pin}\\
$M,\,\A$ & An assignment of pieces to agents, and the pool of candidate assignments & \S\ref{sec:pin}, \S\ref{sec:median}\\
$x^*,\,\hat x^M$ & Agents' values in a certified allocation (the fully exposable vector), and their estimates for candidate assignment $M$ & \S\ref{sec:pin}, \S\ref{sec:median}\\
$y$ & Lower medians of the candidates' estimates, checked by the matching test & \S\ref{sec:median}\\
\bottomrule
\end{tabularx}
\end{table}

\section{The matching test}\label{app:matching}
This appendix proves Lemma~\ref{le:which}, which is used to filter candidates in Algorithm~\ref{alg:median}. Both searches use standard bipartite matching. We show what each one certifies about the trial values using only the table at $y$.

\begin{proof}[Proof of Lemma~\ref{le:which}]
For a piece $j$ let $N_x(j)$ be the set of nodes joined to $j$ in the tie graph at $x$: the agents tight on $j$, and the pin if $j$ is whole. Write $N_x(J)=\bigcup_{j\in J}N_x(j)$ for a set $J$ of pieces. We use the strict Hall inequality~\eqref{eq:strict-hall} at $x^*$.

Let $E$ consist of the pin and the agents $i$ with $y_i=x^*_i$, the nodes whose values are already exact. Since $y\le x^*$, every piece is at most as long at $y$ as at $x^*$. Whenever $N_y(j)$ meets $E$, piece $j$ therefore has the same length at $x^*$ as at $y$, and strict monotonicity of the marks gives $N_{x^*}(j)=N_y(j)\cap E$. We use this observation in both cases.

Build the tie graph at $y$, delete the pin, and take a maximum matching of pieces to agents. All lengths are computed before the pin is deleted.

\emph{An unmatched piece remains.} Search from one unmatched piece $j_0$ along tie-graph edges to agents and back along matching edges. Write $J'$ and $I'$ for the reached pieces and agents. Every agent in $I'$ is matched, since a free reached agent would give an augmenting path. Also $I'\ne\varnothing$: otherwise $j_0$ is joined only to the pin, and then $N_{x^*}(j_0)=\{o\}$, contradicting \eqref{eq:strict-hall}.

For $\varnothing\ne T\subseteq I'$, let $J_T=\{j\in J':N_y(j)\cap T\ne\varnothing\}$. The pieces matched to $T$ already give $|T|$ members. A search path entering $T$ for the first time supplies one more, since the piece preceding its first agent in $T$ is unmatched or matched to an agent outside $T$. Hence $|J_T|\ge |T|+1$. Take $T=I'\cap E$. If it were nonempty, every piece in $J_T$ would keep its length, and every agent tight on it at $x^*$ would lie in $T$. Allowing for the pin, this would give
\[
      |J_T|+1\ \le\ |N_{x^*}(J_T)|\ \le\ |T|+1\ \le\ |J_T|,
\]
where the first inequality is \eqref{eq:strict-hall}. Thus $I'\cap E=\varnothing$, and every $k\in I'$ satisfies $y_k<x^*_k$.

\emph{All pieces are matched.} Restore the pin, which is now free, and search from it along tie-graph edges to pieces and along matching edges back to agents. Let $A$ consist of the pin and the reached agents. These are exactly the nodes that a matching of all pieces along tie-graph edges at $y$ can leave uncovered. Switching the matching along a search path proves one direction, and superposing the current matching with a matching that leaves a given node uncovered proves the converse.

If such a matching left uncovered a node outside $E$, it would cover all $|E|$ nodes of $E$. The pieces $J$ matched to them would keep their lengths and satisfy $N_{x^*}(J)\subseteq E$, hence $|J|+1\le|N_{x^*}(J)|\le |E|=|J|$. This proves $A\subseteq E$. If $T=E\setminus A$ were nonempty, let $J$ be the $|T|$ pieces matched to it. No such piece is joined at $y$ to a node of $A$, since that edge would make its matched agent reachable. Each keeps its length through its matched agent in $T$, so $N_{x^*}(J)\subseteq T$, and again $|J|+1\le|N_{x^*}(J)|\le|T|=|J|$. Therefore $A=E$. If $A$ contains every agent, then $y=x^*$ and the current matching is a supporting assignment. Otherwise every agent $k\notin A$ satisfies $y_k<x^*_k$.
\end{proof}

\section{Zero-valued intervals and empty pieces}\label{sec:ties}
The positive solver uses the same suffix position for every agent and compares one agent's values across pieces. On a zero-mass interval several positions can give the same value, so breaking ties independently on each agent--piece pair can destroy a comparison. We simulate the agents' suffix curves $V_{ij}$ themselves; the marks used by the median algorithm are their inverses. \citet{YeBai26} also use ordered keys to break ties.

\begin{lemma}[Realization and projection of a trimming transcript]\label{le:transcript}
Let $F_{ij}(p)$ be agent $i$'s mass of the suffix at physical position $p$ on ordered piece $j$. As the suffix grows from empty to whole, $F_{ij}$ is continuous and nondecreasing along every component. An empty piece has one position. An isolated endpoint component has positions before and after its inclusion, with equal mass. Given component records, each forward or inverse query to a virtual strictly increasing suffix curve costs at most one physical query. Every finite adaptive transcript is realized by a single positive profile. For a deterministic solver correct on positive inputs, registering its returned suffix lengths for all agents takes at most $n^2$ final evaluations. Its returned allocation then projects to a physical certified allocation with the same assignment and certifying order.
\end{lemma}
\begin{proof}
For each agent--piece pair $(i,j)$, let $T_{ij}$ be the finite table of registered virtual length--amount pairs. The simulation maintains three invariants on the registered keys.
\begin{enumerate}[label=\textup{(\roman*)},leftmargin=*,nosep]
\item For each piece $j$, a single nondecreasing map $\phi_j$ takes virtual lengths in $[0,1]$ to physical suffix positions. Every agent uses this map on $j$.
\item For each agent $i$, a single nondecreasing map $\psi_i$ takes virtual amounts to true values. She uses this map on every piece.
\item For each pair $(i,j)$, the length--amount pairs $(t,x)$ in $T_{ij}$ are strictly increasing in both coordinates in the same order, and satisfy
\begin{equation}\label{eq:realize}
                  F_{ij}(\phi_j(t))=\psi_i(x) \qquad ((t,x)\in T_{ij}).
\end{equation}
\end{enumerate}
Distinct keys may have the same physical image. Even an empty piece can therefore have a strictly increasing virtual curve.

\emph{Initialization.} Map $0,1$ on each piece to its empty and whole positions, even when these coincide. For each agent $i$, sort her values of the whole pieces, using a fixed label order to resolve ties. Give them distinct ranks $w_{ij}\in\{1,\ldots,n\}$ in this order. Set $\psi_i(0)=0$, map each rank to its true whole amount, and start $T_{ij}$ with $(0,0),(1,w_{ij})$. All three invariants hold.

\emph{Insertion in an ordered map.} Let $h$ denote either $\phi_j$ or $\psi_i$ on its registered keys. Suppose two existing keys $a<b$ bracket a requested image $u$ in the sense that $h(a)\le u\le h(b)$. Among the registered keys from $a$ to $b$ there are consecutive keys $d<e$ with $h(d)\le u\le h(e)$. This follows by inspecting the first image at least $u$, using the first adjacent pair if $h(a)=u$. Insert the fresh midpoint $(d+e)/2$ with image $u$. This preserves order, including when all three images are equal. Conversely, to insert a prescribed fresh key, any image between the images of its two global neighbors is admissible. Fix the first eligible gap and the lower neighboring image for these choices.

\emph{Forward query at $t$.} Return the stored amount if $T_{ij}$ already contains $t$. Otherwise its adjacent entries $(t^-,x^-),(t^+,x^+)$ bracket $t$. If $t$ has no image under $\phi_j$, insert that prescribed key as above. Evaluate $u=F_{ij}(\phi_j(t))$. By monotonicity and \eqref{eq:realize},
$\psi_i(x^-)\le u\le\psi_i(x^+)$. Use the insertion rule on $\psi_i$ between $x^-,x^+$ to choose a fresh $x$ with image $u$. Add $(t,x)$ to $T_{ij}$ and return $x$.

\emph{Inverse query at $x$.} Return the stored length if $x$ is in the second column of $T_{ij}$. Otherwise bracket it by $(t^-,x^-),(t^+,x^+)$. If necessary insert the prescribed key $x$ into $\psi_i$, between the images of its global neighbors. Put $u=\psi_i(x)$. Request a physical suffix of amount $u$ on $j$, and clamp the returned position to $[\phi_j(t^-),\phi_j(t^+)]$. This preserves its amount. For example, a returned position left of the bracket has amount $u$, while the left endpoint has amount at most $u$ by \eqref{eq:realize} and at least $u$ by monotonicity. The other endpoint is symmetric. Insert the resulting position as the image of a fresh key $t$ in the global map $\phi_j$, between $t^-,t^+$. Add $(t,x)$ and return $t$. Both rules operate only inside the virtual endpoint pairs. Outside this range use the known slope-one continuation $w_{ij}+t-1$ or its inverse without a physical query.

\emph{One physical query suffices.} Each update preserves the registered pairs and inserts the new pair strictly between its neighbors in both coordinates. Identity~\eqref{eq:realize} holds by construction, including for any valid cut answer on a plateau. The component records locate the boundary interval, so a new virtual query needs at most one ordinary evaluation or cut there.

\emph{Realizing the entire transcript.} Fix a finite transcript, including any evaluations made to read the output. For every pair $(i,j)$, join successive entries of $T_{ij}$ by straight lines and append its prescribed tail. The resulting strictly increasing curve agrees with every forward and inverse answer for that pair. Its increments on the virtual piece define a positive nonatomic measure. Adding these measures across the disjoint virtual pieces gives one valuation profile realizing the entire transcript. If the simulated solver exceeded its uniform query bound, that finite prefix would therefore occur on a positive input as well. The same argument transfers its operation bound, apart from maintaining the finite tables.

\emph{Projecting the certificate.} Register each suffix length returned by the solver for every agent, using at most $n^2$ final evaluations. On every positive profile realizing this extended transcript, determinism gives the same returned lengths, assignment and certifying order. By (i), each chosen length denotes one physical suffix for everyone. By (ii) and \eqref{eq:realize}, a comparison $x\le x'$ across two pieces for agent $i$ gives $\psi_i(x)\le\psi_i(x')$, and a witness equality gives equal physical values. Whole virtual pieces still project to whole pieces. If another suffix also projects to a whole piece, it already satisfies the whole-piece clause. Thus the same assignment is physically envy-free, and the same order certifies every physical trim.
\end{proof}
\begin{example}[A virtual tie on two zero-valued pieces]
Suppose one agent sees masses $0$, $0$, and $1$ on pieces $A,B,C$, where $A$ is empty and $B$ is nonempty but worthless. With label order $A,B,C$, initialization gives $\psi_i(0)=\psi_i(1)=\psi_i(2)=0$ and $\psi_i(3)=1$. Asking for amount $1$ on $B$ returns an interior virtual length between $0$ and $1$, since $T_{iB}$ has endpoint amounts $0$ and $2$. Its physical image can be anywhere among the admissible zero-mass positions on $B$. It has exactly the same true amount as whole $A$, whose virtual amount is also $1$. This is the equality a certificate must preserve.
\end{example}

\section{General capped squares}\label{app:square}
The arguments of Sections~\ref{sec:pin} and~\ref{sec:median} and of Appendix~\ref{app:matching} use the marks only as continuous strictly increasing bijections of $[0,\infty)$ onto itself, and the lengths only through the minimum~\eqref{eq:recover-cake}, with cap one. This appendix states the same results for general curves of this kind. Appendices~\ref{sec:trim-reductions} and~\ref{app:market} use them for chores and assignment markets.

A \emph{capped square} has $q$ rows and $q+1$ columns: $q$ \emph{ordinary} columns and one \emph{spare} column $o$, the pin. Write $\cK$ for the set of all columns. A vector $x\in[0,\infty)^{\cK}$ assigns a \emph{level} to every column. Each row $i$ and column $j$ have a \emph{cost curve} $f_{ij}$, a continuous strictly increasing bijection of $[0,\infty)$ onto itself. A forward query returns $f_{ij}(x)$, and an inverse query returns $f^{-1}_{ij}(z)$. At levels $x$ the cost of row $i$ is
\[
 z_i(x)=\min_{j\in\cK}f_{ij}(x_j),
\]
and the columns attaining the minimum are its \emph{tight} columns. The pin is fixed at level $x_o=1$, and row $i$ has the linear curve $f_{io}(t)=c_i t$ on it. Thus the pin offers row $i$ a fixed alternative of cost $c_i>0$, its \emph{cap}. A vector of ordinary levels is a \emph{capped equilibrium} if a perfect matching between rows and ordinary columns assigns each row a cheapest ordinary column, at a cost at most its cap. A vector $x$ with $x_o=1$ is \emph{fully exposable} if, for every column $j\in\cK$, the pin included, the tight edges can match all rows while leaving $j$ uncovered. At a fully exposable vector, a tight matching of all rows that leaves the pin uncovered is a \emph{supporting assignment}.

Certified cake trimming is the capped square with $q=n$, the pieces as rows, the agents as ordinary columns, the marks $f_{ji}=V_{ij}^{-1}$ as curves, and every cap equal to one. The cost of row $j$ is then the length $t_j(x)$ of~\eqref{eq:recover-cake}, a row is tight with an agent or with the pin exactly when it is joined to it in the tie graph, capped equilibria are the envy-free value vectors, and full exposability and supporting assignments are as in Section~\ref{sec:pin}.

Conversely, divide each row's curves by its cap, setting $f'_{ij}=f_{ij}/c_i$. This gives cap one and preserves the minimizing columns in every row at every vector. A query to a normalized curve or its inverse costs one original query and one arithmetic operation. Regard row $i$ as a cake piece and ordinary column $j$ as an agent. On physical lengths $0\le t\le1$, set $V_{ji}(t)=(f'_{ij})^{-1}(t)$. These continuous strictly increasing functions start at zero, so their increments define positive nonatomic valuations on the pieces. Extend $V_{ji}$ with slope one as in Section~\ref{sec:pin}. For targets above $V_{ji}(1)$, both the original and the new mark exceed one, so the capped lengths and tight edges agree. After recording the $q^2$ whole-piece values, each evaluation or mark costs at most one original query.

Proposition~\ref{pr:pinned} now gives a unique fully exposable vector $x^*$, whose ordinary levels form the coordinatewise greatest capped equilibrium. In the tie graph at $x^*$, every supporting assignment gives each row a path of ties to a row attaining its cap. Lemmas~\ref{le:median} and~\ref{le:which} and Theorem~\ref{th:pinned} apply to these value curves. For $q\ge2$, Algorithm~\ref{alg:median} therefore computes $x^*$ and a supporting assignment in $O(q^5\log q)$ forward and inverse queries. With one row, its sole ordinary level is $f_{11}^{-1}(c_1)$.

\section{Polynomial reductions between cake, chores, and equilibrium}
\label{sec:trim-reductions}
Inverting suffix functions and exchanging the roles of agents and pieces gives polynomial reductions in both directions. One direction needs an extra agent and piece to normalize the endpoints.

Assume that each piece consists of one or more intervals of positive length and that every agent assigns positive value or cost to every interval of positive length. Parameterizing each suffix by its retained length in $[0,1]$ gives continuous, strictly increasing suffix functions vanishing at zero. We use Robertson--Webb queries, exact arithmetic, and the component records of Section~\ref{sec:preliminaries}. Thus \emph{polynomial time} means polynomially many oracle calls and additional exact operations. This is not a bit-complexity claim for unrestricted real functions. The reductions below require strict growth; Appendix~\ref{sec:ties} handles zero values for certified trimming itself.

\subsection{Two-way reductions for certified trimming}
\label{sec:cake-chore-reduction}
For chores, each envy-free agent receives a least-cost suffix; certification orders the holders so that every proper suffix's holder ties an earlier holder's suffix with her own. For both signs, the certified suffix vector is the coordinatewise greatest envy-free vector, and every envy-free matching of it is certified. For cake this follows from Proposition~\ref{prop:greatest}, Lemma~\ref{lem:test} and Section~\ref{sec:pin}. For chores, use a capped square (Appendix~\ref{app:square}) in which agents are rows, pieces are ordinary columns, the curves are the costs $f_{ij}$, and the caps are $c_i=\min_j f_{ij}(1)$. Every physical envy-free matching satisfies the caps, and in any capped equilibrium the row assigned to each ordinary column bounds its level by one. At the greatest vector $x^*$, a holder attaining her cap ties a whole piece. Indeed, if $f_{ik}(1)=c_i$, then $c_i\le f_{ik}(x^*_k)\le c_i$, so $x^*_k=1$. Proposition~\ref{pr:pinned}, in the general form of Appendix~\ref{app:square}, leads every holder along ties to such a row, certifying every ordinary tight matching. Conversely, in a physical chore certificate every holder reaches a whole-piece holder. That holder attains her cap. A switch along the path ending at the pin vacates any chosen ordinary column, which characterizes the same greatest vector.\looseness=-1

\begin{proposition}[Polynomial-time oracle equivalence]
\label{prop:trim-equivalence}
Certified chore trimming on $n$ agents and pieces reduces to certified cake trimming on $n$ agents and pieces. The reverse reduction uses at most $n+1$ agents and pieces. Each uses one target-solver call, $O(n^2)$ preprocessing and recovery queries, $O(1)$ original queries per simulated query, and polynomial additional work in the available records. Both recover the greatest suffix vector and a certified assignment.
\end{proposition}
\begin{proof}
\emph{From chores to cake.}
Let $f_{ij}(x)$ be chore agent $i$'s cost of the suffix of length $x$ on piece $j$, and set $b_i=\min_j f_{ij}(1)>0$. Create one cake agent for each chore piece $j$, and one cake piece for each chore agent $i$, with suffix values
\begin{equation}\label{eq:inverse-transpose}
 G_{ji}(s)=f_{ij}^{-1}(b_i s),\qquad 0\le s\le1.
\end{equation}
The inverse is defined because $b_i\le f_{ij}(1)$. Each $G_{ji}$ is continuous, strictly increasing, and zero at zero, so it specifies a valid nonatomic valuation on the new piece. Figure~\ref{fig:inverse-transpose} shows how a matched pair and its suffix values transform.

The following maps convert envy-free allocations in both directions, reversing their matchings:
\begin{equation}\label{eq:trim-correspondence}
 x_j=\max_i G_{ji}(s_i),
 \qquad
 s_i=\frac{\min_j f_{ij}(x_j)}{b_i}.
\end{equation}
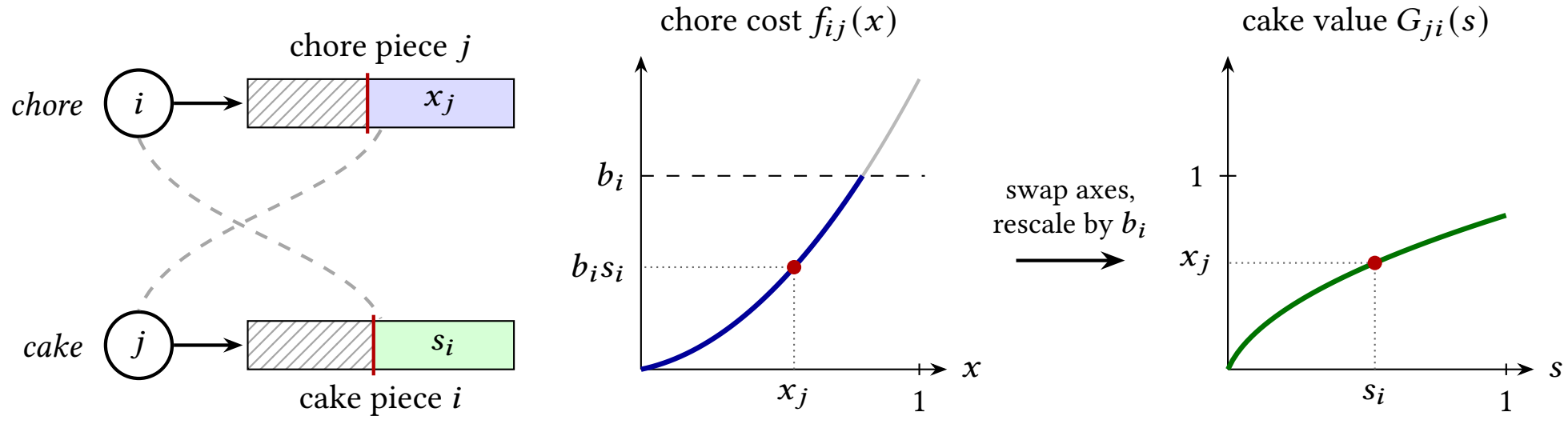
\begin{figure}[htbp]
  \centering
  \begin{tikzpicture}[font=\footnotesize,
    ag/.style={circle,draw,thick,fill=white,minimum size=5.5mm,inner sep=0pt},
    hold/.style={-{Stealth[length=1.6mm]},thick},
    swap/.style={gray!70,dashed,thick},
    guide/.style={densely dotted,gray!80!black},
    axis/.style={-{Stealth[length=1.4mm]}},
    pt/.style={circle,fill=red!70!black,inner sep=1.2pt}]
    \begin{scope}
      \node[anchor=east,font=\footnotesize\itshape] at (-0.35,2.2) {chore};
      \node[anchor=east,font=\footnotesize\itshape] at (-0.35,0.2) {cake};
      \node[ag] (ci) at (0,2.2) {$i$};
      \fill[pattern=north east lines,pattern color=gray!70] (0.9,2.0) rectangle (1.89,2.4);
      \fill[blue!14] (1.89,2.0) rectangle (3.1,2.4);
      \draw (0.9,2.0) rectangle (3.1,2.4);
      \draw[red!70!black,thick] (1.89,1.95)--(1.89,2.45);
      \node at (2.495,2.2) {$x_j$};
      \node[above] at (2.0,2.4) {chore piece $j$};
      \draw[hold] (ci)--(0.86,2.2);
      \node[ag] (kj) at (0,0.2) {$j$};
      \fill[pattern=north east lines,pattern color=gray!70] (0.9,0.0) rectangle (1.94,0.4);
      \fill[green!16] (1.94,0.0) rectangle (3.1,0.4);
      \draw (0.9,0.0) rectangle (3.1,0.4);
      \draw[red!70!black,thick] (1.94,-0.05)--(1.94,0.45);
      \node at (2.52,0.2) {$s_i$};
      \node[below] at (2.0,0.0) {cake piece $i$};
      \draw[hold] (kj)--(0.86,0.2);
      \draw[swap] (ci.south)..controls (0.3,1.2) and (1.7,1.0)..(2.0,0.42);
      \draw[swap] (2.0,1.98)..controls (1.7,1.4) and (0.3,1.2)..(kj.north);
    \end{scope}
    \begin{scope}[shift={(4.15,0)},x=2.3cm,y=1.6cm]
      \draw[axis] (0,0)--(1.1,0) node[right] {$x$};
      \draw[axis] (0,0)--(0,1.62);
      \node[above] at (0.5,1.62) {chore cost $f_{ij}(x)$};
      \draw[gray!55,thick,domain=0.7964:1,samples=20] plot (\x,{1.2*\x*\x+0.3*\x});
      \draw[blue!60!black,very thick,domain=0:0.7964,samples=40] plot (\x,{1.2*\x*\x+0.3*\x});
      \draw[dashed] (0,1)--(1.02,1);
      \node[left] at (0,1) {$b_i$};
      \draw (1,0.03)--(1,-0.03) node[below] {$1$};
      \draw[guide] (0.55,0)--(0.55,0.528)--(0,0.528);
      \node[below] at (0.55,0) {$x_j$};
      \node[left] at (0,0.528) {$b_is_i$};
      \node[pt] at (0.55,0.528) {};
    \end{scope}
    \draw[-{Stealth[length=2mm]},thick] (7.25,0.9)--(8.15,0.9);
    \node[above,align=center,font=\scriptsize] at (7.7,0.95) {swap axes,\\ rescale by $b_i$};
    \begin{scope}[shift={(9.0,0)},x=2.3cm,y=1.6cm]
      \draw[axis] (0,0)--(1.1,0) node[right] {$s$};
      \draw[axis] (0,0)--(0,1.62);
      \node[above] at (0.5,1.62) {cake value $G_{ji}(s)$};
      \draw[green!45!black,very thick,domain=0:1,samples=60] plot (\x,{(-0.3+sqrt(0.09+4.8*\x))/2.4});
      \draw (1,0.03)--(1,-0.03) node[below] {$1$};
      \draw (0.03,1)--(-0.03,1) node[left] {$1$};
      \draw[guide] (0.528,0)--(0.528,0.55)--(0,0.55);
      \node[below] at (0.528,0) {$s_i$};
      \node[left] at (0,0.55) {$x_j$};
      \node[pt] at (0.528,0.55) {};
    \end{scope}
  \end{tikzpicture}
  \caption{The chore-to-cake construction~\eqref{eq:inverse-transpose} for a matched pair. On the left, chore agent~$i$ holds a suffix of length $x_j$ on piece~$j$, and cake agent~$j$ holds a suffix of length $s_i$ on piece~$i$. Dashed lines show the exchange of roles, and the trimmed fronts are hatched. The highlighted part of $f_{ij}$ runs up to $b_i=\min_k f_{ik}(1)$. Swapping the axes and dividing costs by $b_i$ gives the cake value $G_{ji}(s)=f_{ij}^{-1}(b_i s)$. The marked points illustrate $f_{ij}(x_j)=b_i s_i$ and $G_{ji}(s_i)=x_j$ from~\eqref{eq:trim-correspondence}, with $b_i=1$ in the plots.}
  \Description{Left: a chore agent i holds a suffix of chore piece j. Below it, cake agent j holds a suffix of cake piece i. Dashed lines show that agent i becomes piece i and piece j becomes agent j. Middle: an increasing chore cost curve with a dashed level b. The part below that level is thick and contains a marked point. Right: the inverse curve, the cake value, through the corresponding marked point with the axes swapped.}
  \label{fig:inverse-transpose}
\end{figure}

For the first map, suppose cake agent $j$ holds cake piece $i$. Envy-freeness gives $G_{ji}(s_i)=x_j$, so $f_{ij}(x_j)=b_i s_i$. For every other chore label $k$, the definition of $x_k$ gives $f_{ik}(x_k)\ge b_i s_i$. Giving chore piece $j$ to agent $i$ is therefore envy-free. Conversely, in a chore allocation let $d_i=\min_j f_{ij}(x_j)$. Since $x_j\le1$, we have $0\le d_i\le b_i$. With $s_i=d_i/b_i$, every $G_{ji}(s_i)\le x_j$, with equality when $i$ holds $j$. Thus the reversed matching is cake-envy-free.

The matched equalities also show that the two maps in~\eqref{eq:trim-correspondence} are mutual inverses on envy-free suffix vectors. Both maps are coordinatewise nondecreasing. They therefore carry the greatest envy-free vector of either instance to the greatest vector of the other. By the characterization above, a certified cake output recovers a certified chore allocation.

\emph{From cake to chores: normalize the whole-piece values.}
Let $v_{ij}(s)$ be the input cake values. Put $M=2\max_{i,j}v_{ij}(1)$ and $h_{ij}=v_{ij}/M$. This common scaling changes no comparison or certificate. Add a dummy agent $0$ and piece $0$. With $\delta_i=\tfrac12\min_j h_{ij}(1)>0$, define
\begin{equation}\label{eq:dummy-normalization}
 H_{ij}(s)=h_{ij}(s)\ (i,j\ne0),\qquad
 H_{i0}(s)=\delta_i s\ (i\ne0),\qquad
 H_{0j}(s)=s\ (\text{all }j).
\end{equation}
Consequently $\max_i H_{ij}(1)=1$ for every piece $j$.

This padding preserves the certified solution. To see this, let $s^*$ be the original certified vector. Some original piece $k$ is whole. By envy-freeness, each real agent's own value is at least $h_{ik}(1)\ge2\delta_i$, so she does not envy the whole dummy piece. The dummy agent can hold that piece without envy, since every suffix has length at most one. Put her first in the original certificate: this certifies $(s^*,1)$ in the enlarged instance. Uniqueness forces every certified output there to have these same lengths. In any such output, no real agent can hold piece $0$, because she would envy the whole original piece $k$. Removing agent and piece $0$ therefore leaves an envy-free matching of $s^*$, which is certified by Lemma~\ref{lem:test}.

Extend each $H_{ij}$ beyond $1$ by $H_{ij}(s)=H_{ij}(1)+s-1$ for $s\ge1$. Define the chore instance, again transposing agents and labels, by
\begin{equation}\label{eq:cake-to-chore}
 F_{ji}(x)=H_{ij}^{-1}(x),\qquad 0\le x\le1.
\end{equation}
Here $\min_i F_{ji}(1)=1$: all whole-piece cake values are at most one and the dummy agent's are exactly one. Applying~\eqref{eq:inverse-transpose} to this chore instance therefore returns precisely the enlarged cake instance on $[0,1]$. The bidirectional correspondence already proved converts any certified chore output to its certified cake output; deleting the dummy agent and piece solves the original instance.

\emph{Oracle cost.}
Evaluating $G_{ji}$ uses one original suffix cut; its inverse $G_{ji}^{-1}(z)=f_{ij}(z)/b_i$ uses one suffix evaluation. The reverse construction adds only explicit scaling and linear functions. Whole-piece records and output recovery cost $O(n^2)$ queries. With these records, arbitrary virtual evaluations and cuts use $O(1)$ boundary queries, summing known values for intervening whole pieces. Searching records, reversing the matching, and recovering a certificate by reachability take polynomial work.
\end{proof}

\subsection{Connection to minimum-price equilibrium}
\label{sec:cake-minimum-equilibrium}
Create one buyer per cake agent, one indivisible good per piece, and an outside option of utility zero. Buyer $i$'s utility for good $j$ at price $p\ge0$ is
\begin{equation}\label{eq:cake-market}
 U_{ij}(p)=
 \begin{cases}
 v_{ij}(1-p),&0\le p\le1,\\
 1-p,&p\ge1.
 \end{cases}
\end{equation}
These utilities are continuous, strictly decreasing, and need not be quasilinear. A competitive equilibrium assigns each buyer a utility-maximizing good or the outside option, and every unassigned good has price zero \citep{AJM16}.

Every equilibrium matches all buyers: otherwise an unmatched buyer strictly prefers an unassigned zero-price good to the outside option. Individual rationality implies $p_j\le1$. Equilibria therefore correspond exactly to envy-free suffix allocations through $s_j=1-p_j$, with the same matching. By Proposition~\ref{prop:greatest}, the certified vector $s^*$ corresponds to the coordinatewise minimum equilibrium prices $p_j^{\min}=1-s_j^*$. This gives polynomial-time oracle reductions in both directions between certified cake trimming and minimum-price equilibrium for the class~\eqref{eq:cake-market}. If only prices are returned, $O(n^2)$ utility evaluations and a perfect matching recover the allocation. Utility and inverse-utility queries translate into suffix evaluations and cuts with constant overhead.

These reductions carry the polynomial query guarantee to the displayed class of nonlinear markets, with utility and inverse-utility oracles. Appendix~\ref{app:market} applies the median algorithm of Theorem~\ref{th:pinned} directly to a more general assignment market.

\section{Buyer-optimal assignment with nonlinear utilities}\label{app:market}
The capped square of Appendix~\ref{app:square} also gives a query bound in a unit-demand assignment market with item reserves and buyer outside options. \citet{AJM16} characterize extremal equilibria for general utility curves. \citet{DHW15} study assignment with explicitly represented general utility curves. Here each curve is given by exact forward and inverse queries. The conclusion concerns the number of such queries.

There are $b\ge1$ buyers and $m\ge1$ items. Buyer $i$ has outside utility $\omega_i$ and utility $U_{ij}(P_j)$ when she receives item $j$ at price $P_j$. Item $j$ has reserve $r_j$. A competitive equilibrium consists of a partial matching and prices $P_j\ge r_j$. A matched buyer receives an item maximizing her utility over all items and her outside option. An unmatched buyer attains her maximum at the outside option. Every unsold item has price $r_j$. Reserves in this model depend only on the item.

\begin{corollary}[Polynomial queries for buyer-optimal assignment]\label{co:assignment}
Suppose every $U_{ij}$ on $[r_j,\infty)$ is continuous and strictly decreasing, and falls strictly below $\omega_i$ at some finite price. A forward query returns $U_{ij}(P)$ and an inverse query returns the price attaining a specified utility in the curve's range. With these oracles a deterministic algorithm computes the componentwise minimum equilibrium price vector and a supporting matching in
\[
                     O((b+m)^5\log(2(b+m)))
\]
queries. The returned equilibrium simultaneously maximizes every buyer's equilibrium utility. The direct implementation uses $(b+m)!\,(b+m)^{O(1)}$ exact operations.
\end{corollary}
\begin{proof}
\emph{Normalize and extend the curves.}
Put $p_j=P_j-r_j$ and $V_{ij}(p)=U_{ij}(r_j+p)-\omega_i$ for $p\ge0$. Outside utility and reserves are now zero. Extend each $V_{ij}$ to negative prices by $V_{ij}(p)=V_{ij}(0)-p$ for $p<0$. Write $w_{ij}(u)$ for its inverse at $u\ge0$. The inverse exists because the extension is unbounded above and the original curve eventually drops below zero. If $u>V_{ij}(0)$ its price is $V_{ij}(0)-u<0$, computed arithmetically. Otherwise one original inverse query suffices. Thus the extension introduces no assumption about preferences at negative prices in the original market.

\emph{Make every assignment full.}
Add $m$ dummy buyers and $b$ dummy items, giving $N=b+m$ buyers and $N$ items. Every pair involving a dummy has utility $-p$ and outside utility zero. An original equilibrium extends to a full matching in the padded market. Pair unmatched real buyers with dummy items, unsold real items with dummy buyers, and then pair the remaining dummies. Keep the original prices on real items and give every dummy item price zero. Existence for continuous strictly decreasing utilities and finite willingness to pay follows from the competitive-equilibrium theorem of \citet{AJM16}.

Conversely, in any full individually rational equilibrium of this padded market, every dummy item is priced zero. Its holder has utility $-p\ge0$ and its price is nonnegative. Every dummy buyer also receives utility zero and can hold only a zero-price item. On deleting the dummies, unmatched real buyers consequently have utility zero and unsold real items have price zero. All remaining demand inequalities persist, so the restriction is an original equilibrium.

\emph{Encode the market as a capped square.}
Let items be the rows and buyers be the ordinary columns. Choose
\[
 c_j=1+\max\bigl(\{0\}\cup\{w_{ij}(0):i\text{ a padded buyer}\}\bigr).
\]
For a buyer-column level $x\ge1$ define the cost
\[
              f_{ji}(x)=c_j-w_{ij}(x-1).
\]
On $0\le x\le1$ use the line $x(c_j-w_{ij}(0))$. These functions are continuous strictly increasing bijections of the nonnegative ray, since $c_j>w_{ij}(0)$ and $w_{ij}(u)\to-\infty$ as $u\to\infty$. Give row $j$ cap $c_j$, implemented by the pin. Every forward or inverse query to these cost curves uses at most one original utility or inverse-utility query, apart from the initial values. For example, above the linear segment an inverse query at cost $z$ returns $x=1+V_{ij}(c_j-z)$. Its argument is at most $w_{ij}(0)$, so the required utility is nonnegative. Negative arguments use the known extension.

A padded equilibrium with utilities $u_i\ge0$ and prices $p_j\ge0$ satisfies
\[
      w_{ij}(u_i)\le p_j\quad\text{for every }i,j,
      \qquad w_{ij}(u_i)=p_j\quad\text{on matched pairs}.
\]
Hence $x_i=1+u_i$ and the reversed matching form a capped equilibrium, with row cost $c_j-p_j\le c_j$. By Proposition~\ref{pr:pinned} in the general form of Appendix~\ref{app:square}, the greatest capped equilibrium $x^*$ dominates this one. In particular every $x^*_i\ge1$. Apply Theorem~\ref{th:pinned} in the general form of Appendix~\ref{app:square}, reverse its ordinary matching, and put
\[
 u_i^*=x^*_i-1,\qquad
 p_j^*=c_j-\min_i f_{ji}(x^*_i)
       =\max_i w_{ij}(u_i^*).
\]
The cap gives $p_j^*\ge0$. The displayed equations give exactly the demand inequalities and equality on the full matching. The preceding dummy argument recovers an original equilibrium after deletion and restoration of the reserves and outside utilities.

Every original equilibrium embeds in the same capped square. Greatestness of $x^*$ makes all returned buyer utilities weakly larger than in any original equilibrium. Each $w_{ij}$ is decreasing, so the formula for $p_j^*$ makes each returned item price weakly smaller. This proves both optimality assertions. The $O(bm)$ initial evaluations and reservation-price inversions, and the arithmetic curves on dummy pairs, fit within the stated bound.
\end{proof}

\section{Complete envy-free chore allocation}\label{app:chores}
This appendix explains how the trimming and repair bounds of this paper give the $O(n^7\log n)$ query bound for complete envy-free chore allocation stated in Section~\ref{sec:intro}, using the chore protocol of \citet{Aziz26chores}. Write $c_i$ for agent $i$'s cost measure. A chore allocation $X$ with residue $R$ is envy-free if $c_i(X_i)\le c_i(X_j)$ for all $i,j$, and agent $i$ dominates $j$ if $c_i(X_i)+c_i(R)\le c_i(X_j)$.

\paragraph{Trimming.}
In our construction, cake and chores differ at the witness for a trim: for cake an earlier holder is indifferent to a later suffix, and for chores the later holder is indifferent to an earlier one. Likewise, every outsider dominates the group that divides the remainder for cake, and the group dominates every outsider for chores. For positive inputs, Proposition~\ref{prop:trim-equivalence} transfers the trimming bound to chores with constant overhead per query. For arbitrary nonnegative costs, apply the reduction to the virtual positive curves of Appendix~\ref{sec:ties}. Projection preserves each holder's weak cost comparisons and equality with the earlier suffix that witnesses a chore trim.

\paragraph{Repair.}
The repair step in \citet{Aziz26chores} uses the same prescribed shares and total deficit as Lemma~\ref{lem:shares}. Applying that lemma to the normalized cost measures constructs the $m^2$ shares and buffers for $m$ active agents in $O(m^4\log(m^4/\alpha))$ queries. Since $\log(1/\alpha)=O(m\log m)$, a round, including its trims, costs $O(m^5\log m)$ queries.

\paragraph{Retained bundles and the number of rounds.}
The members of the group that divides the remaining chores can also retain their bundles, as in the \hyperref[par:retained-bundles]{proof of Theorem~\ref*{thm:main}}. For chores, each surviving agent dominates each agent whose bundle is frozen. Write $X,R$ and $W,R'$ for the allocations and residues before and after an update. The new holder $\sigma(j)$ of label $j$ receives $W_{\sigma(j)}=X_j\cup B_j$ with $B_j\subseteq R\setminus R'$. For an active agent $i$, envy-freeness among the active agents gives
\[
 c_i(W_i)+c_i(R')\le c_i(W_{\sigma(i)})+c_i(R')\le c_i(X_i)+c_i(R).
\]
The same bound holds for a frozen agent, whose bundle stays fixed. Thus if $i$ dominated $j$, the right-hand side is at most $c_i(X_j)\le c_i(W_{\sigma(j)})$, so $i$ still dominates the holder of label $j$. In particular, active agents still dominate frozen agents. A frozen agent continues to prefer her own bundle because each active label can only acquire more chores. Each repaired round creates a new domination as in \citet{Aziz26chores}. Counting dominated pairs over the original population gives at most $n^2$ repaired rounds in total and hence $O(n^7\log n)$ queries.

\section{Queries on intervals with established endpoints}
\label{app:query-endpoints}
We justify the implementation described after Theorem~\ref{thm:main}. Let $S$ consist of $0$, $1$, and every point returned by a cut query so far, including trial marks that are not retained in the allocation. Every evaluation is on a single interval $[a,b]$ with $a,b\in S$. Every cut is requested within such an interval, and its returned point is then added to $S$. The endpoints need not be consecutive in $S$.

\emph{Queries on ordered pieces.} Maintain the component records of Section~\ref{sec:preliminaries}, with all component endpoints in $S$. For an ordered piece $P=(I_1,\ldots,I_k)$, write $I_h=[a_h,b_h]$ and $s_r=\sum_{h\le r}v_i(I_h)$, with $s_0=0$. To mark a prefix of value $w$, use an existing component boundary if $w=s_r$ for some $r$. Otherwise choose the component $I_r$ with $s_{r-1}<w<s_r$ and ask agent $i$ to cut from $a_r$ at value $w-s_{r-1}$. Every valid reply lies inside $I_r$, since the requested value lies strictly between zero and $v_i(I_r)$. A suffix mark is obtained by using the complementary prefix value. If a suffix starts at an established point $p\in I_r$, its value is
\[
 v_i([p,b_r])+\sum_{h>r}v_i(I_h).
\]
Thus one evaluation on $[p,b_r]$ suffices, with the other terms already recorded. Empty and whole pieces require no new boundary. These rules also apply when some component values are zero.

\emph{The trimming queries.} In Algorithm~\ref{alg:median}, every candidate suffix length $\hat t_j^M$ is the minimum of the whole-piece length and inverse-query replies on piece $j$. Hence its boundary is an existing component endpoint or a returned cut mark. The length grid selects entries by rank, so every forward query on the physical piece uses an established boundary. The matching test and output recovery use the same inverse queries and minimum operation. Values outside the physical range use the known slope-one extension and require no physical query.

For nonnegative valuations, the simulation in Appendix~\ref{sec:ties} has the same property. Initially the physical images of virtual length keys are the empty and whole positions. A prescribed new key is assigned the physical image of its lower neighboring key. An inverse query obtains a physical cut by the component rule above and, if necessary, clamps its reply to an already registered physical position. Inductively, every physical image is therefore an established boundary. Virtual key insertion introduces no new physical endpoint, and all forward queries obey the required restriction.\looseness=-1

\emph{The remaining operations and the query count.} The cutter's equal shares in \Core{} and the final cut-and-choose step use the same component rule. In \Repair{}, Lemma~\ref{le:weighted} chooses one of the agents' half-value marks, so the selected boundary has already been returned by a cut query. Concatenation, reassignment, and the fractional bookkeeping do not introduce new physical endpoints. At a retained cut, the evaluations recording the new components also have endpoints in $S$.

Adding trial marks to $S$ does not require subdividing the component records at every such mark. A query interval may contain other points of $S$ in its interior. The implementation therefore uses exactly the component evaluations and boundary queries already charged in the analysis, with no additional queries for maintaining $S$. This proves the claimed restriction within the same $O(n^7\log n)$ query bound, for every choice of valid cut answers.

\section{A sixth-power query bound for cake}\label{app:batched-cake}

The outer loop can use several allocations from the same cutter at once.
Exchanging suffixes across these allocations creates a dominated group after
one repair. A second repair, when needed, lets us divide the residue among
the smaller side of the resulting partition.

\begin{proposition}\label{prop:batched-cake}
Under the assumptions of Theorem~\ref{thm:main}, a deterministic protocol
returns a complete, exactly envy-free cake allocation in
$O(n^6\log n)$ Robertson--Webb queries for every choice of valid cut answers.
\end{proposition}
\begin{proof}
One or two agents divide the cake directly. With $m\ge3$ agents, fix a
cutter $r$ and run $\Core$ on $s=m-1$ successive residues with this cutter.
Write $Y_i^t$ for agent $i$'s share in call $t$ and $U$ for the final
residue. The union of these allocations is envy-free. If everyone values
$U$ at zero, give it to any agent and finish. If only some do, take
$W_i=\bigcup_tY_i^t$ and $R=U$. The positive valuers form a dominated
group $B$, so we can proceed directly to the recursion below.

\emph{Exchanging suffixes.}
Suppose every $v_i(U)>0$. Write
$g^t_{ij}=(v_i(Y_i^t)-v_i(Y_j^t))/v_i(U)\ge0$.
For each $t$ choose a noncutter $h_t$ with $g^t_{rh_t}\ge1/s$.
Lemma~\ref{lem:core} gives this because the residue just after call $t$
contains $U$. Put $K=8m^2s$. At threshold $z$ form a bipartite graph
$H(z)$ between the $s$ calls and the $s$ noncutters, joining $t$ to $b$
when $g^t_{bh_t}\le z$. Each $t$ is joined to $h_t$ at every threshold.
The maximum matching size is therefore between $1$ and $s$ and increases
weakly with $z$. At two consecutive thresholds in
$K^{-1},\ldots,K^{-(s+1)}$ the sizes agree.
Denote the smaller threshold by $x$ and the larger by $Kx$.\looseness=-1

In call $t$ draw the arc $i\to j$ when $g^t_{ij}\le x$.
If one of these directed graphs has a proper source strongly connected
component $B$, keep every share with its holder and set
$\alpha=x/(4m^2)$. In that call every outsider has gap greater than $x$
to every member of $B$. The other calls contribute nonnegative gaps.
Thus the resulting bundles $D_i=\bigcup_t Y_i^t$ are envy-free, and
every outsider $i$ has advantage greater than $4m^2\alpha v_i(U)$ over
every $b\in B$.

Otherwise all the directed graphs are strongly connected. A maximum
matching of $H(x)$ is also maximum in $H(Kx)$. If it is perfect, let $B$
be all noncutters. Otherwise search in $H(Kx)$ from an unmatched call,
following edges to noncutters and matched edges back to calls. Let $B$
be the reached noncutters. This set is nonempty, and every member is
matched, since a path to a free noncutter would augment the matching.
In either case the matched calls $t(b)$, $b\in B$, are distinct, and all
neighbors of $t(b)$ in $H(Kx)$ lie in $B$.

In call $t=t(b)$ let $b$ take the share of $h_t$. If $b\ne h_t$, close
the arc $b\to h_t$ by a simple directed path from $h_t$ back to $b$ and
give each agent on this cycle her successor's share. Keep all other
shares with their holders. Every agent loses at most $xv_i(U)$ per call
relative to her original share. For each $b\in B$, every outsider $i$
has gap greater than $Kx$ to the share now on label $b$ in call $t(b)$.
This follows from the neighborhood of $t(b)$ for a noncutter and from
$1/s>Kx$ for the cutter. Set $\alpha=sx$ and combine the shares on
each label into $D_j$. In agent $i$'s eyes every $D_j$ is worth at most
$\sum_t v_i(Y_i^t)$, and her own bundle is within $\alpha v_i(U)$ of
this sum. For an outsider $i$ and $b\in B$, label $b$ falls short of
the sum by more than $Kxv_i(U)$. Thus $i$'s advantage over $b$ exceeds
$(K-s)xv_i(U)>4m^2\alpha v_i(U)$.

\emph{Creating domination.}
In both cases $0<\alpha\le1/(8m^2)$, $\log(1/\alpha)=O(m\log m)$, and
the input envy is at most $\alpha v_i(U)$ for each $i$. Apply
Lemma~\ref{lem:repair} to obtain $W_{\sigma(j)}=D_j\cup Z_j$ and
residue $R$. For $i\notin B$ and $b\in B$, envy-freeness gives
$v_i(W_i)\ge v_i(W_{\sigma(i)})$. The unequal additions cost at most
$2m\alpha v_i(U)$ of the advantage just proved, so
\[
 v_i(W_i)-v_i(W_{\sigma(b)})
 >(4m^2-2m)\alpha v_i(U)\ge v_i(R).
\]
No outsider can therefore hold a label in $B$. By bijectivity
$\sigma(B)=B$, and every outsider dominates every member of $B$.

\emph{Dividing the residue.}
Let $F=N\setminus B$, where $N$ is the current agent set. If
$|B|\le |F|$, divide $R$ recursively among $B$ and add the shares to
their current bundles. Every outsider can tolerate any member of $B$
taking all of $R$. Otherwise put $P=\{i\in B:v_i(R)>0\}$ and $p=|P|$.
If $p\le1$, give $R$ to the sole positive valuer in $B$, or to any member
if $p=0$, and finish. For $p\ge2$, apply $\Repair$ to $R$ alone with
participants $P$, empty input bundles and tolerance $1/(8p^3)$.
Write $A_i$ for the envy-free additions by holder, taking
$A_i=\varnothing$ on $B\setminus P$, and $R'$ for the residue. For $i\in P$,
Lemma~\ref{lem:repair} gives $v_i(R')\le v_i(R)/(4p)$, so
\[
 v_i(A_i)\ge \frac{v_i(R)-v_i(R')}{p}\ge v_i(R').
\]
The same final inequality holds for $i\in B\setminus P$. Adding these
shares preserves envy-freeness within $B$ and makes every member of $B$
dominate every member of $F$. Each $f\in F$ still dominates each $b\in B$,
since $A_b$ and $R'$ are disjoint parts of $R$. We can now divide $R'$
recursively among $F$. Each recursive call divides the new residue alone
and its shares are added to the existing allocation.

\emph{Counting queries.}
Write $T(m)$ for a nondecreasing query bound for one certified trim,
including its retained-boundary records. By Lemmas~\ref{lem:core}
and~\ref{lem:repair}, the cutter calls cost $s(T(m)+O(m^2))$ and the
first repair costs $T(m)+O(m^5\log m)$. The repair on $P$, when used,
costs $T(p)+O(p^4\log p)$. An extra evaluation per other agent at
each newly retained boundary suffices to record its component values
for the next call. Even charging these evaluations at every cut of
the repair on $P$ gives a total stage cost of $O(mT(m)+m^5\log m)$.
All graph and exchange steps use recorded values. Fixed orders resolve
all choices, and all physical queries use the component rules of
Appendix~\ref{app:query-endpoints}.

There is at most one recursive call, with at most $\lfloor m/2\rfloor$
agents. Each stage uses at most $m+1$ trims, giving $O(n)$ over the whole
execution. Summing the stage costs over the halving populations gives
$O(nT(n)+n^5\log n)$ queries, including the initial value records.
Lemma~\ref{lem:trim} gives $T(n)=O(n^5\log n)$, as required.
\end{proof}

\end{document}